\documentclass[twocolumn]{article}

\usepackage[
left=0.7in,
right=0.7in,
top=0.7in,
bottom=0.7in,
]{geometry}

\usepackage[utf8]{inputenc}
\usepackage[T1]{fontenc}
\usepackage[english]{babel}
\usepackage[colorlinks=true,linkcolor=blue,citecolor=blue,urlcolor=blue]{hyperref}

\usepackage{graphicx}
\usepackage{amsmath}
\usepackage{booktabs}
\usepackage{subcaption}
\usepackage{multirow}
\usepackage{float}
\usepackage{tikz}
\usetikzlibrary{calc}
\usepackage{makecell}
\usepackage{nicematrix}
\usepackage{xcolor,colortbl}
\usepackage{arydshln}
\usepackage{placeins}
\usepackage{orcidlink}
\usepackage{todonotes}

\usepackage[switch]{lineno}
\definecolor{my-gray}{RGB}{220,220,220}
\definecolor{lightgray}{gray}{0.9}
\usepackage{titlesec}

\titleformat{\section}
{\normalfont\bfseries\large}   
{\thesection}{0.5em}{}
\titleformat{\subsection}
{\normalfont\bfseries\normalsize}
{\thesubsection}{0.5em}{}

\DeclareUnicodeCharacter{0394}{$\Delta$}

\usepackage[
backend=bibtex,
style=nejm,
citestyle=numeric-comp,
sorting=none
]{biblatex}

\newcommand{\teaser}[1]{
\begin{center}
#1
\end{center}
\vspace{0.5em}
}

\newcommand*{\belowrulesepcolor}[1]{%
\noalign{
\kern-\belowrulesep
\begingroup
\color{#1}\hrule height\belowrulesep
\endgroup
}
}

\newcommand*{\aboverulesepcolor}[1]{%
\noalign{
\begingroup
\color{#1}\hrule height\aboverulesep
\endgroup
\kern-\aboverulesep
}
}

\title{TreeON: Reconstructing 3D Tree Point Clouds from \\ Orthophotos and Heightmaps}

\author{
\parbox{\textwidth}{\centering
\large
Angeliki Grammatikaki$^{1,2}$\orcidlink{0000-0001-5779-2182},
Johannes Eschner$^{1}$\orcidlink{0009-0001-6784-8503},
Pedro Hermosilla$^{1}$\orcidlink{0000-0003-3586-4741}, \\[4pt]
Oscar Argudo$^{3}$\orcidlink{0000-0003-3943-1839},
Manuela Waldner$^{1}$\orcidlink{0000-0003-1387-5132}
}
\\[6pt]
\parbox{\textwidth}{\centering
\small\itshape
$^1$TU Wien, Institute of Visual Computing and Human-Centered Technology, Vienna, Austria\\[2pt]
$^2$VRVis GmbH, Vienna, Austria\\[2pt]
$^3$Universitat Polit\`ecnica de Catalunya, Department of Computer Science, Barcelona, Spain
}
}

\date{}

\begin{document}

\twocolumn[
\maketitle

\teaser{
\includegraphics[width=\textwidth]{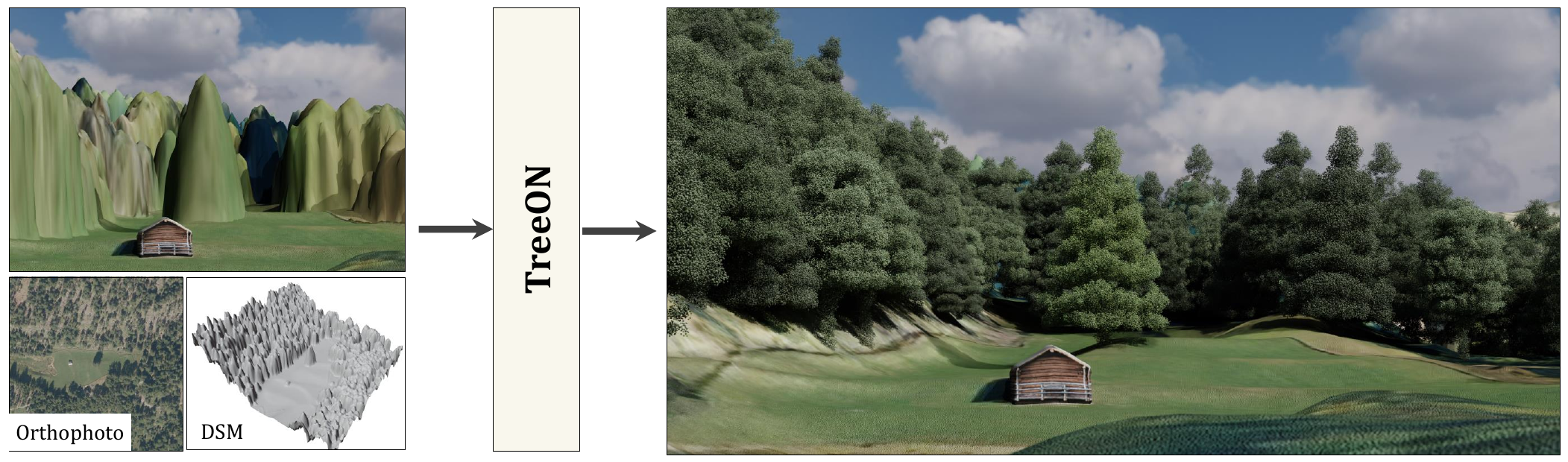}
\captionof{figure}{Our neural-based framework, TreeON, reconstructs coherent tree point clouds for trees in real datasets using only an orthophoto and a Digital Surface Model (DSM).}
\label{fig:teaser}
}

\begin{abstract}
We present TreeON, a novel neural-based framework for reconstructing detailed 3D tree point clouds from sparse top-down geodata, using only a single orthophoto and its corresponding Digital Surface Model (DSM). Our method introduces a new training supervision strategy that combines both geometric supervision and differentiable shadow and silhouette losses to learn point cloud representations of trees without requiring species labels, procedural rules, terrestrial reconstruction data, or ground laser scans. To address the lack of ground truth data, we generate a synthetic dataset of point clouds from procedurally modeled trees and train our network on it. Quantitative and qualitative experiments demonstrate better reconstruction quality and coverage compared to existing methods, as well as strong generalization to real-world data, producing visually appealing and structurally plausible tree point cloud representations suitable for integration into interactive digital 3D maps. The codebase, synthetic dataset, and pretrained model are publicly available at \url{https://angelikigram.github.io/treeON/}.
\end{abstract}

\vspace{1em}
]

\section{Introduction}
Digital 3D maps are widely used to communicate spatial and environmental information, offering intuitive and immersive representations of real landscapes~\cite{sheppard2005landscape}. Unlike urban settings, where terrestrial scans~\cite{bostrom_urban_2006} or multi-view photogrammetry~\cite{snavely_photo_2006} are often used to generate detailed 3D models, high-resolution datasets remain limited or region-specific in natural environments. In rural areas, digital 3D maps are therefore mostly generated from aerial imagery, including orthophotos combined with Digital Surface Models (DSMs; raster heightmaps encoding terrain elevation and above-ground objects such as vegetation and buildings).. As a result, rural 3D maps are, in fact, often limited to a 2.5D surface representation, which lacks detail in close-up oblique views. This limitation is most evident for vegetation, where fine structural detail is lost (see Figure~\ref{fig:teaser} left) , which is perceived as non-appealing~\cite{Grammatikaki09052025}.

We hypothesize that much of the missing 3D information for vegetation, especially for isolated trees, can be inferred from the currently available 2.5D inputs. This would enable lightweight yet visually appealing reconstructions without requiring additional data collection. Such an approach could address a key design dilemma: how can we create tree models that are both visually compelling and computationally lightweight, and hence scalable enough for large outdoor environments, and effective in data-sparse conditions?

This hypothesis is motivated by findings from studies in virtual environments, which show that trees with distinctive shapes not only improve perceived realism but also facilitate orientation and wayfinding, particularly when they serve as recognizable landmarks~\cite{Jaeger}. Just as buildings and monuments serve as key reference points in urban navigation, trees can provide essential visual anchors in natural terrain. Although simplified vegetation can still serve as landmarks, prior work shows that users strongly prefer the visual appeal of more detailed models and perceive them as more reliable for navigation~\cite{Grammatikaki09052025}.

Prior methods for vegetation reconstruction have relied on data-rich techniques involving dense LiDAR point clouds~\cite{xu_knowledge_2007, du_adtree_2019}, multi-view terrestrial or aerial imagery~\cite{shlyakhter_reconstructing_2001, indirabai_terrestrial_2019, you_tree_2021, Hu}, photogrammetry and remote sensing approaches~\cite{choudhury_photogrammetry_2019}, or biologically-inspired procedural modeling based on species-specific rules~\cite{Xu}. While these approaches can produce structurally detailed results, they tend to be computationally expensive, difficult to scale, and reliant on specialized data collection. Consequently, they are ill-suited for wide-area deployment in natural landscapes, where acquiring such detailed datasets is often infeasible. 
This motivates the need for lightweight, scalable solutions that can generate realistic vegetation models from standard geodata sources alone.

An ideal method for large-scale rural maps should satisfy several requirements:
(1) high visual appeal in close-up oblique views, (2) computational efficiency for large-scale deployment (particularly for rendering), (3) structural plausibility for different tree types, and (4) automation without the need for additional input data, like species information, photographs, or user sketches.
These requirements reveal why existing methods fall short: some achieve realism but are inefficient or data-hungry, while others are lightweight yet lack plausibility. 

To fill this gap, we propose a novel approach that reconstructs visually plausible 3D tree point clouds
from sparse geodata by exploiting multiple complementary cues present in DSMs and orthophotos. 
These cues include canopy shape, overall height, and height-to-width ratio from the DSM, as well as canopy texture and shadows from the orthophoto. 
Our pipeline constructs a 3D point cloud representation of each tree from a single orthophoto and its corresponding DSM.
We adopt a point cloud output as a pragmatic, generator-agnostic representation for large-scale 3D mapping, avoiding inference of parameters tied to a specific procedural model, while still allowing higher-level structure to be recovered from the point cloud when needed (TreeStructor~\cite{10950450}).

Our method integrates geometric and visual cues via a PointNet-style terrain encoder and a CNN-based orthophoto encoder, with a neural decoder predicting occupancy at arbitrary 3D query points. 
The training process is supervised through five complementary loss terms: a binary cross-entropy loss to enforce structural plausibility, a color consistency loss to match orthophoto appearance, and three differentiable perceptual losses~\cite{zhang2018perceptual} that supervise alignment with observed shadows, and top and side view silhouettes. 
This multi-modal supervision enables the model to infer plausible and expressive tree shapes, yielding realistic reconstructions in environments where more detailed 3D data is unavailable.
In summary, our work makes three main contributions:
\begin{enumerate}
    \item A synthetic dataset of procedurally generated trees with paired orthophotos and DSMs,
    \item A network architecture for reconstructing detailed 3D tree point clouds from sparse top-down geospatial data, and 
    \item A training supervision strategy that combines occupancy prediction with shadow- and silhouette-based losses. 
\end{enumerate}
In addition, we demonstrate through quantitative and qualitative evaluation that, with these contributions, our approach fulfills the requirements of realistic and visually appealing tree reconstruction better than alternative architectures and supervision strategies.

\section{Related Work}

Modeling realistic trees has been central to both computer graphics and remote sensing. Traditional methods often depend on dense 3D terrestrial laser scans to extract explicit branch skeletons and generate foliage-rich models \cite{xu_knowledge_2007}. However, such data are typically unavailable in remote or rural areas, where only orthophotos and elevation models exist.
These methods satisfy structural plausibility and visual appeal, but fail scalability and automation due to costly data acquisition.

To overcome the need for rich input data, researchers have explored multi-view image reconstructions, including volumetric and point cloud assemblies \cite{shlyakhter_reconstructing_2001, bradley_image-based_2013}, and sketch-based techniques requiring user guidance \cite{tan_image-based_2007}. Argudo et al.~\cite{argudo_single-picture_2016} developed a single-image method to infer plausible structure, but it struggles under significant occlusion. All these approaches either rely on multi-view imagery or human intervention, limiting scalability to large, sparsely mapped landscapes.
Thus, while they improve realism, they fall short in efficiency and automation.

Procedural generation synthesizes tree forms from parametric or stochastic growth models~\cite{yang2022rulebasedproceduraltreemodeling}, which can be inverted from visual cues~\cite{inverse_procedural_modelling} or placed into landscapes using vegetation classification from aerial images~\cite{ARGUDO201823, rs16030524}. However, such classification is often inaccurate or unreliable from aerial data alone~\cite{Quan31122023, rs16203836}.
As a result, these methods may generate highly detailed but biologically implausible tree models, which risks creating unfounded trust in the reconstructions despite low realism. While valuable for producing training data and large-scale vegetation, their realism often depends on species-specific templates or predefined rules, limiting flexibility compared to data-driven reconstruction.
In terms of requirements, procedural models are efficient and automated, but often lack structural plausibility and visual appeal without species priors.

A few works have addressed tree reconstruction using minimal geospatial input data. Earlier approaches include Mayr et al.~\cite{10.1007/978-3-642-60243-6_12}, who fitted ellipsoidal primitives to DSM data to model isolated rural and leafless urban trees, and Hirschmugl et al.~\cite{HIRSCHMUGL2007533}, who combined DSM-based height peaks with spectral cues from orthophotos for tree detection. 
More recently, Grammatikaki et al.~\cite{Grammatikaki09052025} proposed a fully automatic pipeline that models landmark trees directly from DSM and orthophotos, producing varying levels of detail. 
Though insightful, these approaches either approximate trees with coarse shapes or detect trees rather than reconstructing full per-tree structures. 
They therefore only partially meet structural plausibility and do not provide the visual appeal required for landmark-level detail.

In response to these limitations, general single-view reconstruction and view-synthesis methods, such as NeRF-Diff~\cite{nerfdiff} and EscherNet~\cite{eschernet}, as well as recent text- and image-conditioned generative models for object- and scene-level 3D synthesis~\cite{3dscenesurvey}, including DreamFusion~\cite{dreamfusion}, Magic3D~\cite{magic3d}, ProlificDreamer~\cite{prolificdreamer}, Zero123++~\cite{zero123plus}, 3DTopia-XL~\cite{3dtopiaxl}, and TRELLIS~\cite{trellis}, address related 3D inference problems but assume perspective imagery and object- or scene-centric priors, making them unsuitable for geospatial reconstruction from top-down orthophotos and DSMs.

Recent years have seen the rise of neural network-driven vegetation modeling methods that directly target tree and forest reconstruction. DeepTrees~\cite{zhou2023deeptreemodelingtreessituated} models tree growth using situated latent representations that capture spatial and structural dependencies during a biologically inspired generative process. Foliager~\cite{10.1145/3721250.3743024} procedurally generates forest scenes from natural language and ecological data, combining AI planning with scientific growth rules. In parallel, neural radiance field (NeRF)-based methods have been adapted to vegetation~\cite{huang2023evaluatingpointcloudindividual}. They aim to reconstruct individual canopies from sparse aerial views, though the results often remain noisy and miss fine branching detail. Diffusion-based pipelines, such as Tree-D Fusion~\cite{lee2024treedfusionsimulationreadytree} and SVDTree~\cite{10656708}, produce high-fidelity branch-leaf structures from a single image by leveraging learned semantic voxel encodings and generative priors. 

In contrast, TreeStructor~\cite{10950450} introduces a retrieval-and-ranking approach that assembles forest reconstructions from a library of neural components. While these approaches significantly advance the realism of tree modeling, they often require multi-view imagery, genus-level priors, or dense semantic cues-factors that are unavailable in our single-view, label-free geospatial setting, and, in the case of species classification, unreliable~\cite{FASSNACHT201664}.
While these methods strongly address realism and plausibility, they typically require multi-view inputs, species priors, or dense semantic cues, limiting their scalability and automation from sparse rural geodata.

In contrast, our work addresses this gap by reconstructing spatially explicit 3D tree forms from minimal geospatial input, aiming to satisfy all four requirements simultaneously: visual appeal, efficiency, structural plausibility, and automation, without relying on semantic labels, species priors, or dense image coverage. 

\section{Overview}

Our method integrates synthetic data generation, multi-modal neural networks, tailored supervision, and postprocessing for accurate 3D tree reconstruction. 
We generate a training dataset with aligned DSMs, orthophotos, and ground-truth colored point clouds from procedural trees. 
These inputs are encoded by image and point cloud backbones and decoded into occupancy and per-point color. 
Supervision combines geometric, photometric, and projection-based losses, while postprocessing densifies and renders the final reconstructions.

\subsection{Synthetic Training Dataset Generation Pipeline}

\begin{figure}[h]
    \centering
    \begin{subfigure}[t]{0.23\linewidth}
        \centering
        \includegraphics[width=\linewidth]{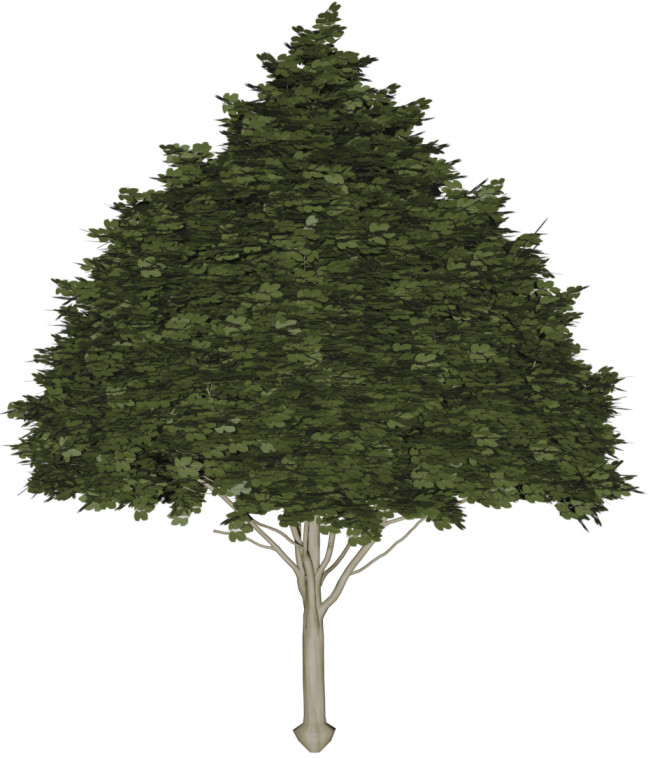}
        \caption{}
        \label{fig:mesh}
    \end{subfigure}
    \begin{subfigure}[t]{0.23\linewidth}
        \centering
        \includegraphics[width=\linewidth]{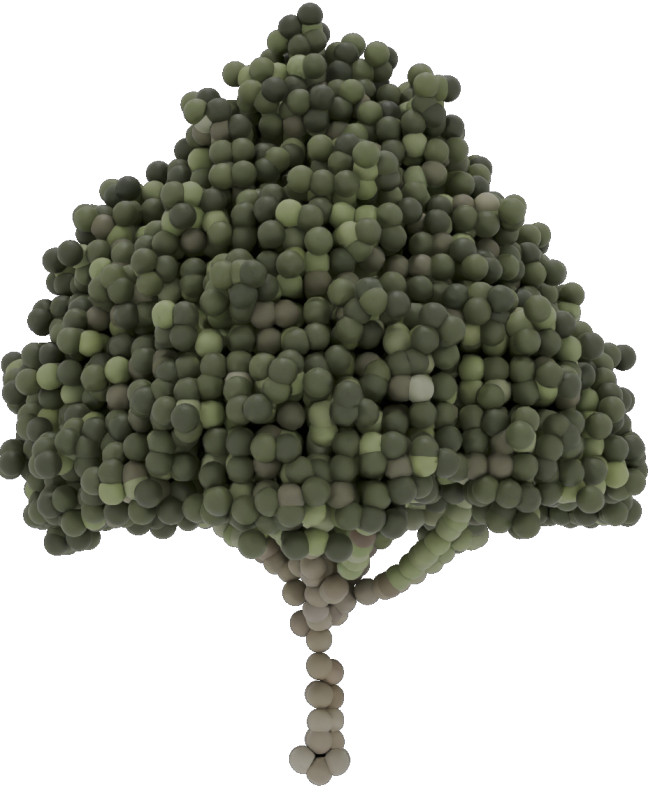}
        \caption{}
        \label{fig:pointcloud}
    \end{subfigure}
    \begin{minipage}[t]{0.02\linewidth}
        \centering
        \begin{tikzpicture}
            \draw[dotted, thick] (0,0) -- (0,3.0); 
        \end{tikzpicture}
    \end{minipage}
    \begin{subfigure}[t]{0.23\linewidth}
        \centering
        \includegraphics[width=\linewidth]{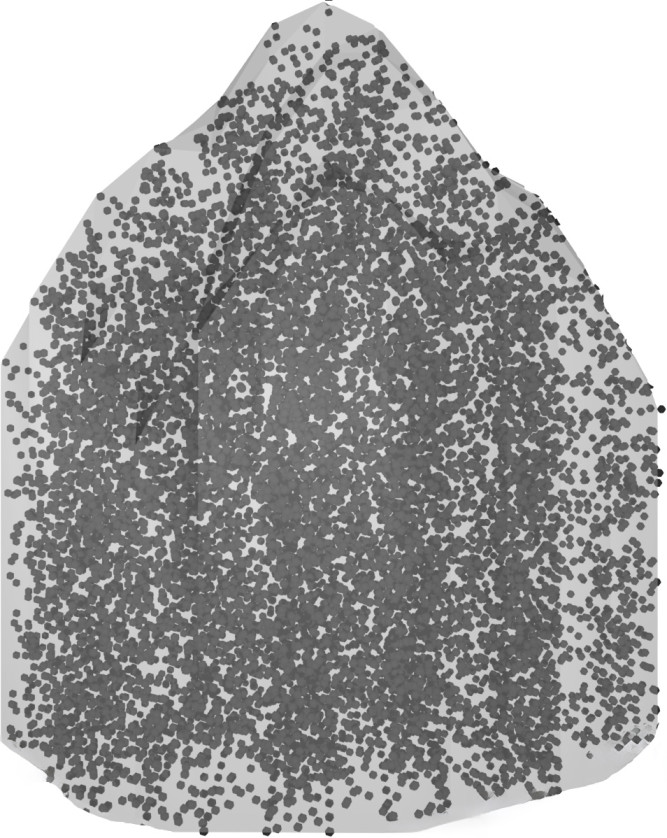}
        \caption{}
        \label{fig:dsm}
    \end{subfigure}
    \begin{subfigure}[t]{0.23\linewidth}
        \centering
        \raisebox{3mm}{\includegraphics[width=\linewidth]{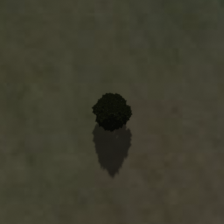}}
        \caption{}
        \label{fig:orthophoto}
    \end{subfigure}

    \caption{Tree generation pipeline starting from (a) procedural mesh: sampled colored point cloud (b), artificial DSM point cloud (visualized as surface with interior points) (c) and orthophoto (d).}
    \label{fig:data_gen_pipeline}
\end{figure}

Whilst our method relies on a multi-modal supervision, such paired data is rarely available in real-world settings.
Therefore, we construct a diverse synthetic dataset tailored for our 3D tree reconstruction. 
Each training instance includes three aligned modalities: a colored point cloud representing a 3D tree (Figure~\ref{fig:data_gen_pipeline}), a Digital Surface Model (DSM), and a top-down orthophoto. 

\paragraph*{Tree Generation} 

Our tree generation pipeline begins with a procedurally generated tree mesh created using Blender's Grove add-on~\cite{grove} (Figure~\ref{fig:data_gen_pipeline} (a)), although other modeling tools could be used. Each tree mesh contains inline vertex colors, which encode species-specific bark and crown appearance. For training, we require point clouds that are compact enough for efficient learning yet still representative of detailed tree geometry and color. We therefore apply Poisson-disk sampling~\cite{10.1145/1278780.1278807} to the mesh vertices, which yields uniformly distributed samples across the crown and trunk. This reduces each mesh to approximately $K \approx 6{,}000$ points, which we found to be a good trade-off between geometric fidelity and computational efficiency. Larger values of $K$ increased memory usage without improving reconstruction quality, while smaller values reduced structural coverage. The resulting colored point cloud serves as the \emph{ground truth target} in training (Figure~\ref{fig:data_gen_pipeline} (b)).

\begin{figure}[b!]
  \centering
\includegraphics[width=\linewidth]{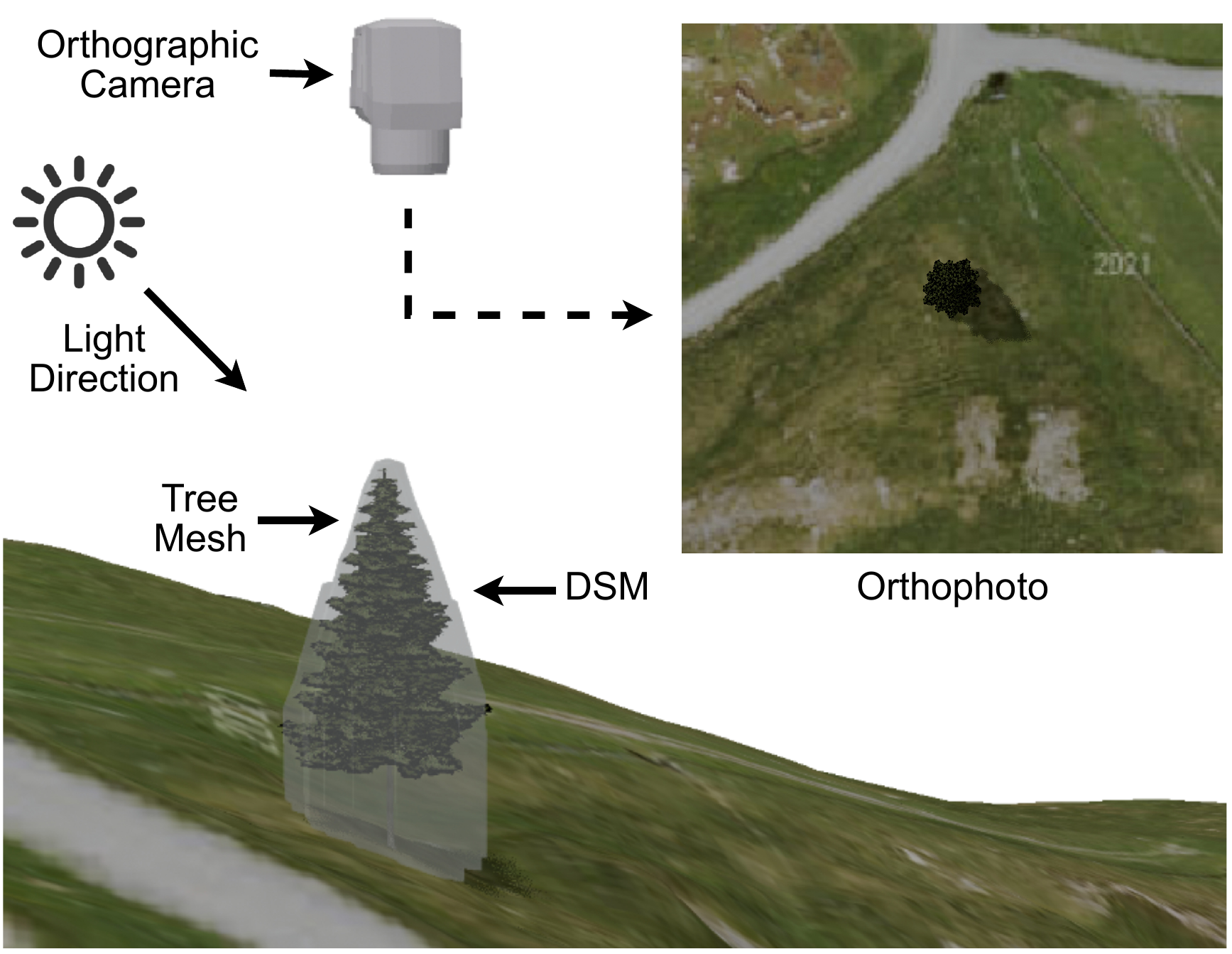}
    \caption{Orthophoto and DSM generation from a top-down orthographic camera. Orthophotos include trees and shadows under directional lighting, while DSMs DSM (visualized as a surface for illustration) are derived from per-pixel depth. Sun positions vary in azimuth and elevation.}
  \label{fig:dsm_ortho_gen}
\end{figure}

\paragraph*{Orthophoto and DSM Rendering} 
For each synthetic tree, we aim to replicate the data modalities commonly available from aerial imagery and airborne LiDAR.
Therefore, we render orthophotos and DSMs from the detailed tree mesh (Figure~\ref{fig:data_gen_pipeline} (a)).
As illustrated in Figure~\ref{fig:dsm_ortho_gen}, we generate data by rendering the processed 3D mesh in Blender from a controlled top-down view. 
For realism, trees are placed on terrain patches sampled from high-resolution DEMs, ensuring diverse contexts such as flatland, hilly, or forested areas. Locations were selected such that terrain features (e.g., valleys, clearings) support plausible vegetation growth, while regions with unsuitable land cover (e.g., water bodies, snow) were avoided. 
To avoid shadow mismatches during supervision, any visible surrounding vegetation is removed using an inpainting neural network~\cite{yu2023inpaint}.
In some scenes, we placed additional synthetic trees around the central target tree to create more realistic orthophotos with surrounding vegetation. 
A top-down orthographic camera is then positioned directly above the target tree, aligned along the $Z$ axis, and centered at the tree's bounding box origin. 
The orthographic scale is fixed by the camera height, centering the tree crown while including surrounding terrain for context. 
We sample ten light directions per tree to diversify shadow cues during training. 
Azimuth angles are drawn from $[0^\circ, 360^\circ)$, while elevation angles are restricted to $[40^\circ, 70^\circ]$, consistent with typical orthophoto acquisitions around mid-day in summer, yet providing sufficient variation for robust shadow supervision. 

The \textbf{orthophotos} are rendered at a configurable resolution consistent with aerial imagery standards.
These images simulate top-down remote sensing captures (e.g., aerial photography or satellite imagery). 
For each condition, the orthographic top-down camera renders the tree together with its cast shadow on the ground plane. 
Shadows are generated in Blender's Cycles renderer with a directional light source, where the sun's angular diameter $\alpha \sim U(0.5, 10)$ controls the penumbra size, producing both hard and soft shadows.
Shadow masks encode occlusion intensity ($0 =$ fully shadowed, $1 =$ illuminated), with Gaussian noise, channel variations, and clipping applied for realism following Morales et al.~\cite{10.1007/978-3-030-11680-4_31}.
The resulting orthophotos thus capture naturalistic shadow transitions and crown textures under varying illumination.
For each rendered orthophoto, the corresponding sun direction vector is also recorded and stored in a metadata file. 
This allows us to consistently reproduce shadows under the same lighting setup during supervision.

To produce the \textbf{DSM}, we emulate a LiDAR-like surface scanning process. 
Specifically, we extract the $Z$-buffer from the same orthographic camera view and convert per-pixel depth values into real-world elevation. 
This is done by subtracting the depth from the camera's world-space height, resulting in a per-pixel heightmap that captures the vertical profile of all visible geometry.
The heightmap is normalized by subtracting the local ground elevation and scaling heights to the unit range [0,1], ensuring consistent DSM encoding while preserving relative crown geometry.
Each DSM effectively represents the tree's canopy shape and elevation in image form, matching the modality of publicly available DSMs. The resolution of the DSM rendering is configurable and reported for each dataset in Section~\ref{sec:experiments}.

For use in the network, we back-project DSM pixels into 3D, yielding a point cloud representation. This preserves sharp height discontinuities and explicit 3D geometry, and aligns naturally with point-based encoding and occupancy supervision at arbitrary 3D query locations.
We sample uniformly within the DSM volume until reaching about $K \approx 6{,}000$ points per tree to obtain a fixed-size and stable PointNet input. Fewer points led to unstable gradients, while larger sets increased memory and computation without measurable performance gains.

\begin{figure*}[t]
    \centering
    \includegraphics[width=0.95\textwidth]{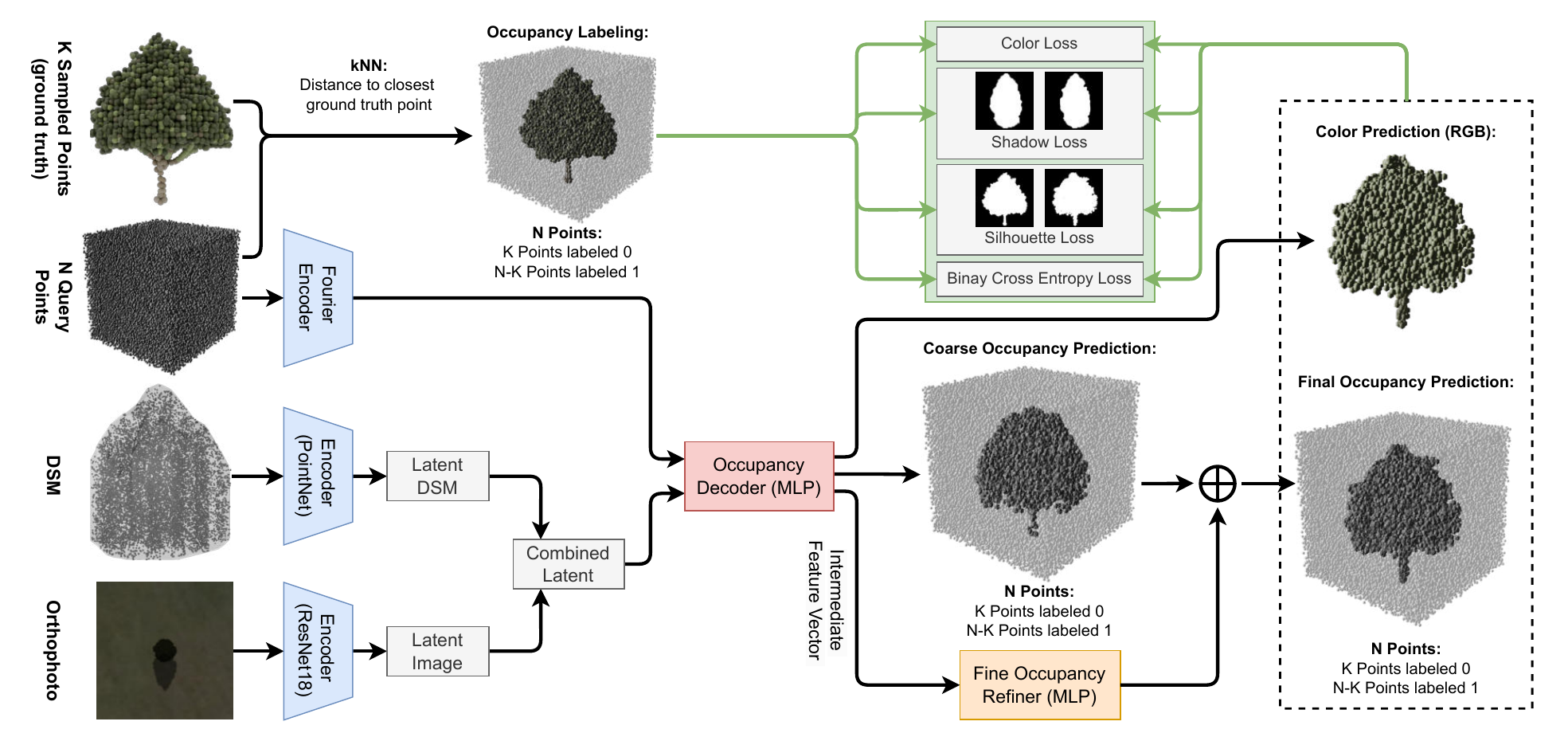} 
    \caption{Overview of the network architecture and training supervision (green).}
    \label{fig:network_architecture}
\end{figure*}

\subsection{Network Architecture}

The model processes three inputs: an RGB orthophoto of size $3 \times 224 \times 224$, a DSM-based point cloud of shape $K \times 3$, and a set of $N$ 3D query points sampled within a unit cube $[0,1]^3$ (see Figure~\ref{fig:network_architecture}). 
We fix $K \approx 6{,}000$ points, following the voxel-based downsampling in our synthetic dataset (Section~3.1). 
The number of query points is set to $N=25{,}000$, which we found sufficient to balance coverage of inside/outside regions with computational tractability.
For each query point, the model predicts whether it lies inside or outside the tree surface, while also estimating per-point RGB appearance. 
This implicit formulation avoids using a fixed voxel grid and allows for continuous shape reconstruction. 
To ensure supervision across both occupied and free-space regions, query points are sampled to balance inside/outside labels rather than concentrating only near the DSM surface.

The orthophoto is encoded using a ResNet-18~\cite{he2015deepresiduallearningimage} backbone pre-trained on ImageNet. 
Intermediate features are pooled and passed through a projection head with batch normalization~\cite{ioffe2015batchnormalizationacceleratingdeep} and LeakyReLU activation~\cite{Maas2013RectifierNI}, yielding a 1024-dimensional image embedding normalized via $\ell2$ norm. 
Simultaneously, the DSM input represented as a sparse 3D point cloud is encoded using a PointNet-style~\cite{qi2017pointnetdeeplearningpoint} architecture with 1D convolutions (64, 128, 1024 channels), batch normalization, and LeakyReLU activations. 
The resulting per-point features are aggregated via max-pooling into another 1024-dimensional latent vector, also $\ell2$-normalized.
This design leverages the strengths of each encoder: ResNet extracts texture and shadow cues from 2D orthophotos, while PointNet captures the unordered 3D structure of DSM point clouds, aligning with the occupancy formulation. 
Then, the image and point cloud embeddings are concatenated into a 2048-dimensional latent representation. 
The fused image--DSM latent is injected into every decoder layer together with Fourier-encoded query points, conditioning local occupancy predictions on global context.

Each 3D query point is passed through a Fourier positional encoder~\cite{tancik2020fourierfeaturesletnetworks} with 25 frequency bands, producing a 153-dimensional representation. 
Both the global latent vector from the image and point cloud encoders and the Fourier-encoded query points are then provided as inputs to the occupancy decoder, where they are combined internally, yielding an effective per-point input size of 2201. 
The occupancy decoder is a deep multilayer perceptron with six fully connected layers, each incorporating Layer Normalization~\cite{ba2016layernormalization}, dropout~\cite{dropout}, LeakyReLU activation, and residual connections~\cite{he2015deepresiduallearningimage}.  It branches into two output heads: one predicts a scalar occupancy logit for each query point, and the other predicts a 3-channel RGB color vector by combining the learned features with the original Fourier-encoded input. 

\begin{figure}[h]
  \centering
  \begin{subfigure}{0.2\linewidth}
    \includegraphics[width=\linewidth, height=3.0cm]{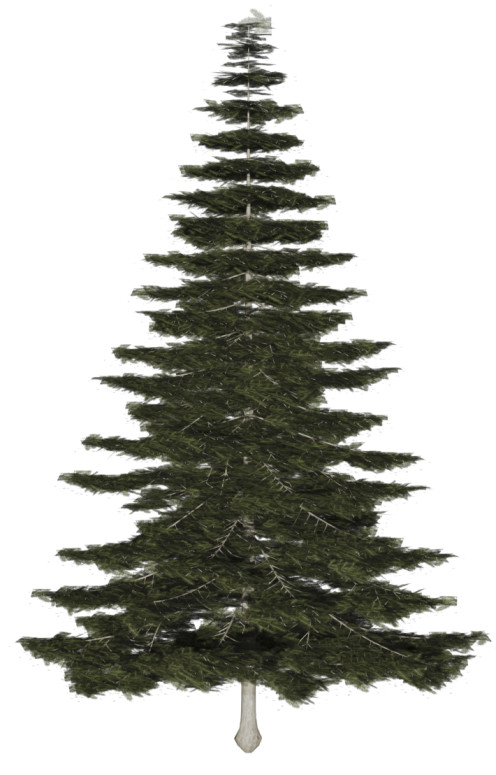}
    \caption{}
  \end{subfigure}
  \hfill
  \begin{subfigure}{0.2\linewidth}
    \includegraphics[width=\linewidth, height=3.0cm]{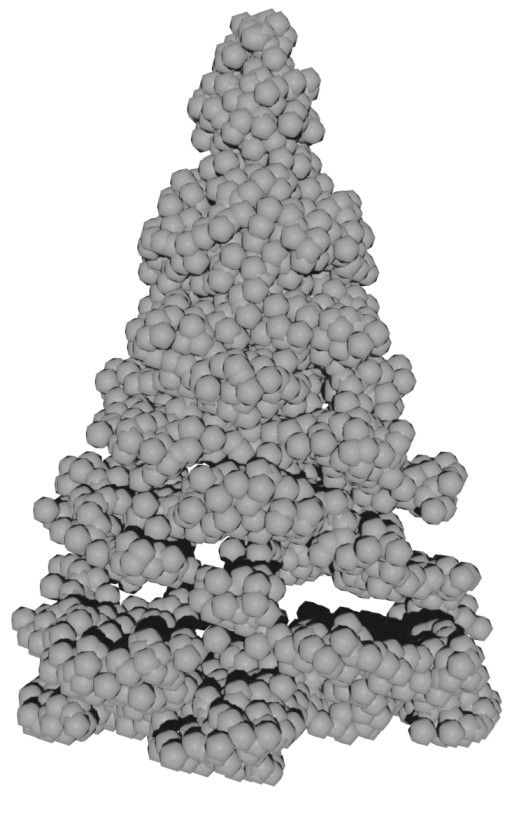}
    \caption{}
  \end{subfigure}
  \hfill
  \begin{subfigure}{0.2\linewidth}
    \includegraphics[width=\linewidth, height=3.0cm]{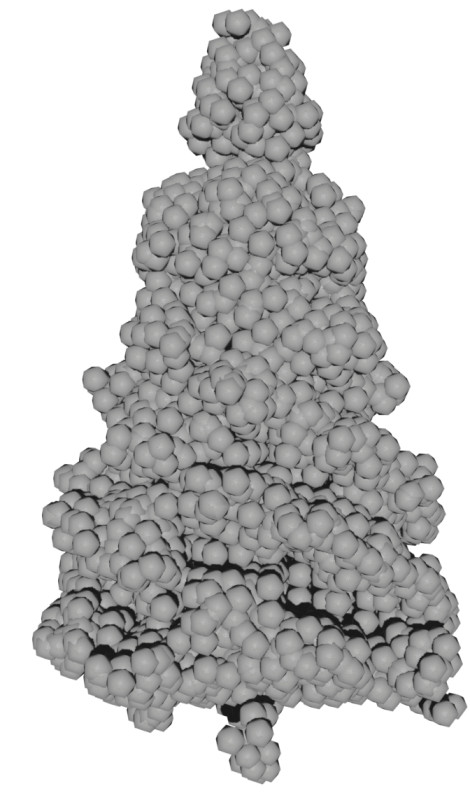}
    \caption{}
  \end{subfigure}
    \caption{(a) Ground-truth tree, (b) reconstruction without refinement, and (c) reconstruction with refinement.}
  \label{fig:refinement_ablation}
\end{figure}

During initial experiments, we observed that the coarse MLP decoder tended to produce overly smooth occupancy transitions at the tree surface, failing to capture crisp details such as fine-branch boundaries (see Figure~\ref{fig:refinement_ablation}). 
To address this, we integrated a lightweight refinement module: three small residual MLP layers that compute a corrective delta added to the coarse occupancy logits. 
This module acts as a local corrector: it reduces false positives and negatives near the surface boundary and recovers fine-scale geometric details that may be lost in the coarse prediction, while keeping the main decoder compact. 
We refine only geometry since the main bottleneck is surface sharpness, while color already benefits from dense per-point supervision and does not suffer from the same over-smoothing problem. 
Similar architectures have been adopted in related contexts, e.g., attention-based refinement modules improve surface fidelity in human reconstruction tasks~\cite{du_modeling_2024}, and progressive refinement was proposed for occupancy prediction in diffusion-style decoders~\cite{wang2024occgengenerativemultimodal3d}.

The final outputs of the model are the refined occupancy logits over the query points and per-point RGB color predictions. 
This architecture supports end-to-end learning of both geometry and appearance, conditioned on visual input and terrain structure. 
An overview is shown in Figure~\ref{fig:network_architecture}.

\subsection{Supervision Strategy}

We employ a combination of point-level, color, and projection losses to jointly supervise geometry and appearance. 
Point-level losses enforce correct occupancy at sampled 3D coordinates, while a color loss supervises the predicted per-point RGB values. 
Projection-based supervision uses differentiable shadow and silhouette rendering, complemented with a perceptual LPIPS loss~\cite{zhang2018perceptual} on the silhouettes to capture structural differences beyond pixel-level matching.

We supervise the model at the point level using a binary cross-entropy (BCE) loss $\mathcal{L}_{\text{occ}}$, a standard choice for implicit occupancy networks~\cite{mescheder2019occupancy, peng2020convolutionaloccupancynetworks}. 
This loss penalizes discrepancies between predicted occupancy logits and ground-truth labels, guiding the network to distinguish occupied (inside) from free-space (outside) points.
To obtain ground-truth occupancy labels for this loss, first, we compute the distance to the nearest neighbor in the reference point cloud. 
Points closer than a threshold to a ground-truth sample are then labeled as occupied (1), while all others are labeled as free space (0). 
Given the flat structure of leaves and the thinness of branches, tree meshes typically lack significant volume, making this a reasonable approximation.
Additionally, we supervise the predicted color of each point with a standard L2 loss.

While BCE is widely adopted in prior point cloud and implicit surface reconstruction methods, our contribution lies in extending this supervision with novel projection-based constraints. 
Specifically, we introduce differentiable shadow and silhouette projections to regularize tree structure in ways not captured by point-level occupancy alone.
For shadow supervision, the occupancy values of query points are projected into the top-down orthophoto plane through a soft differentiable renderer~\cite{liu2019softras}, yielding a shadow that encodes the 3D shape along the illumination direction.
The shadow loss $\mathcal{L}_{\text{shadow}}$ compares using LPIPS the predicted shadow against a ground-truth shadow rendered under the same lighting direction as the orthophoto. 
This encourages alignment with the real shadow observed in the top-down imagery.
For silhouettes, projections are generated from five canonical viewpoints ($0^\circ$, $45^\circ$, $90^\circ$, $135^\circ$, and top view) to constrain the global crown outline across viewpoints, independent of lighting.
The silhouette loss $\mathcal{L}_{\text{silh}}$ constrains the global structure by aligning the predicted and ground-truth projections using the LPIPS loss.

The final training loss $\mathcal{L}_{\text{total}}$ is defined as a weighted sum of these four components:
\begin{equation}
\mathcal{L}_{\text{total}} = \lambda_{\text{occ}} \mathcal{L}_{\text{occ}} + \lambda_{\text{col}} \mathcal{L}_{\text{col}} + \lambda_{\text{sh}} \mathcal{L}_{\text{sh}} + \lambda_{\text{silh}} \mathcal{L}_{\text{silh}}.
\label{eq:total_loss}
\end{equation}
\noindent where $\mathcal{L}_{\text{occ}}$ is the occupancy loss, $\mathcal{L}_{\text{col}}$ is the color loss, and $\mathcal{L}_{\text{silh}}$ is the silhouette loss.
The weights $\lambda_{\text{occ}}, \lambda_{\text{col}}, \lambda_{\text{sh}}, \lambda_{\text{silh}}$ control the relative influence of each term. 
In our experiments, we use $\lambda_{\text{occ}} = 0.14$, $\lambda_{\text{col}} = 0.71$, $\lambda_{\text{sh}} = 0.07$, and $\lambda_{\text{silh}} = 0.08$, since we found them to produce the best trade-off between detail, smoothness, and structural plausibility.
These combined losses guide the model to produce reconstructions that are not only geometrically accurate but also visually consistent with natural shadows and silhouettes. 

\subsection{Postprocessing and Rendering}

While during training, we query random points within the volume, at inference time, we only output a set of points inside the tree surface.
We achieve this by ranking all query points by their predicted occupancy probability and retaining the top-$K$ points (highest logits after sigmoid). 
This yields a compact fixed-size reconstruction aligned with the ground-truth size ($K \approx 6{,}000$), while discarding low-confidence points. 
This step is non-differentiable and used only for visualization and downstream evaluation.
Note that while thresholding on occupancy might be an alternative, it produces highly variable point counts across trees.
Our approach, on the other hand, ensures consistent reconstructions and fair comparisons.

Additionally, to improve visual density while preserving predicted colors, we upsample the sparse set of occupied points. This is done by discretizing the point cloud into a volumetric representation via a voxel grid and then randomly resampling the volume at a higher density. Both the resampling density and the number of voxels are dynamically derived from the reconstructed tree's volume, computed via a convex-hull approximation of the original point cloud. 
This ensures that larger trees receive more points, yielding reconstructions that are both scale-aware and visually consistent. During upsampling, the original \texttt{Color} attribute is propagated: new points inherit interpolated colors from their nearest neighbors in the original set. This preserves the canopy's spatially coherent texture while adapting the point density to the tree's size. 

For rendering, each point is visualized as a shaded sphere with its assigned radius and color, producing a dense and natural appearance that conveys both the reconstructed geometry and its authentic tonal variation. Surface normals used for shading are estimated via local PCA on the upsampled point cloud.
Finally, although higher tree density increases ambiguity from top-down inputs, TreeON reconstructs trees individually and composes them into full scenes (Figure~\ref{fig:teaser}) using the pipeline of Grammatikaki et al.~\cite{Grammatikaki09052025}, which ensures correct scaling, placement, and terrain alignment.

\section{Experiments}
\label{sec:experiments}

We evaluate our method on both synthetic and real-world datasets, using quantitative and qualitative analyses to assess reconstruction accuracy, completeness, and generalization.

\subsection{Datasets}

\begin{figure}[h]
\centering
\includegraphics[width=0.95\linewidth]{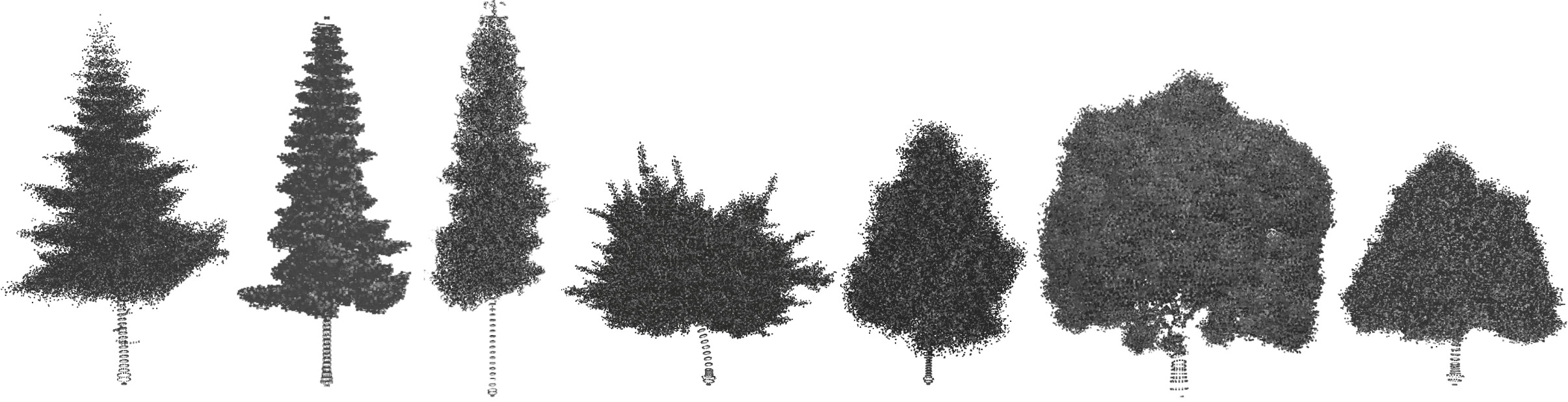}
\caption{Examples of tree species (coniferous and deciduous) generated from Grove, shown as 3D point clouds.}
\label{fig:tree_examples}
\end{figure}

We generated 3,000 synthetic tree models spanning 17 species (Beech, Oak, Birch, Hornbeam, Alder, Aspen, Poplar, Ash, Linden, Elm, Maple, Field Maple, Plane Tree, Fir, Austrian Pine, Scots Pine, and Stone Pine) with proportions based on alpine forestry data~\cite{AustriaForestReport2023}: approximately 60\% conifers and 40\% deciduous species. 
An overview of representative species is shown in Figure~\ref{fig:tree_examples}. 
To simulate diverse environments, each tree is placed on one of 12 sampled terrain patches covering \(3 \times 3\, \mathrm{km}\).  
From each terrain, we crop a localized \(30 \times 30\, \mathrm{m}\) region centered on the tree, ensuring that the tree is embedded in a realistic but computationally tractable environment. 

Orthophotos are rendered at a resolution consistent with aerial imagery standards (30\,cm/pixel in the Austrian dataset, equivalent to Zoom Level 19 in the basemap.at WMTS service~\cite{basemapat}), while DSMs are normalized to 8-bit grayscale images, where black corresponds to ground level and white to the tree apex. 
To ensure realistic sampling density, DSMs are rendered at 1-meter resolution, consistent with BEV's Open Data Austria height models~\cite{datagv}. 
Each DSM is then back-projected into a point cloud of about $K \approx 6{,}000$ points per tree, matching the ground-truth point cloud size for consistency.  

To test generalization, we use a dataset of Austrian landmark trees~\cite{grammatikaki_2024_nsj20-6ka24}, which contains isolated trees with orthophotos, DSMs, and photographs but no 3D ground truth, restricting evaluation to qualitative comparisons. 
Additionally, we validate our reconstructions against data from the French National Institute of Geographic and Forest Information (IGN)~\cite{IGN_LiDARHD}, including orthophotos (20 cm/pixel), dense LiDAR point clouds ($\geq$ 10 pts/m$^2$), and derived elevation models (DSMs and DTMs). 
In selected Pyrenees regions, isolated trees were identifiable in both orthophotos and point clouds.

\subsection{Implementation Details}

We train our model for 700 epochs using the Adam optimizer with an initial learning rate of $10^{-2}$ and a batch size of 16. A learning rate scheduler \texttt{ReduceLROnPlateau} halves the learning rate if the validation loss does not improve for 20 epochs. 
The training/validation split follows an 85/15 ratio over the 2,500 training trees, ensuring no overlap of tree instances between sets. 
The remaining 500 trees are held out for testing, including ablation studies and baseline comparisons. 
All models are trained on a single NVIDIA GeForce RTX 3070 GPU with 8 GB memory, requiring about 24 hours for 700 epochs. 
Inference on a single tree takes about 0.3 seconds.



\subsection{Evaluation Protocol} 

Our evaluation follows the four requirements defined in Section 1. 
Structural plausibility is measured quantitatively through Chamfer Distance (CD), Normalized Chamfer Distance (NCD), and F1-Score, which capture geometric accuracy and completeness. 
Visual appeal is evaluated qualitatively via comparisons to baselines and real imagery, since perceptual realism cannot be fully captured by metrics alone. 
Automation is demonstrated by the fully automatic pipeline requiring no species priors or multi-view input. 
Finally, computational efficiency is measured through training cost, inference speed, and scalability across thousands of trees.

\textbf{Chamfer Distance (CD)} is a widely used metric in 3D shape reconstruction that quantifies the geometric error between p
redicted and ground-truth point clouds~\cite{fan2017point, mescheder2019occupancy}. Given the predicted, $P$, and the ground truth, $Q$, point clouds, the CD is computed as:
\begin{equation}
\text{CD}(P, Q) = \frac{1}{|P|} \sum_{x \in P} \min_{y \in Q} \|x - y\|_2^2 + \frac{1}{|Q|} \sum_{y \in Q} \min_{x \in P} \|y - x\|_2^2.
\end{equation}
Since CD averages squared Euclidean distances, its unit is \(\text{m}^2\) when coordinates are expressed in meters. 
CD values are non-negative, with smaller values indicating closer alignment between prediction and ground truth. 

To enable fair comparison across trees of different sizes, we normalize CD by the cube root of the ground-truth tree volume $v$, yielding the \textbf{Normalized Chamfer Distance (NCD)}~\cite{10350393}:
\begin{equation}
\text{NCD}(P, Q) = \frac{\text{CD}(P, Q)}{v^{2/3}}.
\end{equation}
Here, $v^{1/3}$ provides a characteristic length scale of the tree (with units of meters), and $v^{2/3}$ has units of \(\text{m}^2\). 
Dividing CD (\(\text{m}^2\)) by $v^{2/3}$ (\(\text{m}^2\)) produces a unitless NCD. 
Like CD, NCD takes values $\geq 0$, with lower values indicating higher geometric accuracy.

To evaluate both reconstruction accuracy and completeness, we use the \textbf{F1-Score}, which combines precision and recall within a fixed distance threshold $\varepsilon$~\cite{alonso2025moreefficientpointcloud, Sokolova}. 
In our evaluation, $\varepsilon$ ranges between $0.02$ and $0.04\,\mathrm{m}$, depending on the tree size and sampling density. 
The threshold is tied to the scale of the query point sampling and the dataset's spatial resolution, and is configurable: smaller values enforce stricter matching, while larger values tolerate greater deviation.
The metrics are computed as:
\begin{align}
\text{Precision} &= \frac{|\{x \in P : \min_{y \in Q} \|x - y\| < \varepsilon\}|}{|P|}, \\
\text{Recall} &= \frac{|\{y \in Q : \min_{x \in P} \|y - x\| < \varepsilon\}|}{|Q|}, \\
\text{F1-Score} &= \frac{2 \cdot \text{Precision} \cdot \text{Recall}}{\text{Precision} + \text{Recall}}.
\end{align}
Precision, Recall, and F1-Score each range from $0$ to $1$, with higher values indicating better alignment. 
A score of $1$ indicates that all predicted points match ground-truth points within $\varepsilon$, and vice versa.

To quantify output diversity in the context of multi-sample reconstruction, we employ the \textbf{Coverage Score (COV)}~\cite{achlioptas2018learningrepresentationsgenerativemodels, ZHANG2022103551}. 
Let $\{P_i\}$ denote a set of predicted point clouds and $\{Q_j\}$ the ground-truth set. 
COV is computed as the fraction of unique ground-truth shapes that are closest (under CD) to at least one prediction:
\begin{equation}
\text{COV} = \frac{|\{Q_j : \exists i \text{ such that } Q_j = \arg\min_{j} \text{CD}(P_i, Q_j)\}|}{|\{Q_j\}|}
\end{equation}
COV values also range from $0$ to $1$, with higher values indicating that predictions cover a larger fraction of the ground-truth shape space. 
A value of $1$ means that every ground-truth shape is matched by at least one prediction, while lower values indicate mode collapse or limited diversity.

Finally, to additionally assess appearance fidelity, we report the 
\textbf{CIEDE2000 color difference} ($\Delta$E00)~\cite{Mirjalili_2019}, a perceptual metric that quantifies color discrepancies between predicted and ground-truth reconstructions. $\Delta$E00 values correspond to perceptual thresholds: values below $1$ are imperceptible to the human eye, values of $2$--$3$ are perceptible upon close observation, while values above $5$ are clearly visible differences~\cite{Luo}. 

\section{Results}
We report quantitative and qualitative results. First, we analyze the impact of input modalities and supervision in an ablation study. We then assess generalization on Austrian landmark trees and IGN LiDAR data. Finally, we compare our method against state-of-the-art baselines retrained on our dataset.


\begin{table*}[t]
\centering
\resizebox{\textwidth}{!}{%
\begin{tabular}{llccccccccccccccc}
\toprule
& & \multicolumn{5}{c}{\textbf{DSM}} & \multicolumn{5}{c}{\textbf{Orthophoto}} & \multicolumn{5}{c}{\textbf{DSM + Orthophoto}} \\
\cmidrule(lr){3-7} \cmidrule(lr){8-12} \cmidrule(lr){13-17}
\textbf{Losses} & & CD $\downarrow$ & NCD $\downarrow$ & F1 $\uparrow$ & COV $\uparrow$ & $\Delta$E00 $\downarrow$ & 
CD $\downarrow$ & NCD $\downarrow$ & F1 $\uparrow$ & COV $\uparrow$ & $\Delta$E00 $\downarrow$ & 
CD $\downarrow$ & NCD $\downarrow$ & F1 $\uparrow$ & COV $\uparrow$ & $\Delta$E00 $\downarrow$ \\
\midrule
Shadow & &2.1964 & 0.4850 & 0.1335 & 38.7\% & 10.18 & 
1.9938 & 0.4484 & 0.1835 & 25.4\% & 9.20 & 
2.0699 & 0.4550 & 0.1885 & 25.8\% & 8.08 \\
Silhouettes & & 2.1046  & 0.4624 & 0.1510 & 35.6\% & 8.84 & 
1.9328 & 0.4566 & 0.1428 & 32.3\% & 7.81 & 
2.1061 & 0.4446 & 0.1641 & 35.7\% & \textbf{7.59} \\
Shadow + Silh. & & 2.2125 & 0.4410 & 0.2174 & 30.2\% & 8.83 & 
1.5358 & 0.3723 & 0.2239 & 35.9\% & 7.06 & 
1.9371 & 0.4531 & 0.2969 & 30.9\% & 8.77 \\
BCE & & 1.3553 & 0.2789 & 0.7488 & 70.5\% & 8.77 & 
1.3750 & 0.3427 & 0.5208 & 50.6\% & 8.25 & 
1.3096 & 0.2765 & 0.8010 & 65.7\% & 8.15 \\
BCE + Shadow & & 1.2277 & 0.2618 & 0.7372 & 65.4\% & 8.76 & 
1.4297 & 0.3108 & 0.5879 & 60.2\% & 8.34 & 
1.1738 & 0.2688 & 0.8201 & 75.8\% & 7.97 \\
BCE + Silh. & & 1.1104 & 0.2446 & 0.8569 & 75.2\% & \textbf{8.05} & 
1.4585 & 0.3200 & 0.5301 & 50.8\% & 7.19 & 
1.0754 & 0.2456 & 0.8728 & 78.6\% & 8.83 \\
Mixed & & \textbf{0.9994} & \textbf{0.2296} & \textbf{0.8958} & \textbf{85.9\%} & 8.75 & 
\textbf{1.3371} & \textbf{0.2913} & \textbf{0.6376} & \textbf{70.4\%} & \textbf{6.91} & 
\textbf{0.9699} & \textbf{0.2239} & \textbf{0.8846} & \textbf{90.7\%} & 8.23 \\
\bottomrule
\end{tabular}
} 
\caption{Ablation study across loss functions and input modalities. Each cell reports Chamfer Distance (CD in m), Normalized CD, F1 Score, Coverage Score (COV in \%), and Mean $\Delta$E00 color error ($\Delta$E00). Arrows ($\downarrow$/$\uparrow$) indicate whether lower or higher values are better.}
\label{tab:ablation_metrics_split}
\end{table*}

\begin{table*}[h]
    \centering
    \setlength{\tabcolsep}{2pt}
    \begin{tabular}{
        >{\centering\arraybackslash}m{1.43cm}
        >{\centering\arraybackslash}m{1.6cm}
        >{\centering\arraybackslash}m{0.35cm}
        >{\centering\arraybackslash}m{1.4cm}
        >{\centering\arraybackslash}m{1.4cm}
        >{\centering\arraybackslash}m{1.6cm}
        >{\centering\arraybackslash}m{1.8cm}
        >{\centering\arraybackslash}m{1.8cm}
        >{\centering\arraybackslash}m{1.7cm}
        >{\centering\arraybackslash}m{2.8cm}
    }

    \cline{4-10}
    \textbf{\footnotesize DSM} & \textbf{\footnotesize Orthophoto} & &
    \textbf{\scriptsize BCE} & 
    \textbf{\scriptsize Shadow} &
    \textbf{\scriptsize Silhouettes} &
    \textbf{\scriptsize Shadow+Silh.} &
    \textbf{\scriptsize BCE+Shadow} &
    \textbf{\scriptsize BCE+Silh.} &
    \textbf{\scriptsize BCE+Shadow+Silh.} \\
    \cline{4-10}
    \noalign{\vskip 3pt}

    \raisebox{-0.4cm}{\includegraphics[width=1.42cm, height=1.5cm, keepaspectratio]{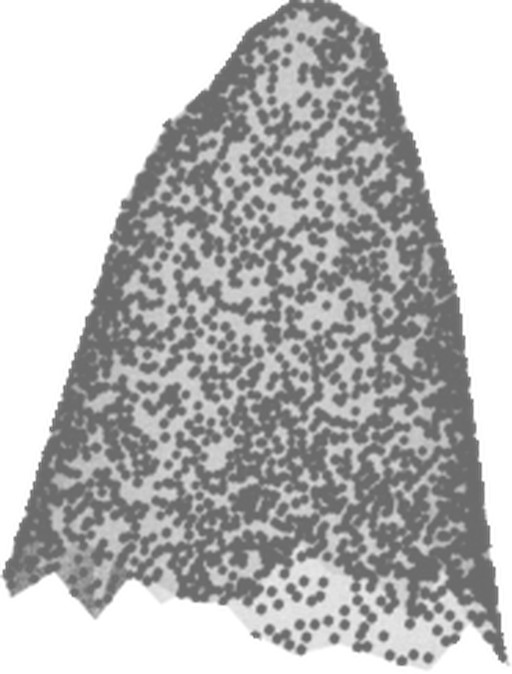}} & 
    \raisebox{-0.4cm}{\includegraphics[width=1.42cm, height=1.5cm, keepaspectratio]{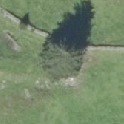}} & 
    \rotatebox[origin=c]{90}{\footnotesize\textit{DSM}} &
    \includegraphics[width=1.4cm, height=1.5cm, keepaspectratio]{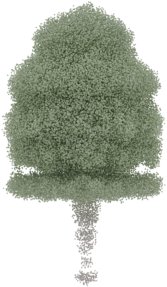} &
    \includegraphics[width=1.4cm, height=1.5cm, keepaspectratio]{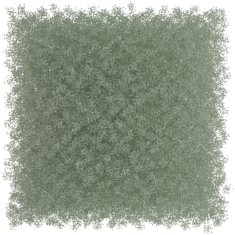} &
    \includegraphics[width=1.4cm, height=1.5cm, keepaspectratio]{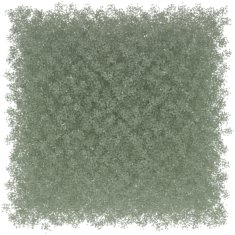} &
    \includegraphics[width=1.4cm, height=1.5cm, keepaspectratio]{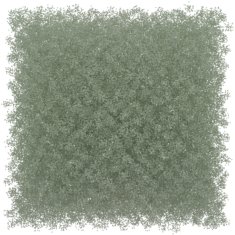} &
    \includegraphics[width=1.4cm, height=1.5cm, keepaspectratio]{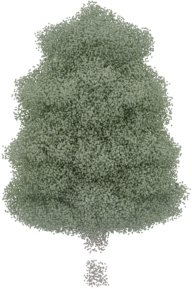} &
    \includegraphics[width=1.4cm, height=1.5cm, keepaspectratio]{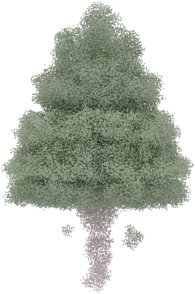} &
    \includegraphics[width=1.4cm, height=1.5cm, keepaspectratio]{RESULTS/ablation/dsm_bce_silhouettes.jpg} \\

    \multicolumn{2}{c}{%
      \multirow{2}{*}{%
        \begin{tabular}{c}
          \textbf{Target} \\[2pt]       
          \includegraphics[width=2.5cm, height=2.5cm, keepaspectratio]{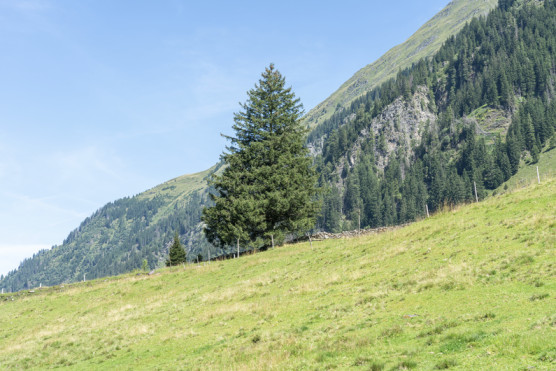}
        \end{tabular}
      }%
    } &
    \rotatebox[origin=c]{90}{\footnotesize\textit{Orthophoto}} &
    \includegraphics[width=1.4cm, height=1.5cm, keepaspectratio]{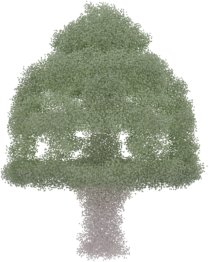} &
    \includegraphics[width=1.4cm, height=1.5cm, keepaspectratio]{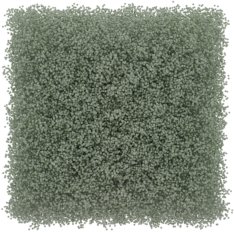} &
    \includegraphics[width=1.4cm, height=1.5cm, keepaspectratio]{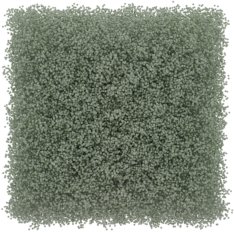} &
    \includegraphics[width=1.4cm, height=1.5cm, keepaspectratio]{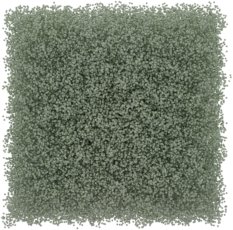} &
    \includegraphics[width=1.4cm, height=1.5cm, keepaspectratio]{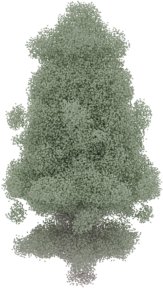} &
    \includegraphics[width=1.4cm, height=1.5cm, keepaspectratio]{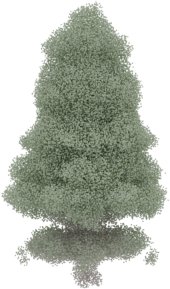} &
    \includegraphics[width=1.4cm, height=1.5cm, keepaspectratio]{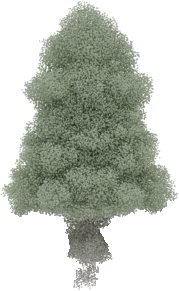} \\

    & &
    \rotatebox[origin=c]{90}{\footnotesize\textit{DSM + Ortho}} &
    \includegraphics[width=1.4cm, height=1.5cm, keepaspectratio]{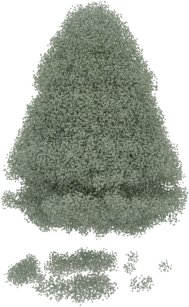} &
    \includegraphics[width=1.4cm, height=1.5cm, keepaspectratio]{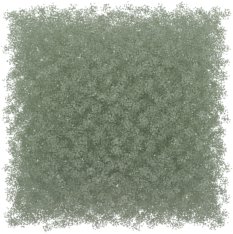} &
    \includegraphics[width=1.4cm, height=1.5cm, keepaspectratio]{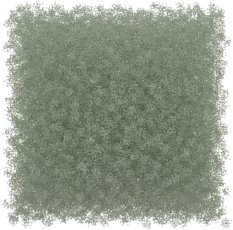} &
    \includegraphics[width=1.4cm, height=1.5cm, keepaspectratio]{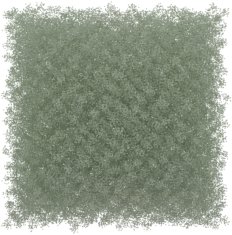} &
    \includegraphics[width=1.4cm, height=1.5cm, keepaspectratio]{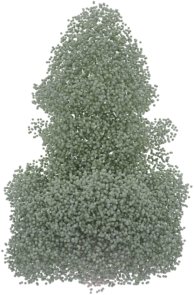} &
    \includegraphics[width=1.4cm, height=1.5cm, keepaspectratio]{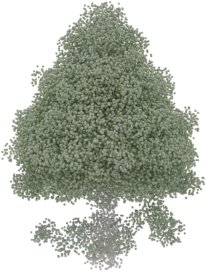} &
    \includegraphics[width=1.4cm, height=1.5cm, keepaspectratio]{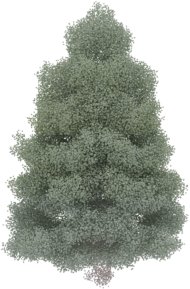} 
    \\

    \end{tabular}
    \caption{Qualitative ablation study of different loss functions (columns) and input modalities (rows). Left: DSM and orthophoto inputs with target photo. Right: Rows show reconstructions from DSM, orthophoto, or both inputs, while columns compare supervision strategies.}
    \label{tab:ablation_full_grid}
\end{table*}

\begin{table*}[h]
\centering
\resizebox{\textwidth}{!}{%
\begin{tabular}{c p{0.12\textwidth} p{0.12\textwidth} p{0.12\textwidth} p{0.12\textwidth} ||
                 c p{0.12\textwidth} p{0.12\textwidth} p{0.12\textwidth} p{0.12\textwidth}}
\toprule
& \multicolumn{4}{c}{\textbf{Typical Reconstructions}} &
& \multicolumn{4}{c}{\textbf{Difficult Cases}} \\
\cmidrule{2-5} \cmidrule{7-10}
& \multicolumn{1}{c}{\textbf{DSM}} & 
\multicolumn{1}{c}{\textbf{Orthophoto}} &
\multicolumn{1}{c}{\textbf{Target}} &
\multicolumn{1}{c}{\textbf{Output}} & 
& \multicolumn{1}{c}{\textbf{DSM}} & 
\multicolumn{1}{c}{\textbf{Orthophoto}} &
\multicolumn{1}{c}{\textbf{Target}} &
\multicolumn{1}{c}{\textbf{Output}} \\
\hline
\textbf{1} & \parbox[c][0.12\textwidth][c]{0.12\textwidth}{\centering\includegraphics[width=\linewidth,height=0.1\textwidth,keepaspectratio]{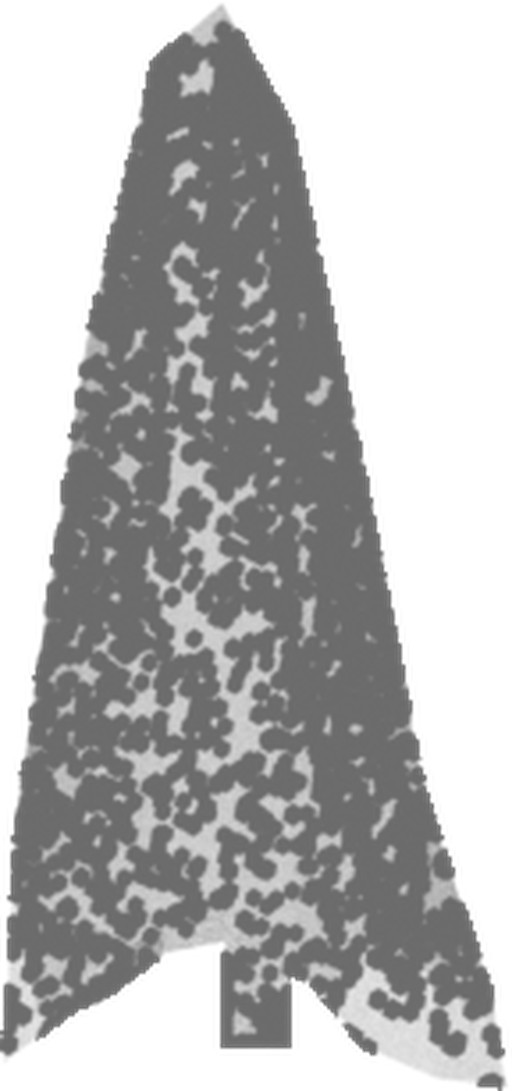}} & 
\parbox[c][0.12\textwidth][c]{0.12\textwidth}{\centering\includegraphics[width=\linewidth,height=0.1\textwidth,keepaspectratio]{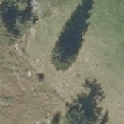}} &
\parbox[c][0.12\textwidth][c]{0.12\textwidth}{\centering\includegraphics[width=\linewidth,height=0.1\textwidth,keepaspectratio]{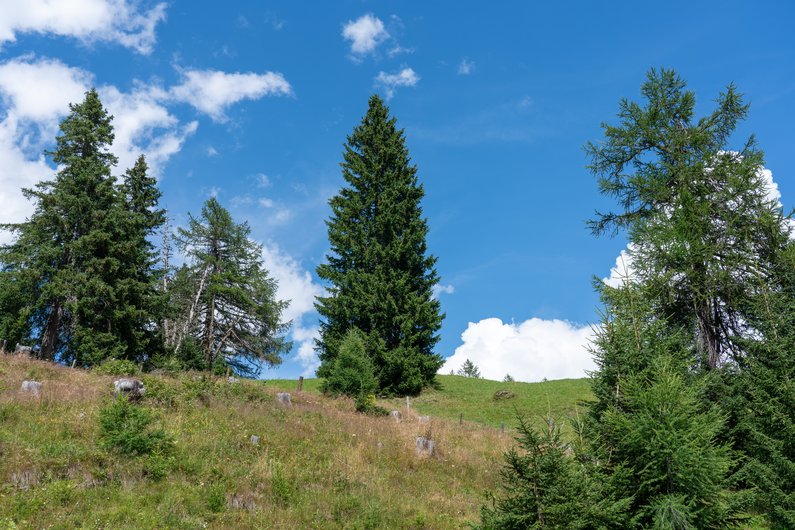}} &
\parbox[c][0.12\textwidth][c]{0.12\textwidth}{\centering\includegraphics[width=\linewidth,height=0.1\textwidth,keepaspectratio]{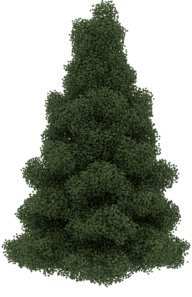}} &
\textbf{1} & \parbox[c][0.12\textwidth][c]{0.12\textwidth}{\centering\includegraphics[width=\linewidth,height=0.1\textwidth,keepaspectratio]{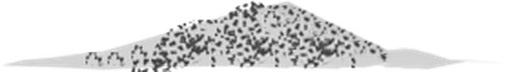}} &
\parbox[c][0.12\textwidth][c]{0.12\textwidth}{\centering\includegraphics[width=\linewidth,height=0.1\textwidth,keepaspectratio]{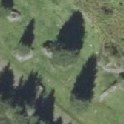}} &
\parbox[c][0.12\textwidth][c]{0.12\textwidth}{\centering\includegraphics[width=\linewidth,height=0.1\textwidth,keepaspectratio]{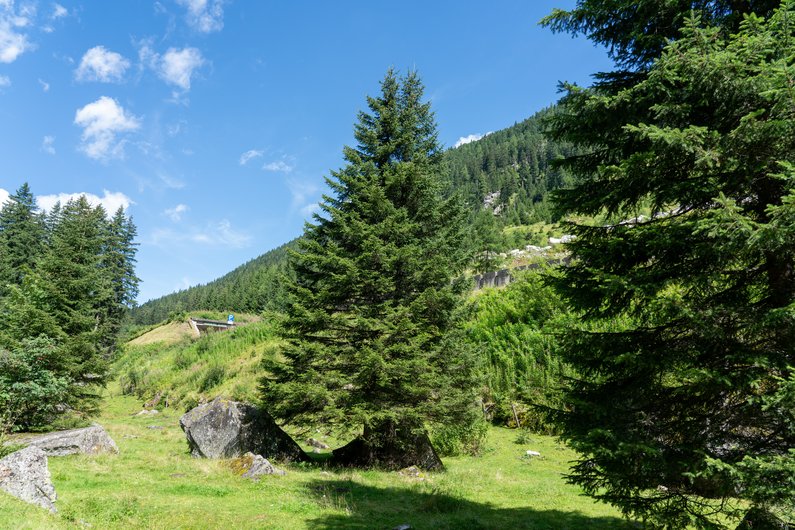}} &
\parbox[c][0.12\textwidth][c]{0.12\textwidth}{\centering\includegraphics[width=\linewidth,height=0.1\textwidth,keepaspectratio]{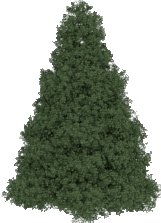}} \\
\hline
\textbf{2} & \parbox[c][0.12\textwidth][c]{0.12\textwidth}{\centering\includegraphics[width=\linewidth,height=0.1\textwidth,keepaspectratio]{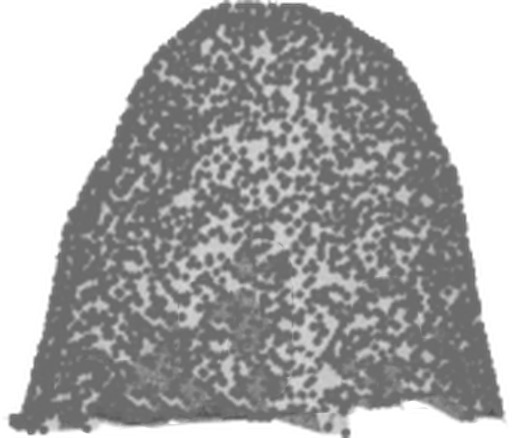}} &
\parbox[c][0.12\textwidth][c]{0.12\textwidth}{\centering\includegraphics[width=\linewidth,height=0.1\textwidth,keepaspectratio]{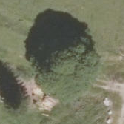}} &
\parbox[c][0.12\textwidth][c]{0.12\textwidth}{\centering\includegraphics[width=\linewidth,height=0.1\textwidth,keepaspectratio]{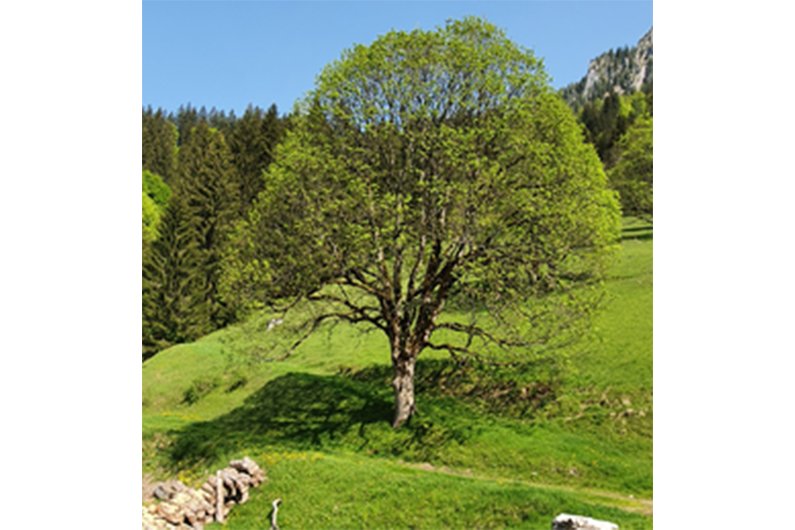}} &
\parbox[c][0.12\textwidth][c]{0.12\textwidth}{\centering\includegraphics[width=\linewidth,height=0.1\textwidth,keepaspectratio]{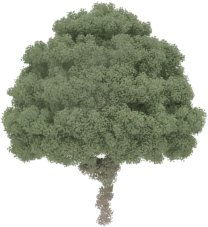}} &
\textbf{2} & \parbox[c][0.12\textwidth][c]{0.12\textwidth}{\centering\includegraphics[width=\linewidth,height=0.1\textwidth,keepaspectratio]{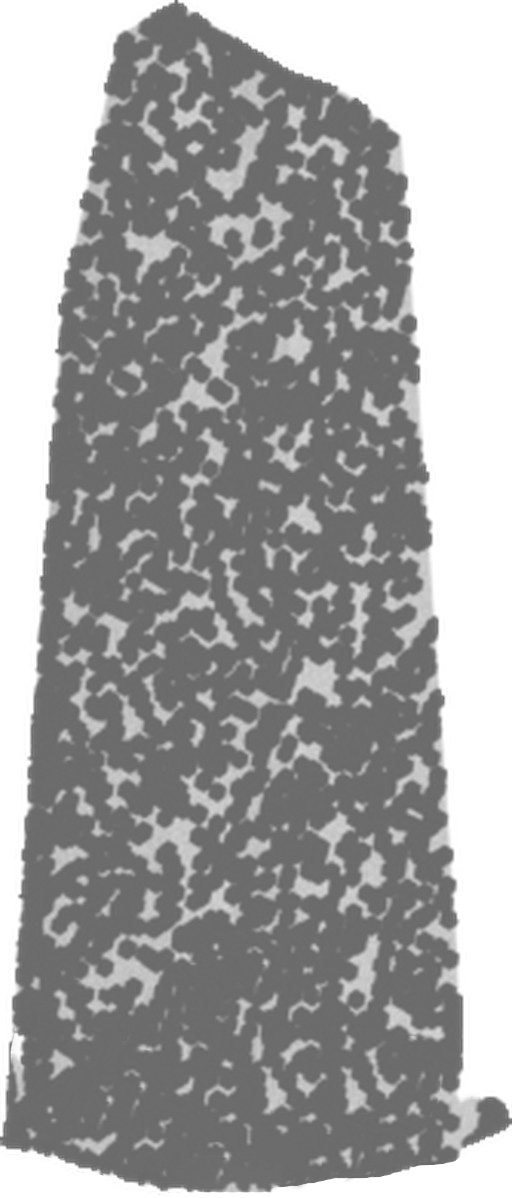}} &
\parbox[c][0.12\textwidth][c]{0.12\textwidth}{\centering\includegraphics[width=\linewidth,height=0.1\textwidth,keepaspectratio]{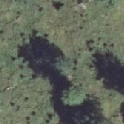}} &
\parbox[c][0.12\textwidth][c]{0.12\textwidth}{\centering\includegraphics[width=\linewidth,height=0.1\textwidth,keepaspectratio]{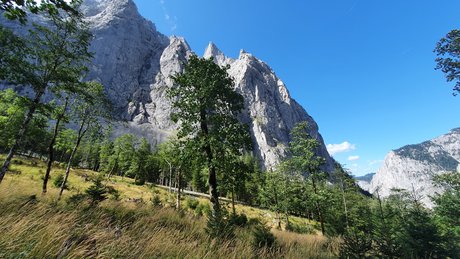}} &
\parbox[c][0.12\textwidth][c]{0.12\textwidth}{\centering\includegraphics[width=\linewidth,height=0.1\textwidth,keepaspectratio]{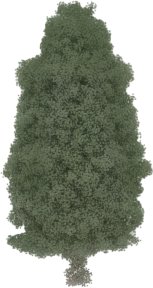}} \\
\hline
\textbf{3} & \parbox[c][0.12\textwidth][c]{0.12\textwidth}{\centering\includegraphics[width=\linewidth,height=0.1\textwidth,keepaspectratio]{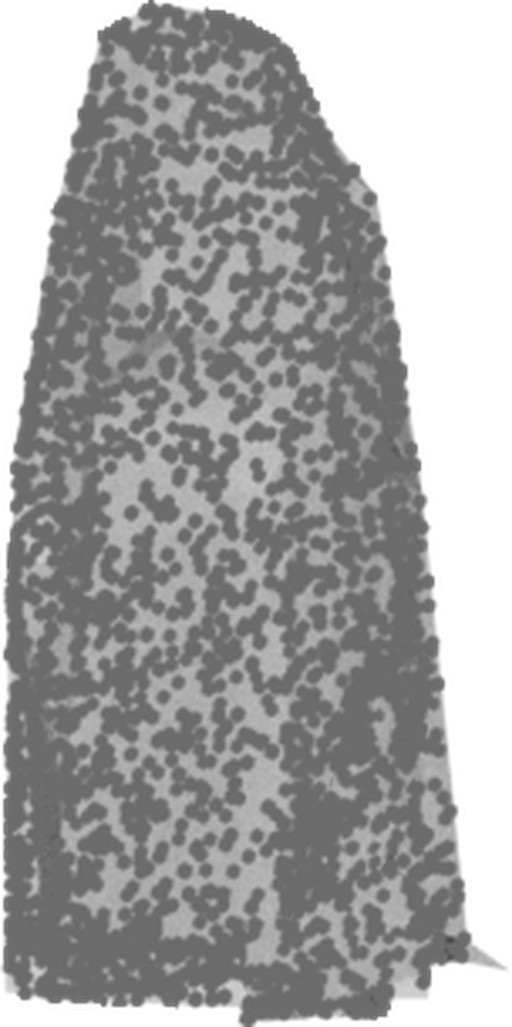}} &
\parbox[c][0.12\textwidth][c]{0.12\textwidth}{\centering\includegraphics[width=\linewidth,height=0.1\textwidth,keepaspectratio]{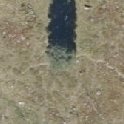}} &
\parbox[c][0.12\textwidth][c]{0.12\textwidth}{\centering\includegraphics[width=\linewidth,height=0.1\textwidth,keepaspectratio]{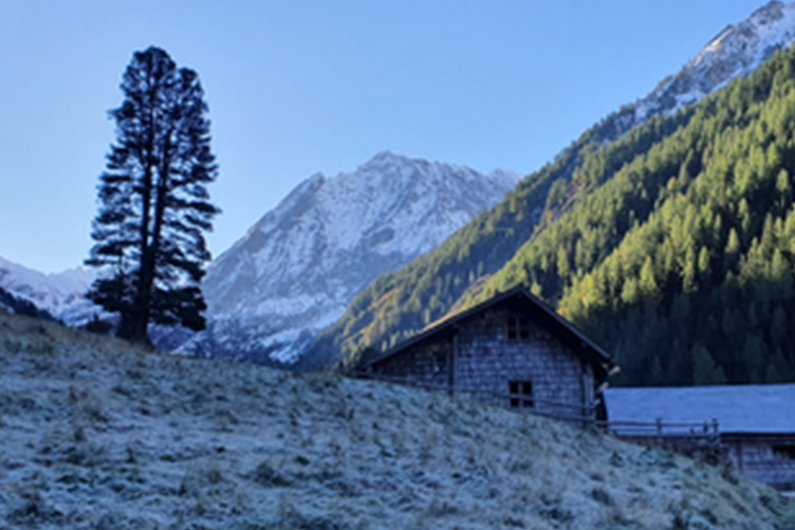}} &
\parbox[c][0.12\textwidth][c]{0.12\textwidth}{\centering\includegraphics[width=\linewidth,height=0.1\textwidth,keepaspectratio]{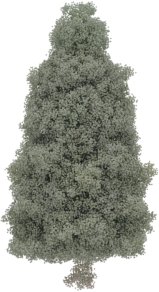}} &
\textbf{3} & \parbox[c][0.12\textwidth][c]{0.12\textwidth}{\centering\includegraphics[width=\linewidth,height=0.1\textwidth,keepaspectratio]{RESULTS/DSM/tree_68.jpg}} &
\parbox[c][0.12\textwidth][c]{0.12\textwidth}{\centering\includegraphics[width=\linewidth,height=0.1\textwidth,keepaspectratio]{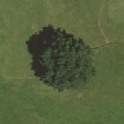}} &
\parbox[c][0.12\textwidth][c]{0.12\textwidth}{\centering\includegraphics[width=\linewidth,height=0.1\textwidth,keepaspectratio]{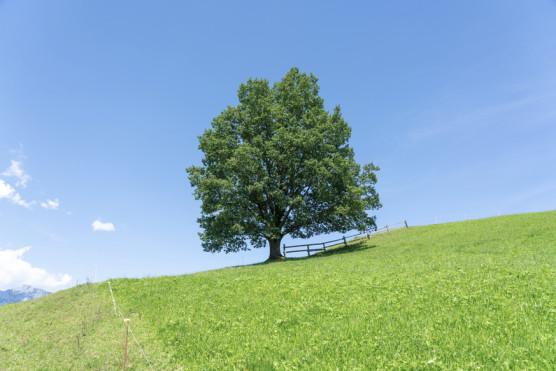}} &
\parbox[c][0.12\textwidth][c]{0.12\textwidth}{\centering\includegraphics[width=\linewidth,height=0.1\textwidth,keepaspectratio]{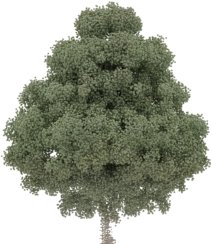}} \\
\hline
\textbf{4} & \parbox[c][0.12\textwidth][c]{0.12\textwidth}{\centering\includegraphics[width=\linewidth,height=0.1\textwidth,keepaspectratio]{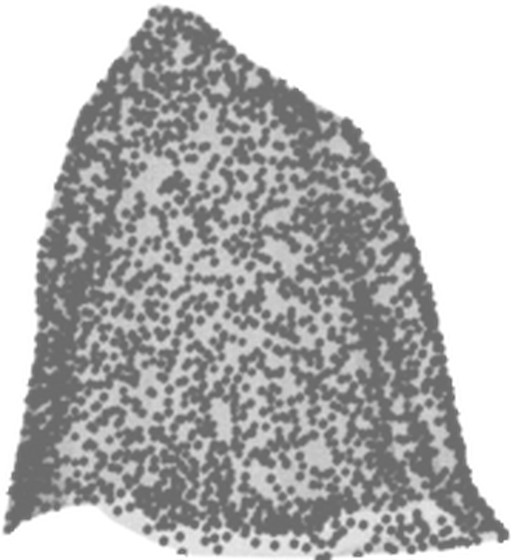}} &
\parbox[c][0.12\textwidth][c]{0.12\textwidth}{\centering\includegraphics[width=\linewidth,height=0.1\textwidth,keepaspectratio]{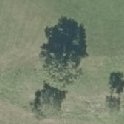}} &
\parbox[c][0.12\textwidth][c]{0.12\textwidth}{\centering\includegraphics[width=\linewidth,height=0.1\textwidth,keepaspectratio]{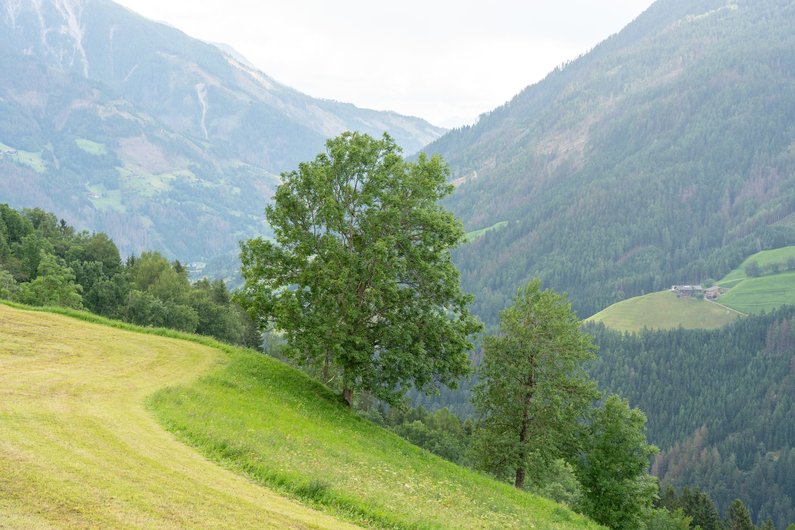}} &
\parbox[c][0.12\textwidth][c]{0.12\textwidth}{\centering\includegraphics[width=\linewidth,height=0.1\textwidth,keepaspectratio]{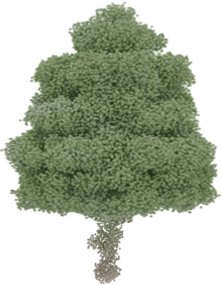}} &
4 &  \parbox[c][0.12\textwidth][c]{0.12\textwidth}{\centering\includegraphics[width=\linewidth,height=0.1\textwidth,keepaspectratio]{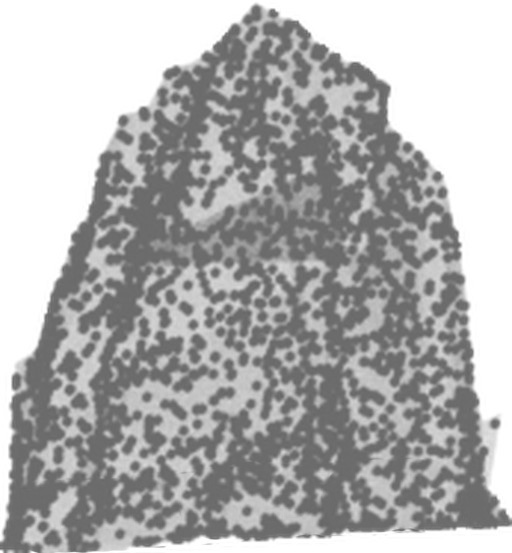}} &
\parbox[c][0.12\textwidth][c]{0.12\textwidth}{\centering\includegraphics[width=\linewidth,height=0.1\textwidth,keepaspectratio]{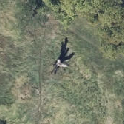}} &
\parbox[c][0.12\textwidth][c]{0.12\textwidth}{\centering\includegraphics[width=\linewidth,height=0.1\textwidth,keepaspectratio]{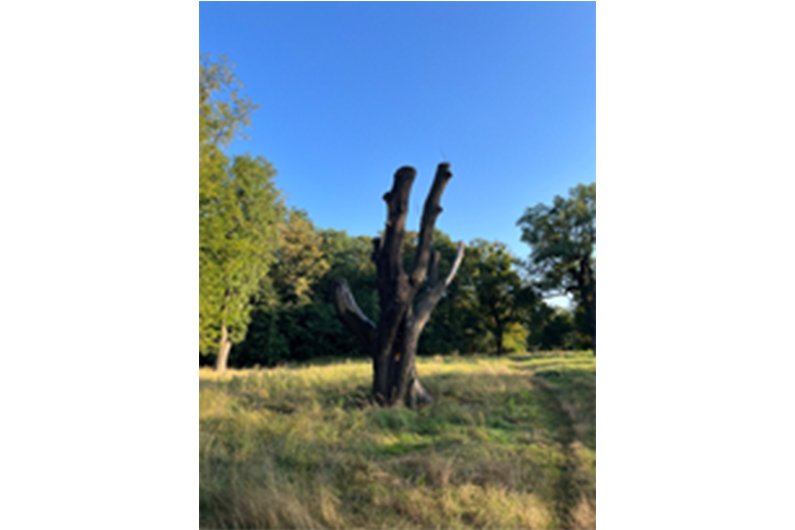}} &
\parbox[c][0.12\textwidth][c]{0.12\textwidth}{\centering\includegraphics[width=\linewidth,height=0.1\textwidth,keepaspectratio]{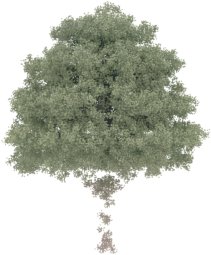}} \\

\textbf{5} &  \parbox[c][0.12\textwidth][c]{0.12\textwidth}{\centering\includegraphics[width=\linewidth,height=0.1\textwidth,keepaspectratio]{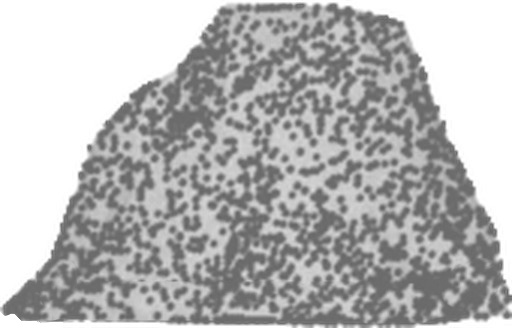}} &
\parbox[c][0.12\textwidth][c]{0.12\textwidth}{\centering\includegraphics[width=\linewidth,height=0.1\textwidth,keepaspectratio]{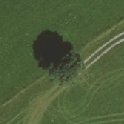}} &
\parbox[c][0.12\textwidth][c]{0.12\textwidth}{\centering\includegraphics[width=\linewidth,height=0.1\textwidth,keepaspectratio]{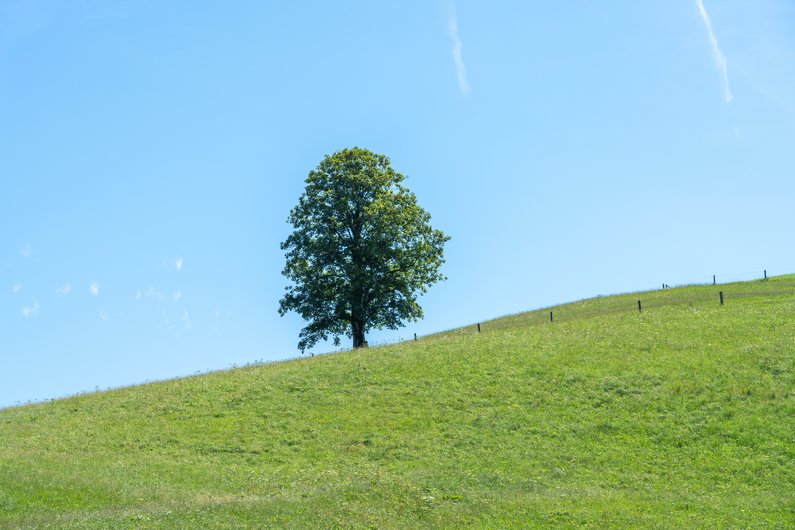}} &
\parbox[c][0.12\textwidth][c]{0.12\textwidth}{\centering\includegraphics[width=\linewidth,height=0.1\textwidth,keepaspectratio]{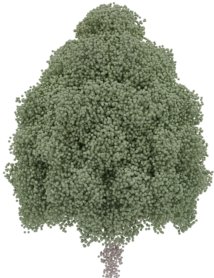}} & 

\textbf{5} &  \parbox[c][0.12\textwidth][c]{0.12\textwidth}{\centering\includegraphics[width=\linewidth,height=0.1\textwidth,keepaspectratio]{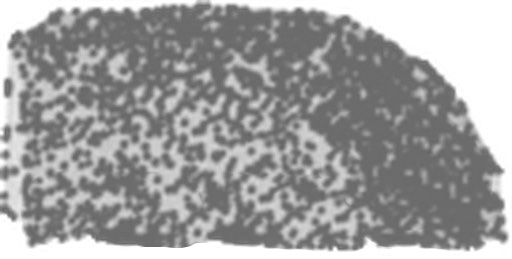}} &
\parbox[c][0.12\textwidth][c]{0.12\textwidth}{\centering\includegraphics[width=\linewidth,height=0.1\textwidth,keepaspectratio]{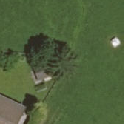}} &
\parbox[c][0.12\textwidth][c]{0.12\textwidth}{\centering\includegraphics[width=\linewidth,height=0.1\textwidth,keepaspectratio]{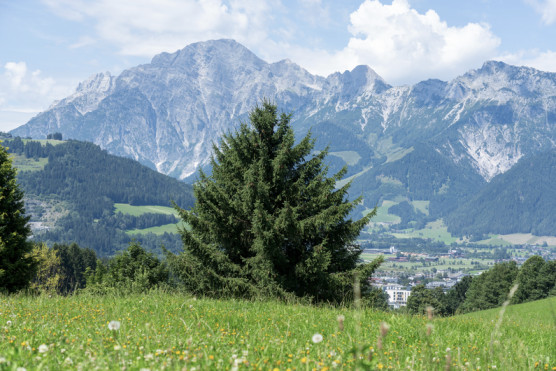}} &
\parbox[c][0.12\textwidth][c]{0.12\textwidth}{\centering\includegraphics[width=\linewidth,height=0.1\textwidth,keepaspectratio]{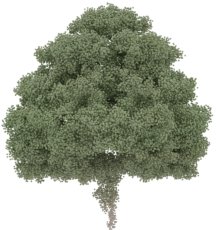}} \\
\bottomrule
\end{tabular}
}
\caption{Qualitative comparison of reconstructed landmark trees against target photograph of the real tree. Left column: reconstructions. Right column: difficult cases: (1) DSM underestimates tree size, (2) orthophotos are cluttered by neighboring trees, (3) shadows are weak or missing, and (4) dead tree in the orthophoto. In each case, the model compensates by leveraging the complementary input signal. (5) Failure case: DSM shape and weak ortho cues mislead the model into reconstructing a conifer instead of the deciduous target.}
\label{tab:landmark_trees}
\end{table*}

\subsection{Ablation Study}

Table~\ref{tab:ablation_metrics_split} presents an ablation study evaluating performance across four key metrics:
Chamfer Distance (CD), Normalized Chamfer Distance (NCD), F1-Score (F1), and Coverage (COV) over different input modalities and loss combinations. Specifically, we evaluate models trained on (1) DSM only, (2) orthophoto only, and (3) both modalities, each under five supervision regimes: (i) BCE only, (ii) BCE + shadow, (iii) BCE + silhouette, (iv) shadow + silhouette, and (v) BCE + shadow + silhouette.

\textbf{Quantitative}
The results presented in Table~\ref{tab:ablation_metrics_split} reveal several important insights into the impact of loss functions and input modalities on tree shape reconstruction performance. Among all configurations, the mixed supervision with DSM and Orthophoto input stands out as the most effective approach, achieving the lowest Chamfer Distance (CD = 0.9699), the highest F1 Score (F1 = 0.8846), and the most complete reconstructions in terms of Coverage (COV = 90.7\%). 

An interesting trend emerges when comparing supervision types. Loss functions based solely on shadows or silhouettes perform the worst across all input configurations. For instance, using only the shadow loss with DSM input yields a CD of 2.1964 and a meager F1 of 0.1335, indicating a high degree of geometric error and almost no structural correctness. This shows that shadows or silhouettes in isolation lack sufficient signal to guide accurate reconstruction, particularly when they are not paired with 3D-aware or pixel-wise losses, as projection-based cues alone under-constrain depth and allow multiple degenerate 3D solutions.

In contrast, introducing BCE supervision --- despite its simplicity --- results in a large performance jump. On DSM input alone, BCE improves CD by nearly 1 meter (from 2.2125 to 1.3553) and increases F1 from 0.2174 to 0.7488. This effect is consistent across all modalities, underscoring BCE's importance in guiding point-level precision. Adding shadow or silhouette supervision on top of BCE further improves results modestly in most cases, especially in terms of Coverage, suggesting that these additional losses help regularize and fill out the reconstructed shape.

Between the two types of projection losses, silhouettes generally prove more beneficial than shadows: they align directly with canopy contours and provide a stable global constraint, whereas shadows vary with lighting direction and length, making them less reliable. For DSM or mixed inputs, silhouettes consistently lead to more complete reconstructions and higher F1 scores. The only exception is with orthophoto input, where BCE and shadow supervision slightly outperforms BCE and silhouette, suggesting that in image-based settings, shadows can provide complementary cues that silhouettes alone may not capture.

Overall, the mixed supervision setting with DSM and Orthophoto input demonstrates the value of combining all available cues: DSM provides geometric structure, orthophotos supply texture and appearance, BCE ensures point-level accuracy, and shadow/silhouette losses add shape variety. This combination not only improves per-sample accuracy but also reduces mode collapse, leading to the highest COV scores (90.7\%) by covering a larger fraction of the ground-truth shape space. The synergies between these modalities and losses are critical, and future work could explore how to further enhance such fusion, possibly with learned weighting strategies or attention-based mechanisms that adaptively prioritize each signal.

A further dimension of the ablation is provided by the color error ($\Delta$E00). As expected, orthophoto input improves color fidelity compared to DSM-only training, with the lowest error observed when using orthophotos with mixed supervision (6.91). In contrast, DSM-only runs consistently exhibit higher color errors ($\geq8.75$), reflecting the lack of spectral information in elevation data. Overall, $\Delta$E00 values range between $6.91$ and $10.18$ across experiments, indicating that while reconstructions capture broad color trends, visible deviations remain due to projection mismatches between orthophoto colors and 3D geometry, as well as smoothing effects in the learned color representation. 

\textbf{Qualitative}
To better understand how different inputs and supervision signals affect reconstruction quality, we conducted a qualitative ablation study shown in Table~\ref{tab:ablation_full_grid}. The previously discussed trends, including the impact of individual inputs and the benefits of combining supervision signals, are visually evident in the table. 

\begin{table*}[h]
\centering
\resizebox{\textwidth}{!}{%
\begin{tabular}{
| p{0.15\textwidth} |
  p{0.15\textwidth} |
  p{0.15\textwidth} ||
  >{\centering\arraybackslash}p{0.15\textwidth} |
  >{\centering\arraybackslash}p{0.15\textwidth} |
  >{\centering\arraybackslash}p{0.15\textwidth} |
  >{\centering\arraybackslash}p{0.15\textwidth} |
  >{\centering\arraybackslash}p{0.15\textwidth} |
  >{\centering\arraybackslash}p{0.15\textwidth} |
  >{\centering\arraybackslash}p{0.15\textwidth} |
  >{\centering\arraybackslash}p{0.15\textwidth} |
  >{\centering\arraybackslash}p{0.15\textwidth} |
  >{\centering\arraybackslash}p{0.15\textwidth} |
}
\cline{1-13} 
\multicolumn{2}{|c|}{\cellcolor{my-gray}\textbf{Inputs}} & \multicolumn{1}{c||}{\multirow{2}{*}{\makebox[0pt][c]{\textbf{Ground Truth}}}} & \multicolumn{10}{c|}{\textbf{Networks}} \\ \cline{1-2} \cline{4-13} \multicolumn{1}{|c|}{\cellcolor{my-gray}\textbf{DSM}} & \multicolumn{1}{c|}{\cellcolor{my-gray}\textbf{Orthophoto}} & & \textbf{OpenLRM} & \textbf{Zero123++} & \textbf{TRELLIS} & \textbf{TMNet} & \textbf{AtlasNet} & \textbf{3DTopia-XL} & \textbf{DPMs} & \textbf{PUGeoNet} & \textbf{RepKPU} & \textbf{Ours} \\ \cline{1-3} 
& & & \cite{openlrm} & \cite{zero123plus} & \cite{trellis} & \cite{tmnet} & \cite{atlasnet} & \cite{3dtopiaxl} & \cite{dpms} & \cite{pugeonet} & \cite{repkpu} & \\ \cline{4-13}
\parbox[c][0.17\textwidth][c]{\linewidth}{\centering\includegraphics[width=\linewidth,height=0.12\textwidth,keepaspectratio]{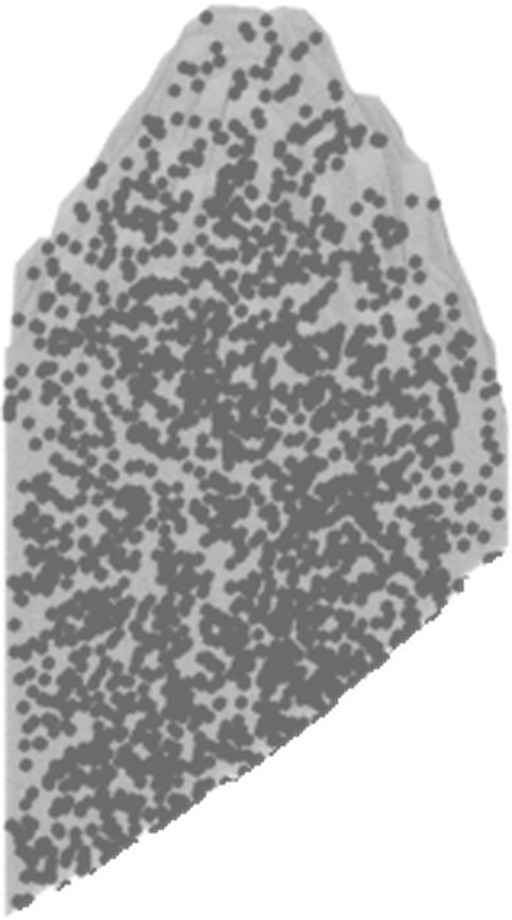}} &
\parbox[c][0.17\textwidth][c]{\linewidth}{\centering\includegraphics[width=\linewidth,height=0.12\textwidth,keepaspectratio]{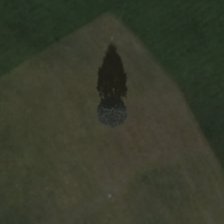}} &
\parbox[c][0.17\textwidth][c]{\linewidth}{\centering\includegraphics[width=\linewidth,height=0.12\textwidth,keepaspectratio]{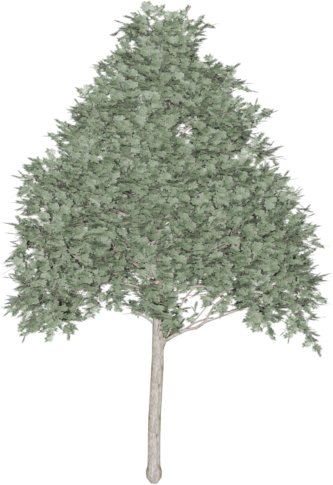}} &
\parbox[c][0.17\textwidth][c]{\linewidth}{\centering\includegraphics[width=\linewidth,height=0.12\textwidth,keepaspectratio]{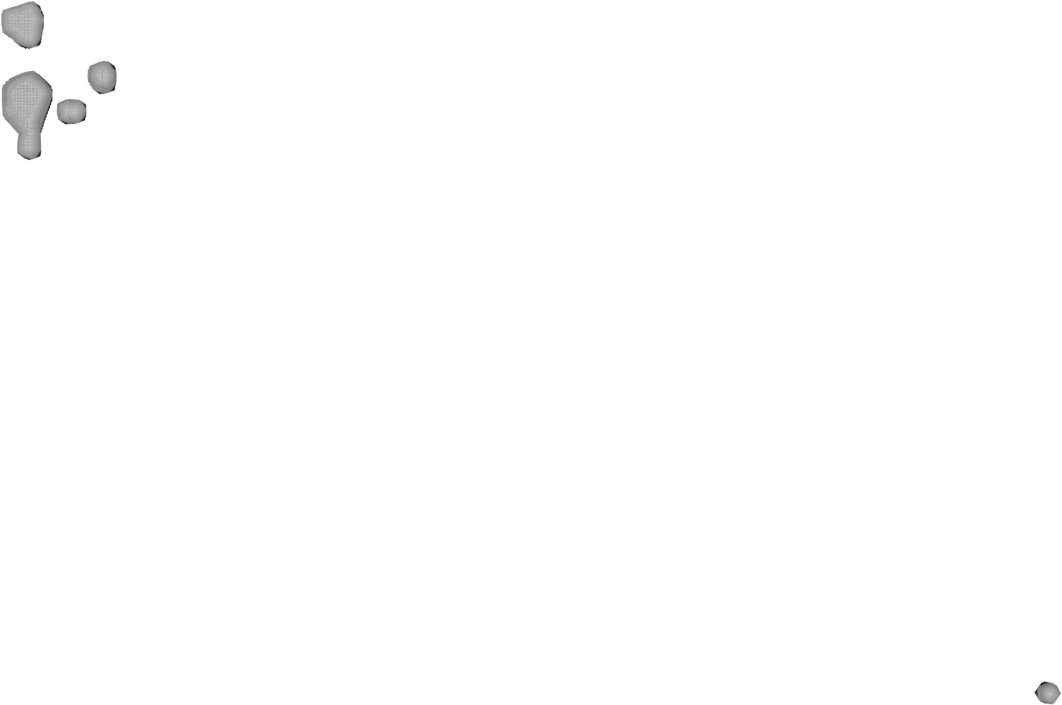}} &
\parbox[c][0.17\textwidth][c]{\linewidth}{\centering\includegraphics[width=\linewidth,height=0.12\textwidth,keepaspectratio]{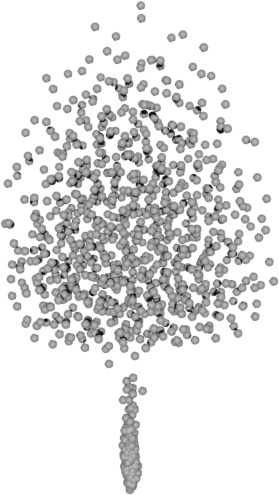}} &
\parbox[c][0.17\textwidth][c]{\linewidth}{\centering\includegraphics[width=\linewidth,height=0.12\textwidth,keepaspectratio]{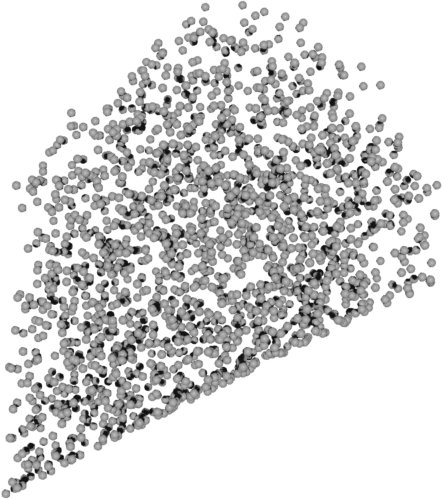}} &
\parbox[c][0.17\textwidth][c]{\linewidth}{\centering\includegraphics[width=\linewidth,height=0.12\textwidth,keepaspectratio]{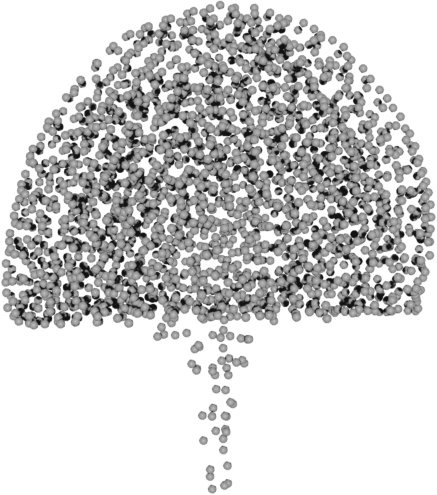}} &
\parbox[c][0.17\textwidth][c]{\linewidth}{\centering\includegraphics[width=\linewidth,height=0.12\textwidth,keepaspectratio]{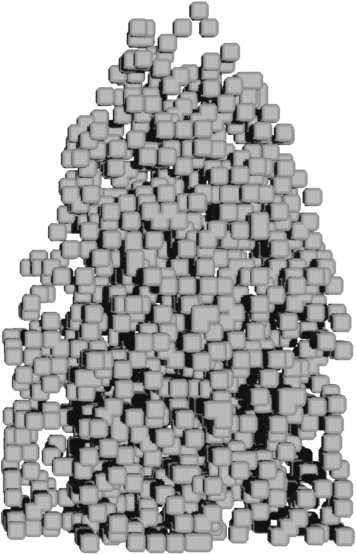}} & 
\parbox[c][0.17\textwidth][c]{\linewidth}{\centering\includegraphics[width=\linewidth,height=0.12\textwidth,keepaspectratio]{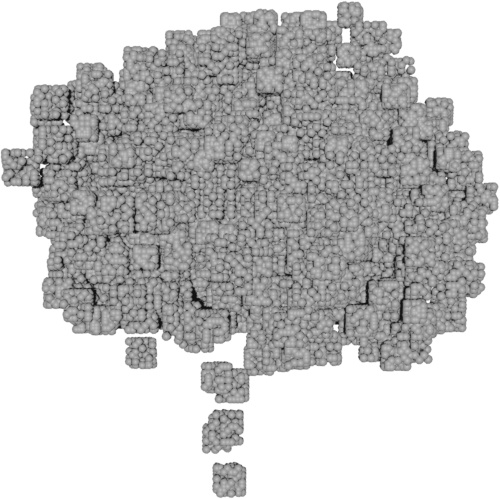}} &
\parbox[c][0.17\textwidth][c]{\linewidth}{\centering\includegraphics[width=\linewidth,height=0.12\textwidth,keepaspectratio]{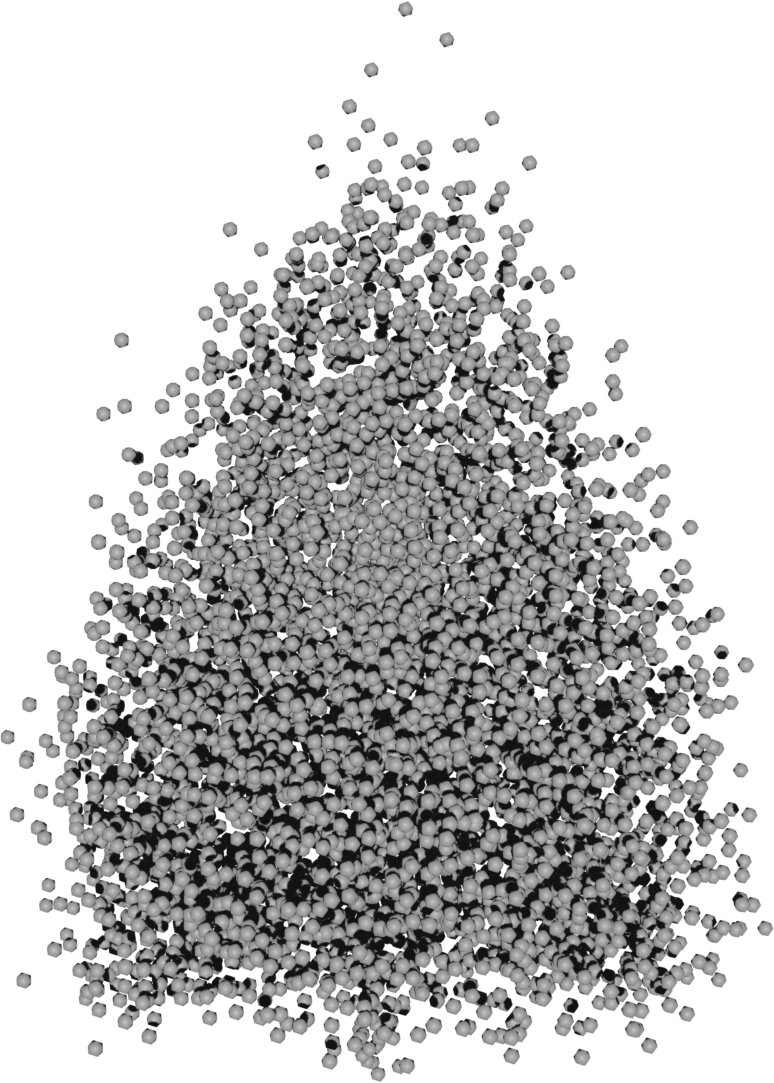}} &
\parbox[c][0.17\textwidth][c]{\linewidth}{\centering\includegraphics[width=\linewidth,height=0.12\textwidth,keepaspectratio]{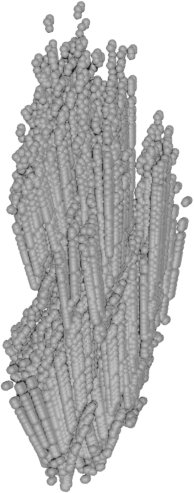}} &
\parbox[c][0.17\textwidth][c]{\linewidth}{\centering\includegraphics[width=\linewidth,height=0.12\textwidth,keepaspectratio]{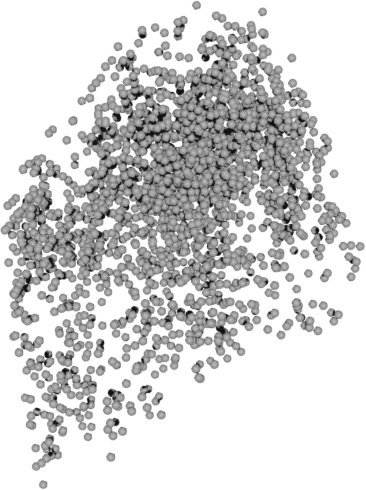}} &
\parbox[c][0.17\textwidth][c]{\linewidth}{\centering\includegraphics[width=\linewidth,height=0.12\textwidth,keepaspectratio]{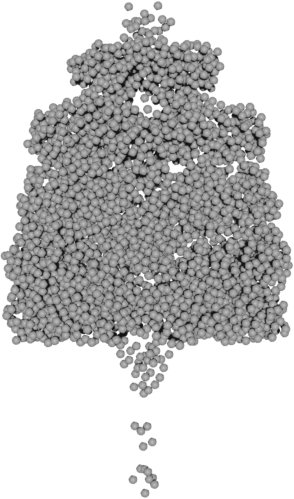}} \\
\hline
\parbox[c][0.17\textwidth][c]{\linewidth}{\centering\includegraphics[width=\linewidth,height=0.12\textwidth,keepaspectratio]{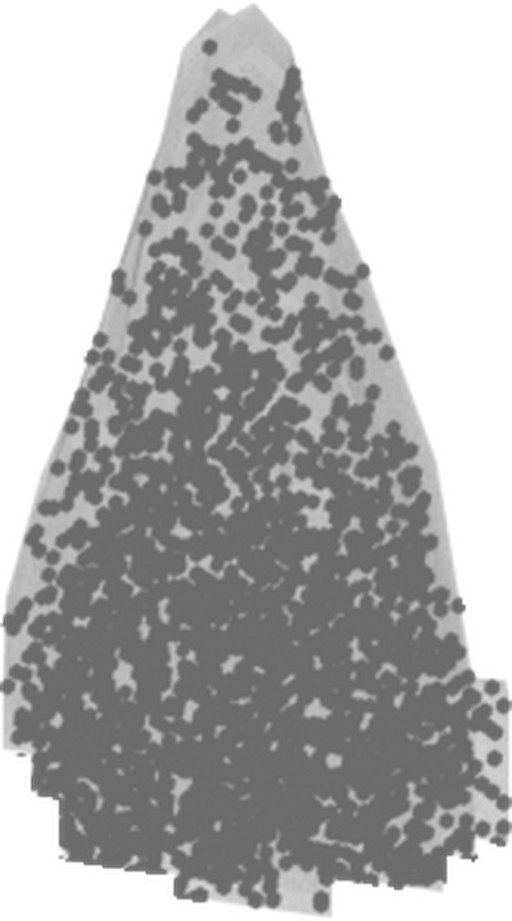}} &
\parbox[c][0.17\textwidth][c]{\linewidth}{\centering\includegraphics[width=\linewidth,height=0.12\textwidth,keepaspectratio]{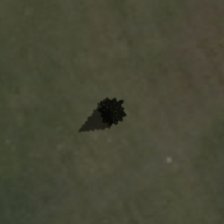}} &
\parbox[c][0.17\textwidth][c]{\linewidth}{\centering\includegraphics[width=\linewidth,height=0.12\textwidth,keepaspectratio]{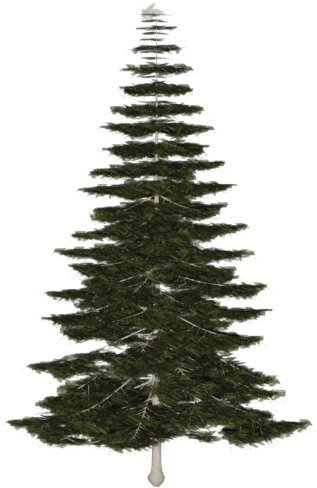}} &
\parbox[c][0.17\textwidth][c]{\linewidth}{\centering\includegraphics[width=\linewidth,height=0.12\textwidth,keepaspectratio]{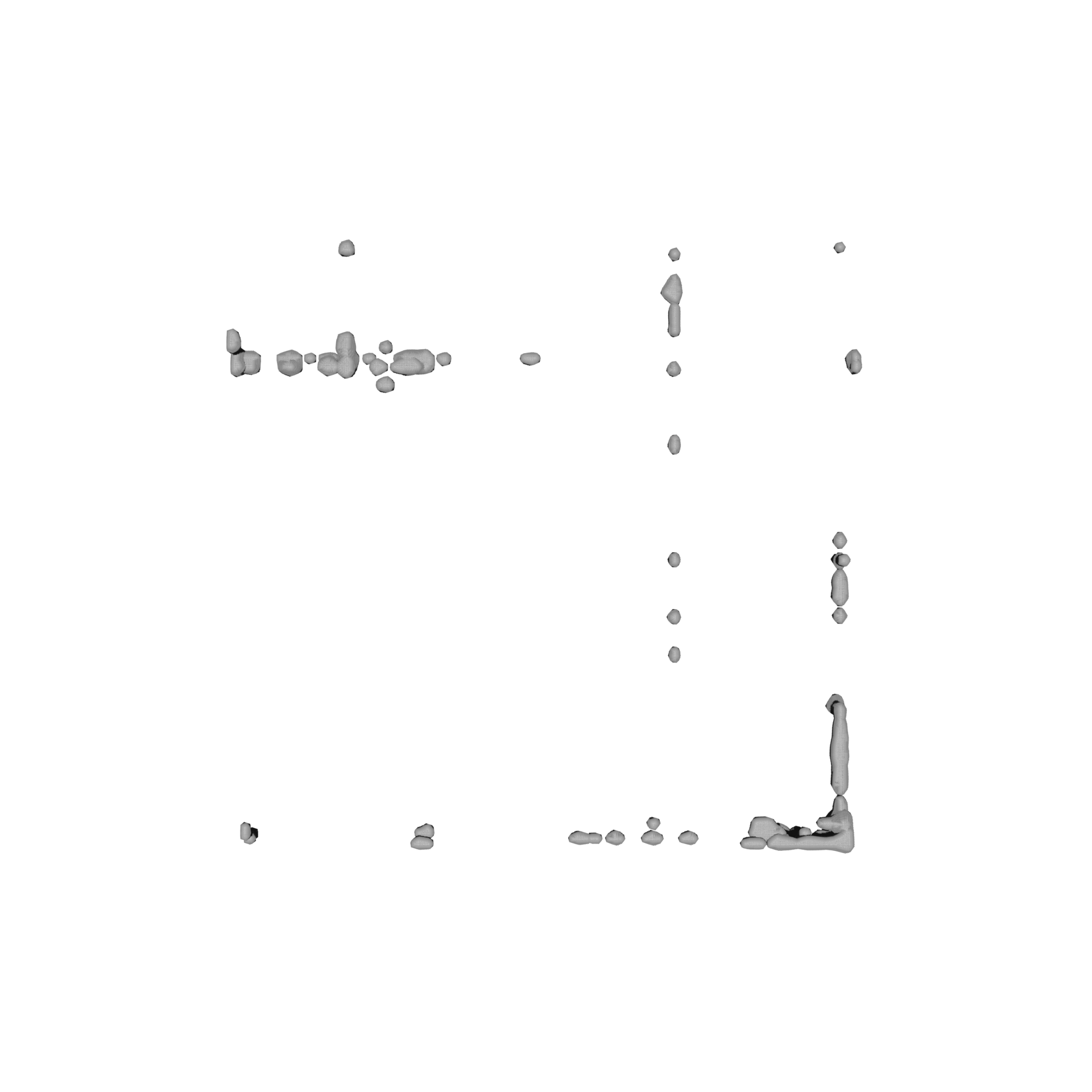}} &
\parbox[c][0.17\textwidth][c]{\linewidth}{\centering\includegraphics[width=\linewidth,height=0.12\textwidth,keepaspectratio]{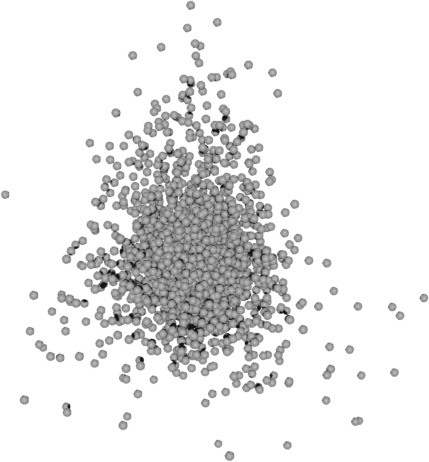}} &
\parbox[c][0.17\textwidth][c]{\linewidth}{\centering\includegraphics[width=\linewidth,height=0.12\textwidth,keepaspectratio]{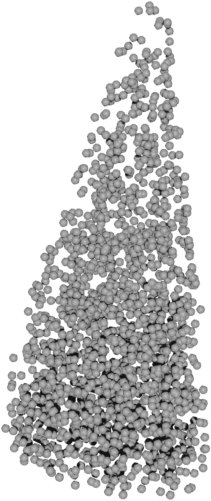}} &
\parbox[c][0.17\textwidth][c]{\linewidth}{\centering\includegraphics[width=\linewidth,height=0.12\textwidth,keepaspectratio]{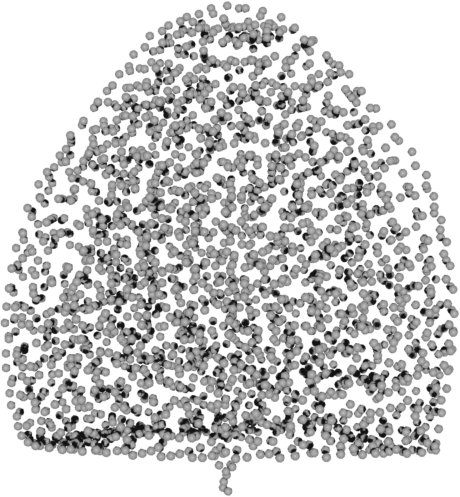}} &
\parbox[c][0.17\textwidth][c]{\linewidth}{\centering\includegraphics[width=\linewidth,height=0.12\textwidth,keepaspectratio]{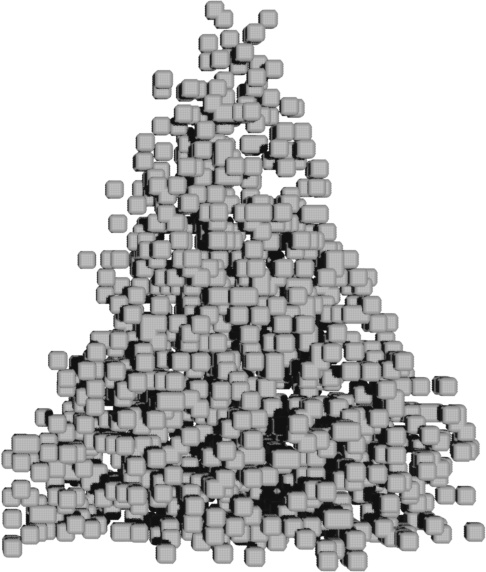}} & 
\parbox[c][0.17\textwidth][c]{\linewidth}{\centering\includegraphics[width=\linewidth,height=0.12\textwidth,keepaspectratio]{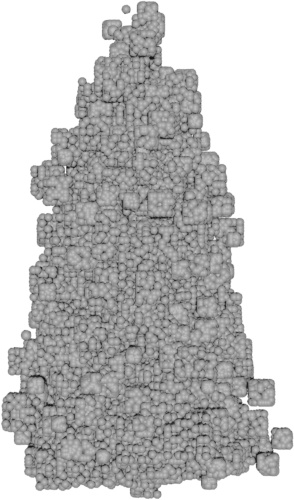}} &
\parbox[c][0.17\textwidth][c]{\linewidth}{\centering\includegraphics[width=\linewidth,height=0.12\textwidth,keepaspectratio]{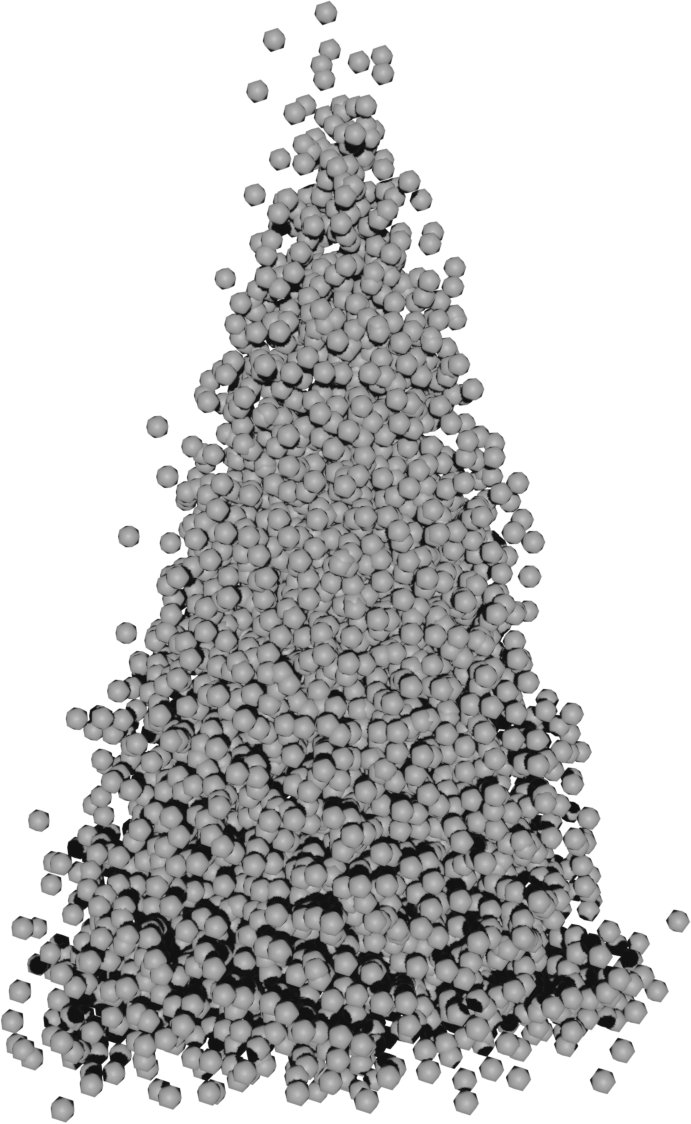}} &
\parbox[c][0.17\textwidth][c]{\linewidth}{\centering\includegraphics[width=\linewidth,height=0.12\textwidth,keepaspectratio]{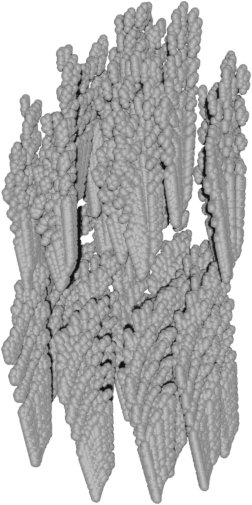}} &
\parbox[c][0.17\textwidth][c]{\linewidth}{\centering\includegraphics[width=\linewidth,height=0.12\textwidth,keepaspectratio]{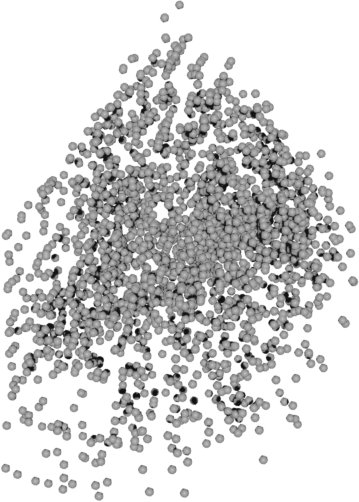}} &
\parbox[c][0.17\textwidth][c]{\linewidth}{\centering\includegraphics[width=\linewidth,height=0.12\textwidth,keepaspectratio]{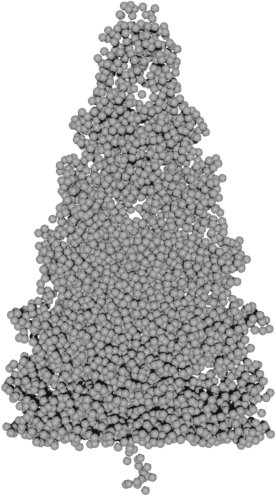}} \\
\hline
\parbox[c][0.17\textwidth][c]{\linewidth}{\centering\includegraphics[width=\linewidth,height=0.12\textwidth,keepaspectratio]{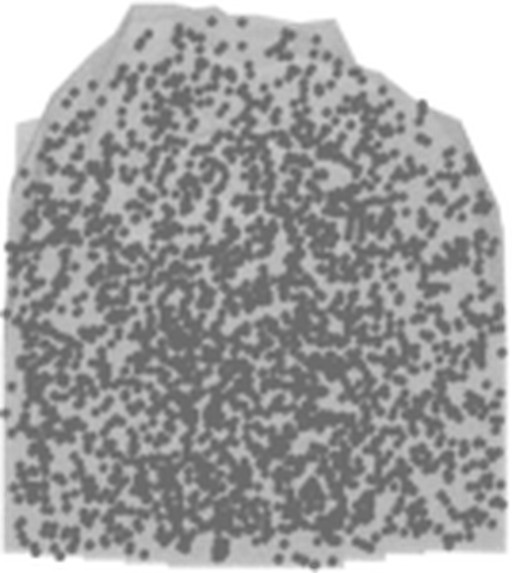}} &
\parbox[c][0.17\textwidth][c]{\linewidth}{\centering\includegraphics[width=\linewidth,height=0.12\textwidth,keepaspectratio]{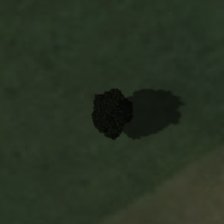}} &
\parbox[c][0.17\textwidth][c]{\linewidth}{\centering\includegraphics[width=\linewidth,height=0.12\textwidth,keepaspectratio]{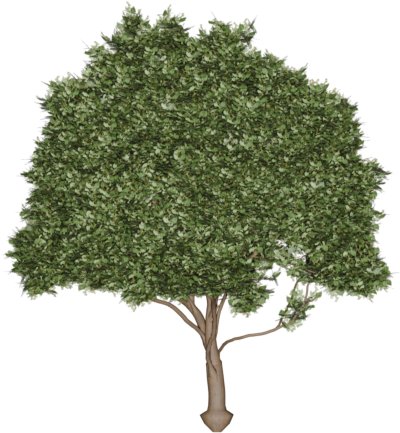}} &
\parbox[c][0.17\textwidth][c]{\linewidth}{\centering\includegraphics[width=\linewidth,height=0.12\textwidth,keepaspectratio]{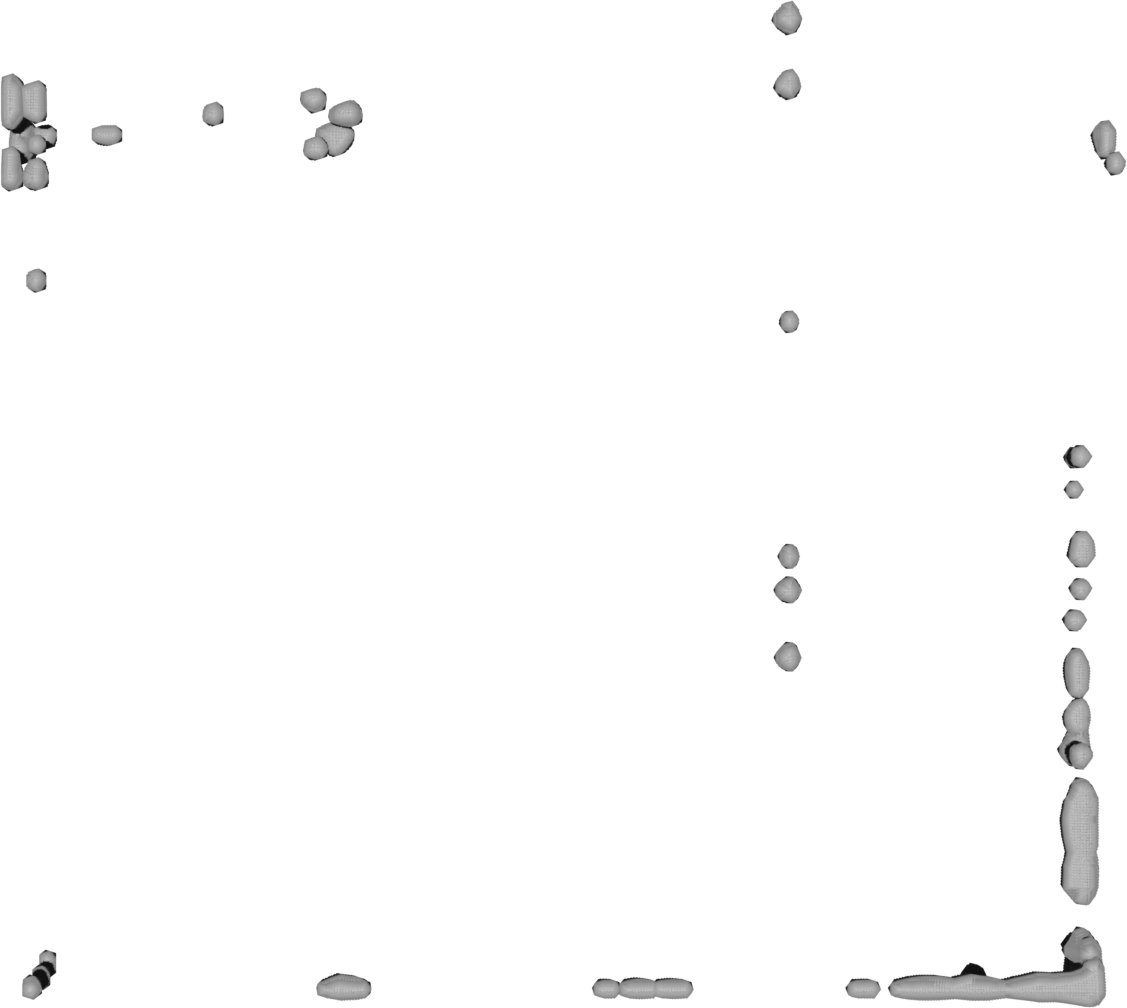}} &
\parbox[c][0.17\textwidth][c]{\linewidth}{\centering\includegraphics[width=\linewidth,height=0.12\textwidth,keepaspectratio]{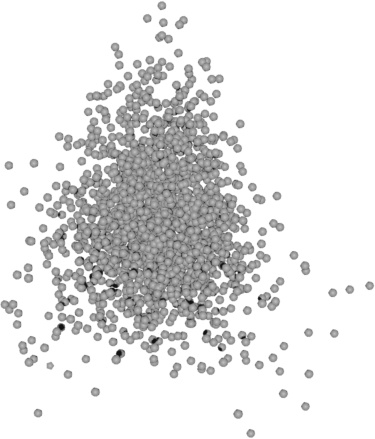}} &
\parbox[c][0.17\textwidth][c]{\linewidth}{\centering\includegraphics[width=\linewidth,height=0.12\textwidth,keepaspectratio]{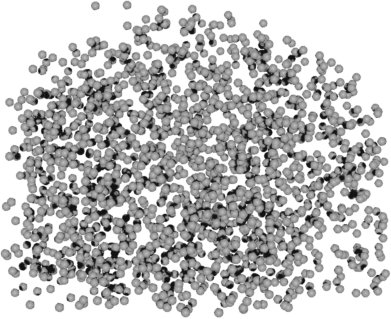}} &
\parbox[c][0.17\textwidth][c]{\linewidth}{\centering\includegraphics[width=\linewidth,height=0.12\textwidth,keepaspectratio]{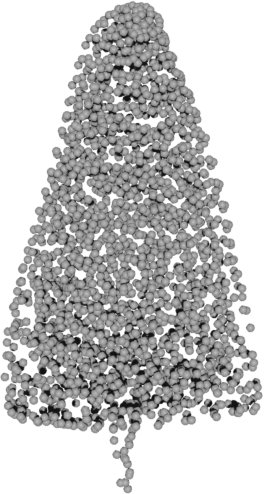}} &
\parbox[c][0.17\textwidth][c]{\linewidth}{\centering\includegraphics[width=\linewidth,height=0.12\textwidth,keepaspectratio]{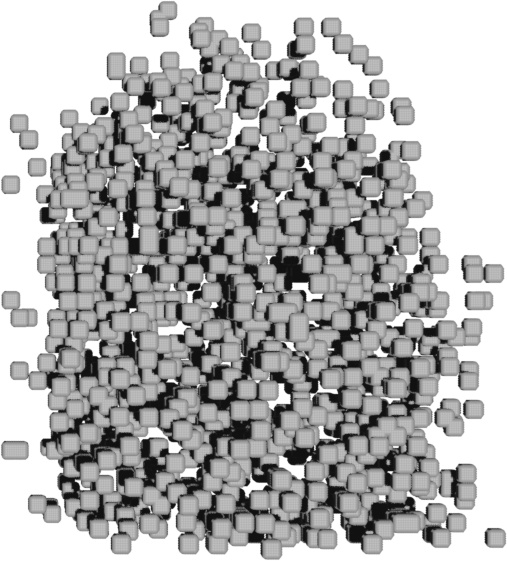}} & 
\parbox[c][0.17\textwidth][c]{\linewidth}{\centering\includegraphics[width=\linewidth,height=0.12\textwidth,keepaspectratio]{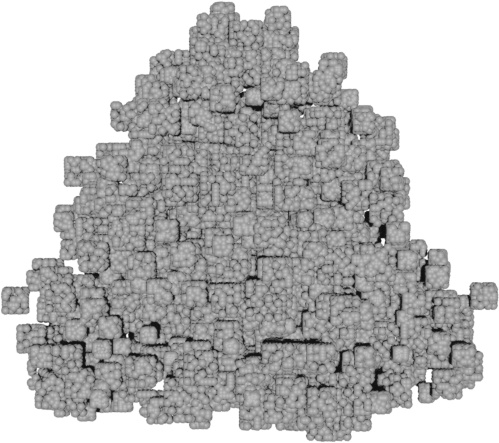}} &
\parbox[c][0.17\textwidth][c]{\linewidth}{\centering\includegraphics[width=\linewidth,height=0.12\textwidth,keepaspectratio]{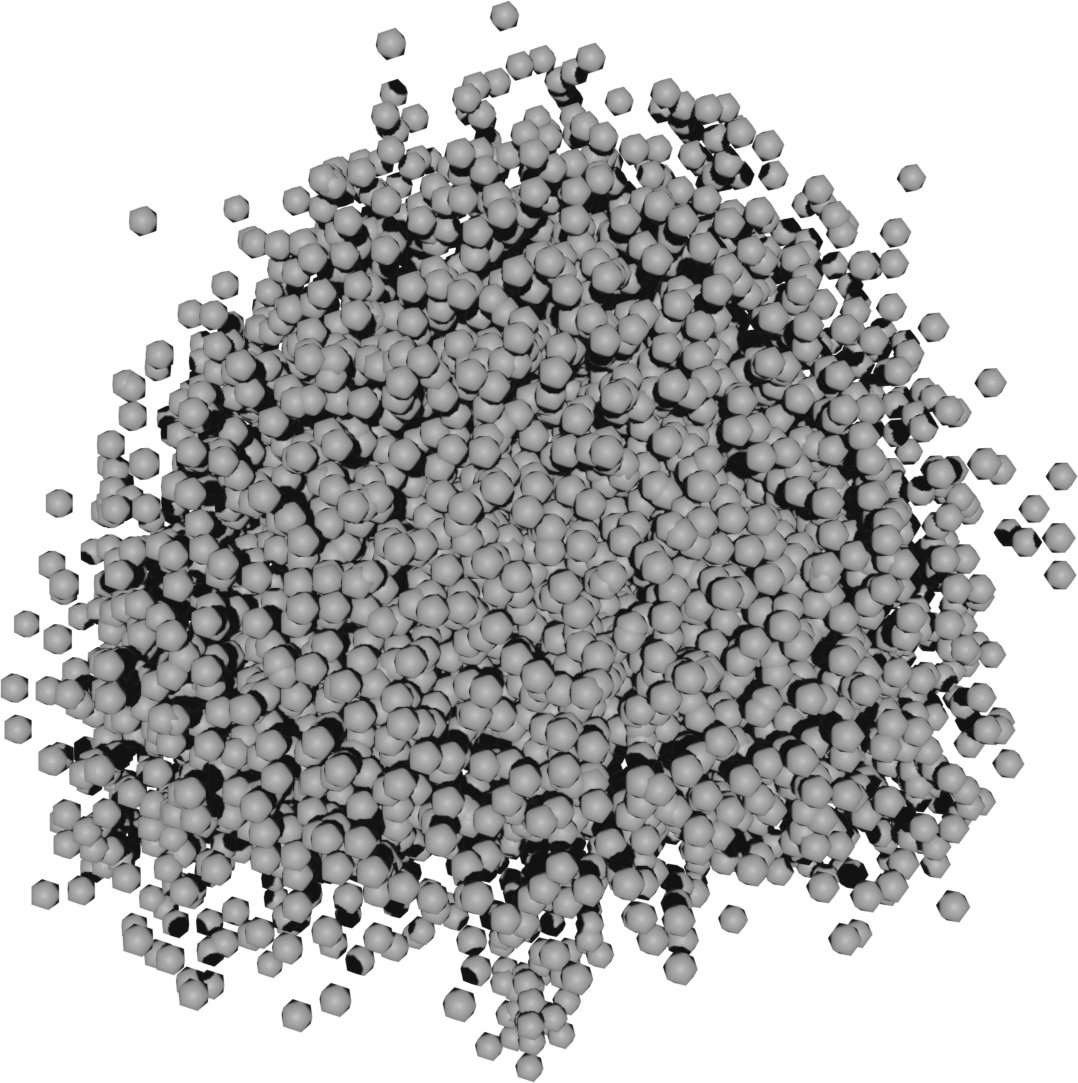}} &
\parbox[c][0.17\textwidth][c]{\linewidth}{\centering\includegraphics[width=\linewidth,height=0.12\textwidth,keepaspectratio]{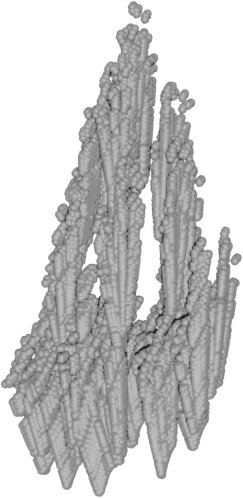}} &
\parbox[c][0.17\textwidth][c]{\linewidth}{\centering\includegraphics[width=\linewidth,height=0.12\textwidth,keepaspectratio]{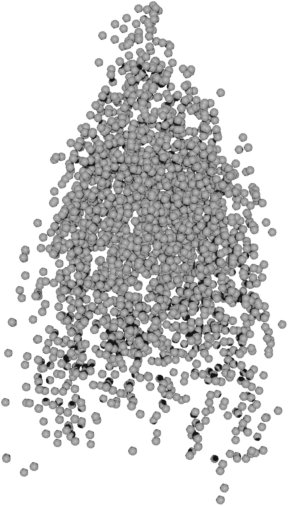}} &
\parbox[c][0.17\textwidth][c]{\linewidth}{\centering\includegraphics[width=\linewidth,height=0.12\textwidth,keepaspectratio]{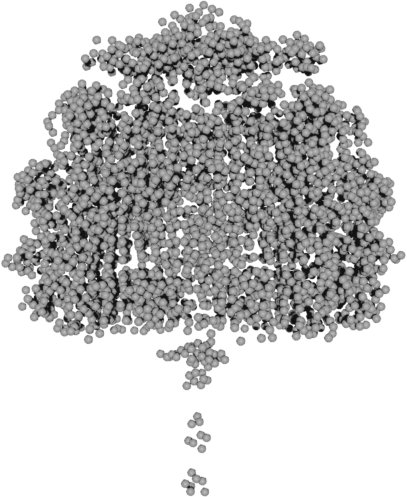}} \\
\hline
\parbox[c][0.17\textwidth][c]{\linewidth}{\centering\includegraphics[width=\linewidth,height=0.12\textwidth,keepaspectratio]{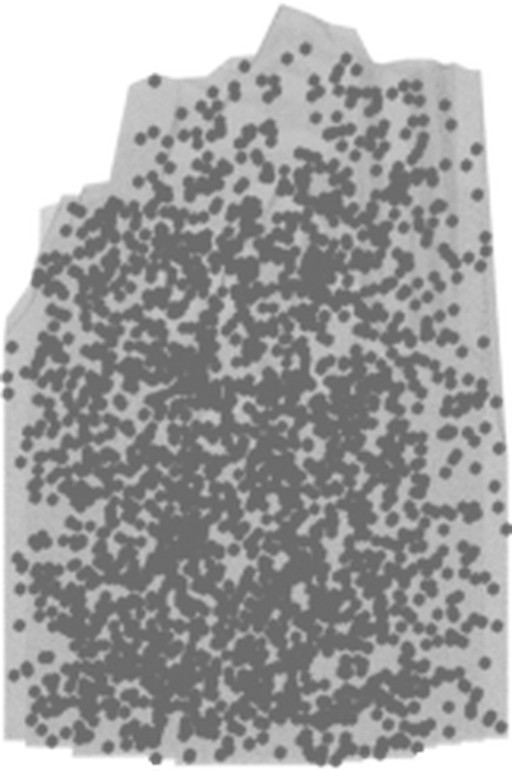}} &
\parbox[c][0.17\textwidth][c]{\linewidth}{\centering\includegraphics[width=\linewidth,height=0.12\textwidth,keepaspectratio]{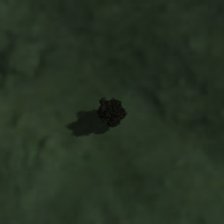}} &
\parbox[c][0.17\textwidth][c]{\linewidth}{\centering\includegraphics[width=\linewidth,height=0.12\textwidth,keepaspectratio]{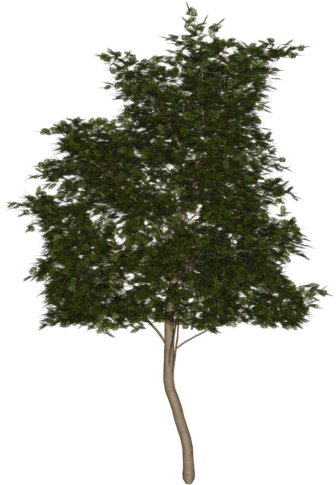}} &
\parbox[c][0.17\textwidth][c]{\linewidth}{\centering\includegraphics[width=\linewidth,height=0.12\textwidth,keepaspectratio]{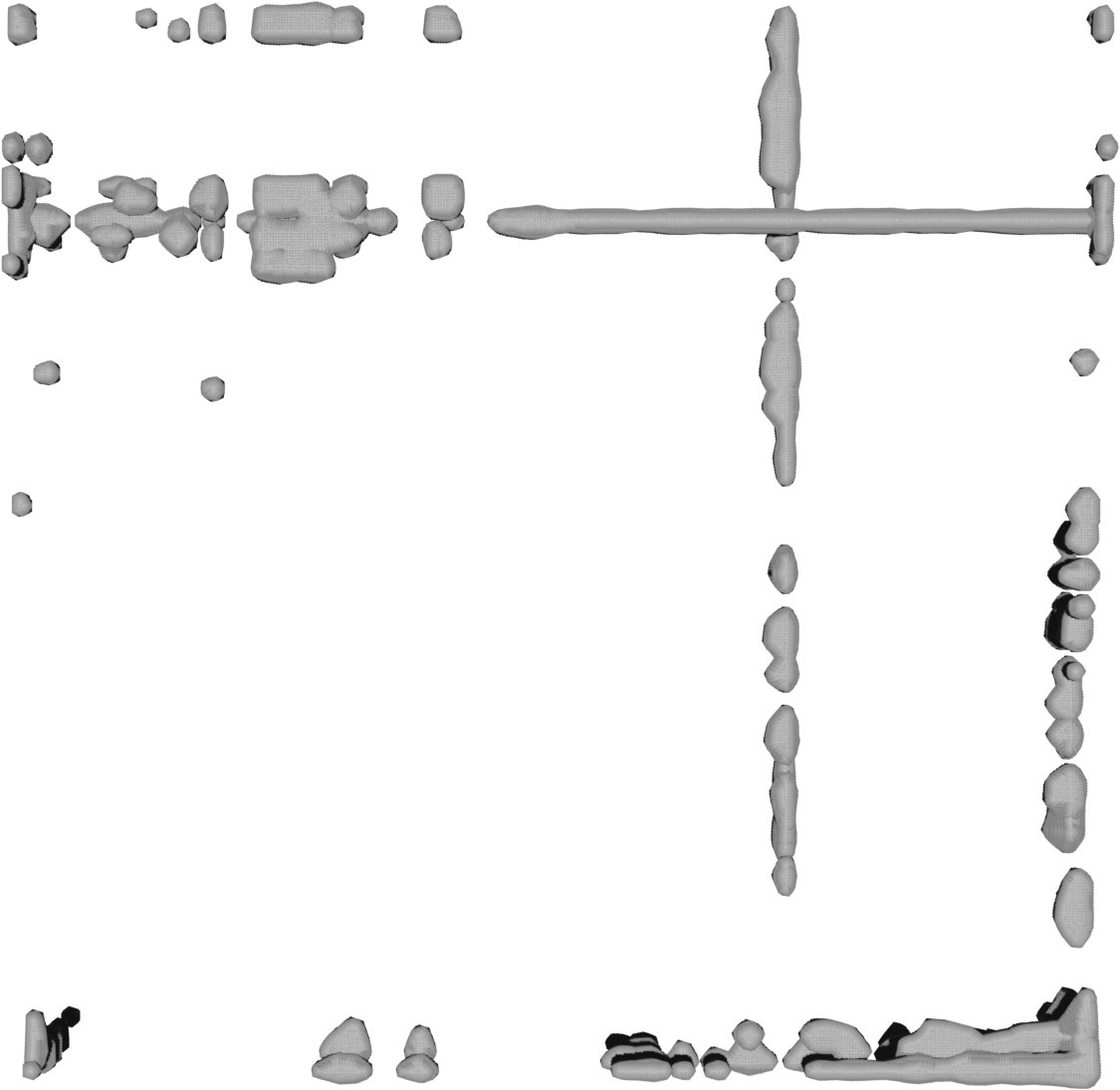}} &
\parbox[c][0.17\textwidth][c]{\linewidth}{\centering\includegraphics[width=\linewidth,height=0.12\textwidth,keepaspectratio]{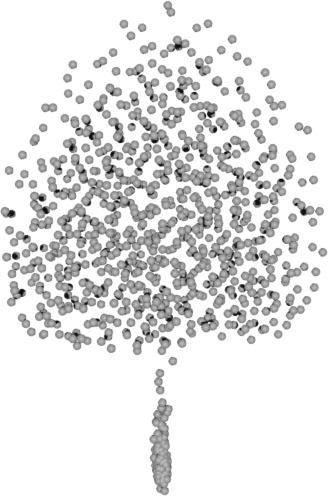}} &
\parbox[c][0.17\textwidth][c]{\linewidth}{\centering\includegraphics[width=\linewidth,height=0.12\textwidth,keepaspectratio]{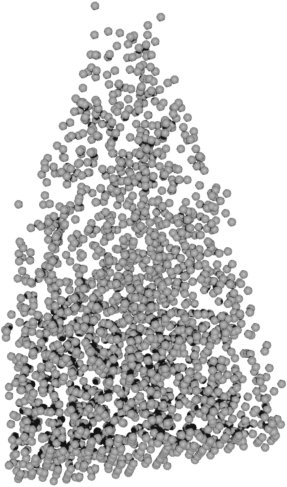}} &
\parbox[c][0.17\textwidth][c]{\linewidth}{\centering\includegraphics[width=\linewidth,height=0.12\textwidth,keepaspectratio]{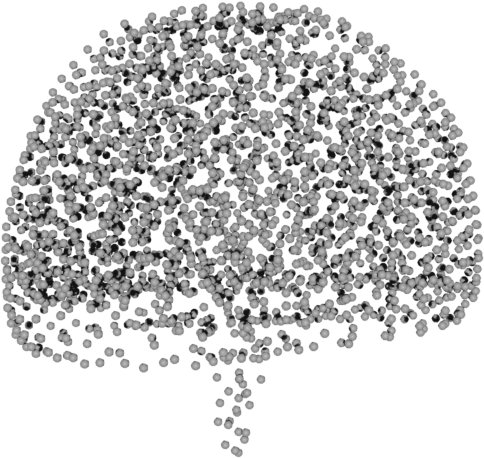}} &
\parbox[c][0.17\textwidth][c]{\linewidth}{\centering\includegraphics[width=\linewidth,height=0.12\textwidth,keepaspectratio]{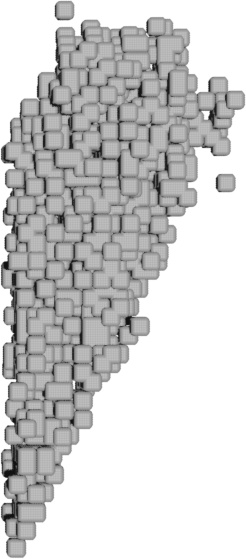}} & 
\parbox[c][0.17\textwidth][c]{\linewidth}{\centering\includegraphics[width=\linewidth,height=0.12\textwidth,keepaspectratio]{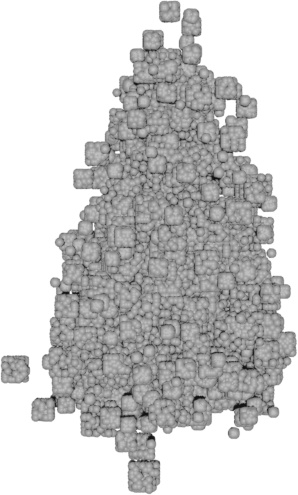}} &
\parbox[c][0.17\textwidth][c]{\linewidth}{\centering\includegraphics[width=\linewidth,height=0.12\textwidth,keepaspectratio]{RESULTS/SOTA/DPM/diffusion-point-cloud_0030_output.jpg}} &
\parbox[c][0.17\textwidth][c]{\linewidth}{\centering\includegraphics[width=\linewidth,height=0.12\textwidth,keepaspectratio]{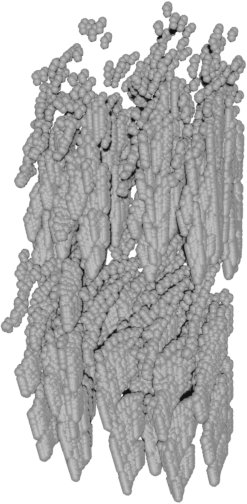}} &
\parbox[c][0.17\textwidth][c]{\linewidth}{\centering\includegraphics[width=\linewidth,height=0.12\textwidth,keepaspectratio]{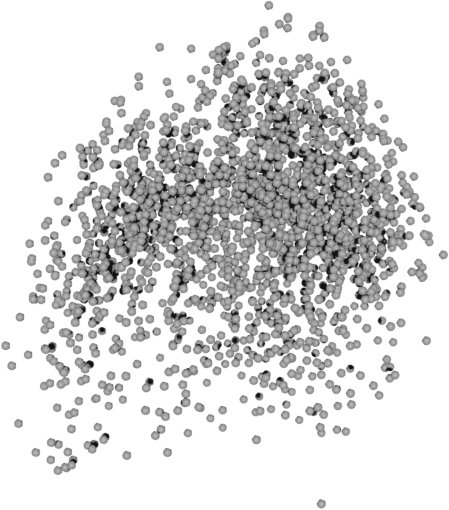}} &
\parbox[c][0.17\textwidth][c]{\linewidth}{\centering\includegraphics[width=\linewidth,height=0.12\textwidth,keepaspectratio]{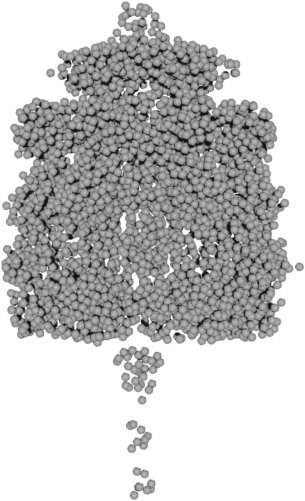}} \\
\bottomrule
\end{tabular}
}
\caption{Visual comparison across baseline models with our approach.}
\label{tab:sota_comparison}
\end{table*}

\subsection{Results on Austrian Landmark Trees}

Individual tree instances (orthophoto and DSM patches) are extracted from large scenes using the automatic segmentation and localization pipeline of Grammatikaki et al.~\cite{grammatikaki_2024_nsj20-6ka24}, ensuring consistent centering and scale across inputs.
To understand how the model generalizes to real world data we validated our approach using the dataset of landmark trees in Austria~\cite{grammatikaki_2024_nsj20-6ka24}. 

\textbf{Qualitative evaluation} In Table~\ref{tab:landmark_trees} each row corresponds to a single tree instance. From left to right, we display the DSM and orthophoto inputs used for reconstruction, a real-world photograph of the corresponding tree, and a rendering of the predicted 3D point cloud.

Despite the sparse input data, our model is able to reconstruct 3D tree shapes that match key morphological traits observed in the target photos-such as tree height, crown size, and overall silhouette. 
In addition, we observe substantial variation in the reconstructed shapes, which reflects natural diversity across species (e.g., coniferous vs. deciduous forms). This suggests that the model generalizes well, even in the absence of species labels or multi-view supervision.

An important strength of our approach is its ability to remain reliable when one of the input signals is degraded. Table~\ref{tab:landmark_trees} illustrates three such cases (1, 2, and 3) in the right column.    
In case 1, the DSM underestimates tree size, but the orthophoto compensates. In case 2, cluttered orthophotos are resolved by relying on the DSM, and in case 3, weak shadows in the orthophoto are balanced by the DSM structure, yielding coherent reconstructions in all cases.
These examples confirm that the dual-input design allows the network to adaptively rely on the most reliable signal, ensuring robust results even when individual inputs are imperfect.

\textbf{Quantitative evaluation}
Table~\ref{tab:landmark_quantitative} reports image-based quantitative metrics for the landmark trees in Table~\ref{tab:landmark_trees}. Silhouette IoU measures crown geometry agreement, while normalized CIELAB color error~\cite{cielab} and LPIPS assess appearance consistency. Typical cases achieve higher IoU and lower appearance error, while difficult cases degrade when input cues are misleading (cases~4–5). Case~2 reflects near-ideal DSM and orthophoto cues, while case~4 corresponds to a dead tree with ambiguous geometry and weak appearance signals, explaining the best and worst performance.

\begin{table}[h]
\centering
\small
\resizebox{\columnwidth}{!}{
\begin{tabular}{c ccc || c ccc}
\toprule
& \multicolumn{3}{c}{\textbf{Typical Reconstructions}} &
& \multicolumn{3}{c}{\textbf{Difficult Cases}} \\
\cmidrule{2-4} \cmidrule{6-8}
\multicolumn{1}{c}{} & \multicolumn{1}{c}{\textbf{IoU}} $\uparrow$ & \multicolumn{1}{c}{\textbf{Color}} $\downarrow$ & \multicolumn{1}{c}{\textbf{LPIPS}} $\downarrow$ &
\multicolumn{1}{c}{} & \multicolumn{1}{c}{\textbf{IoU}} $\uparrow$ & \multicolumn{1}{c}{\textbf{Color}} $\downarrow$ & \multicolumn{1}{c}{\textbf{LPIPS}} $\downarrow$ \\
\midrule
1 & 0.755 & 0.177 & 0.302 & 1 & 0.753 & 0.149 & 0.252 \\
2 & \textbf{0.841} & \textbf{0.132} & \textbf{0.212} & 2 & 0.735 & 0.160 & 0.323 \\
3 & 0.804 & 0.239 & 0.273 & 3 & 0.808 & 0.171 & 0.267 \\
4 & 0.759 & 0.152 & 0.298 & 4 & \textbf{0.470} & \textbf{0.317} & \textbf{0.475} \\
5 & 0.809 & 0.179 & 0.266 & 5 & 0.568 & 0.272 & 0.381 \\
\midrule
Mean & 0.794 & 0.176 & 0.270 &
Mean & 0.667 & 0.214 & 0.340 \\
\bottomrule
\end{tabular}
}
\caption{
Quantitative evaluation for the landmark trees shown in Table~\ref{tab:landmark_trees}. Row indices correspond to Table~3. Each cell reports IoU, normalized Color Error (CIELAB), and LPIPS.
Arrows ($\downarrow$/$\uparrow$) indicate whether lower or higher values are better.
}
\label{tab:landmark_quantitative}
\end{table}

\subsection{Comparison to State-of-the-Art Models}

Table~\ref{tab:sota_comparison} and Table~\ref{tab:baseline_metrics} compare our method with representative baselines from four categories: single-view reconstruction networks (AtlasNet~\cite{atlasnet}, TMNet~\cite{tmnet}, Zero123++~\cite{zero123plus}), point-based upsampling networks (PUGeoNet~\cite{pugeonet}, RepKPU~\cite{repkpu}), generative 3D models, including diffusion-based approaches (Diffusion PC (DPMs)~\cite{dpms}, 3DTopia-XL~\cite{3dtopiaxl}) and rectified-flow-based models (TRELLIS~\cite{trellis}), and large reconstruction models (OpenLRM~\cite{openlrm}).
We selected point-based upsampling methods (PUGeoNet, RepKPU) because they represent a natural baseline for refining sparse 3D inputs such as DSMs; single-view reconstruction networks (AtlasNet, TMNet, Zero123++), as they aim to infer full geometry from a single projected input, analogous to our orthophoto setting;  generative 3D models, including diffusion-based models (DPMs, 3DTopia-XL) and rectified-flow-based models (TRELLIS), as they have recently emerged as a strong paradigm for unconditional or weakly conditioned 3D generation; and large reconstruction models (OpenLRM) that assume multi-view calibrated imagery but represent the current frontier in scaling 3D reconstruction. Together, these categories span the main strategies that could, in principle, be adapted to reconstruct tree geometry from our DSM and orthophoto inputs.

Most of these baselines were not originally developed for DSM or orthophotography data. Instead, they were originally designed for object-centric benchmarks, such as ShapeNet, or synthetic point cloud datasets, where the goal is to reconstruct rigid CAD-like objects from multi-view RGB images with known camera intrinsics. AtlasNet and TMNet are optimized for mesh generation and deformation from single-view images, while Zero123++ performs single-view view synthesis.; PUGeoNet and RepKPU target point cloud upsampling on synthetic shapes; diffusion-based models focus on unconditional or multi-view 3D generation, while rectified-flow-based models use a different flow-based generative paradigm; and large models such as OpenLRM assume multi-view imagery and camera calibration. None of these assumptions holds in our setting, where input comes from top-down orthophotos and DSMs of irregular tree crowns. 

\begin{table}[H]
\centering
\begin{tabular}{lrrrr}
\toprule
\textbf{Method} & \textbf{CD} $\downarrow$ & \textbf{NCD} $\downarrow$ & \textbf{F1} $\uparrow$ & \textbf{COV} $\uparrow$ \\ 
\midrule
OpenLRM        & 10.604      & 0.735     & 0.001     & 75.8\%   \\
Zero123++        & 3.474  & 0.354 & 0.006   & 17.3\% \\ 
TMNet          & 1.436  & 0.338 & 0.016 & 47.6\% \\
AtlasNet       & 2.031  & 0.310 & 0.204   & 85.2\% \\
3DTopia-XL     & 1.351  & 0.301 & 0.212   & 24.4\% \\ 
Diffusion PC   & 1.148 & 0.277 & 0.353 & 32.8\%   \\
TRELLIS        & 1.412  & 0.327 & 0.124 & 33.3\% \\ 
PUGeoNet       & 5.759  & 0.330 & 0.002 & 35.2\% \\
RepKPU         & 1.828  & 0.968 & 0.038  & 31.9\% \\
Ours           & \textbf{0.969}  & \textbf{0.224} & \textbf{0.884}  & \textbf{90.7\%} \\
\bottomrule 
\end{tabular}
\caption{Comparison of baseline models on our dataset.}
\label{tab:baseline_metrics}
\end{table}

To ensure fairness, we retrain all baselines from scratch on our dataset using the same DSM and orthophoto inputs; for methods assuming calibrated or multi-view cameras (e.g., OpenLRM), we use replicated top-down orthophotos with a nadir-view camera.
Nonetheless, their architectural priors remain poorly aligned with our domain: tree crowns are highly irregular, lack rigid symmetries, and cannot be observed from side or front views. As a result, reconstructions often oversimplify canopy geometry, introduce noise, or miss structural details such as trunks. Qualitatively (Table~\ref{tab:sota_comparison}), single-modality baselines (DSM or orthophoto only) produce biased reconstructions, either oversimplifying or missing crown details. Single-view image-based reconstruction and view-synthesis models (AtlasNet, TMNet, Zero123++) reduce Chamfer Distance but oversmooth the canopy, while point-based upsampling methods (PUGeoNet, RepKPU) generate noisy or fragmented shapes, reflected in negligible F1 scores. 
Generative models (Diffusion PC, 3DTopia-XL, TRELLIS) achieve low Chamfer Distance (1.148, 1.351, and 1.412) but suffer from unstable structure and poor coverage (32.8\%, 24.4\%, and 33.3\%).
OpenLRM achieves very high coverage (75.8\%) but almost zero F1, showing that while outputs are diverse, they lack geometric consistency; this is expected, as OpenLRM assumes calibrated multi-view side and front imagery, which is incompatible with our top-down DSM-orthophoto inputs.
In contrast, our approach achieves the best balance (Table~\ref{tab:baseline_metrics}): low CD (0.969), the lowest NCD (0.224), and the highest F1 (0.884), while maintaining strong coverage (90.7\%). This demonstrates that our model captures sharper crowns and more coherent geometry, aligning closely with ground truth.

\subsection{Comparison with high-resolution LiDAR data}
We further validate our reconstructions against IGN LiDAR scans from the Pyrenees. Table~\ref{tab:lidar_comparison} shows that our outputs align well with the LiDAR shapes, capturing overall height and crown spread. 
While fine details (e.g., branching) differ, the consistency across modalities confirms that our method recovers realistic tree geometry from sparse geodata.

\begin{table}[h]
\centering
\resizebox{\columnwidth}{!}{%
\begin{tabular}{c p{0.2\columnwidth} p{0.2\columnwidth} p{0.2\columnwidth} p{0.2\columnwidth}}
\toprule
& \multicolumn{1}{c}{\textbf{DSM}} & 
\multicolumn{1}{c}{\textbf{Orthophoto}} & 
\multicolumn{1}{c}{\textbf{LiDAR}} & 
\multicolumn{1}{c}{\textbf{Ours}} \\
\midrule
\textbf{1} &
\parbox{\linewidth}{\centering\includegraphics[width=\linewidth,height=0.18\columnwidth,keepaspectratio]{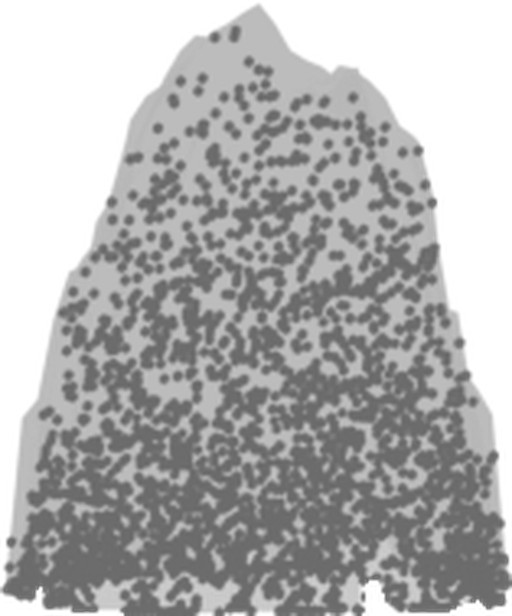}\rule{0pt}{0.18\columnwidth}} &
\parbox{\linewidth}{\centering\includegraphics[width=\linewidth,height=0.18\columnwidth,keepaspectratio]{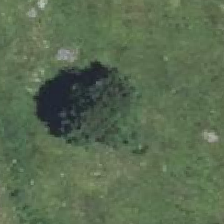}\rule{0pt}{0.18\columnwidth}} &
\parbox{\linewidth}{\centering\includegraphics[width=\linewidth,height=0.18\columnwidth,keepaspectratio]{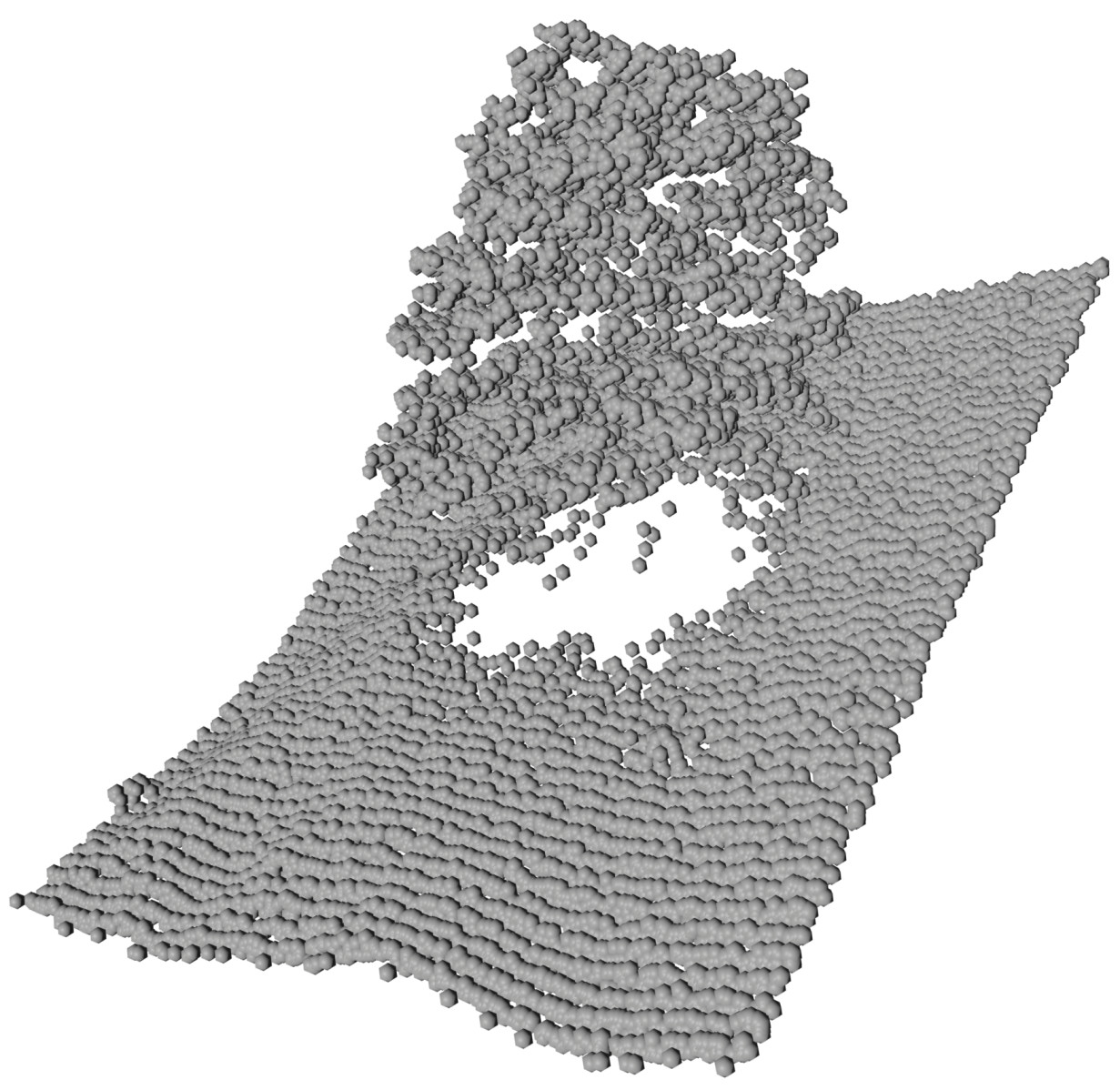}\rule{0pt}{0.18\columnwidth}} &
\parbox{\linewidth}{\centering\includegraphics[width=\linewidth,height=0.18\columnwidth,keepaspectratio]{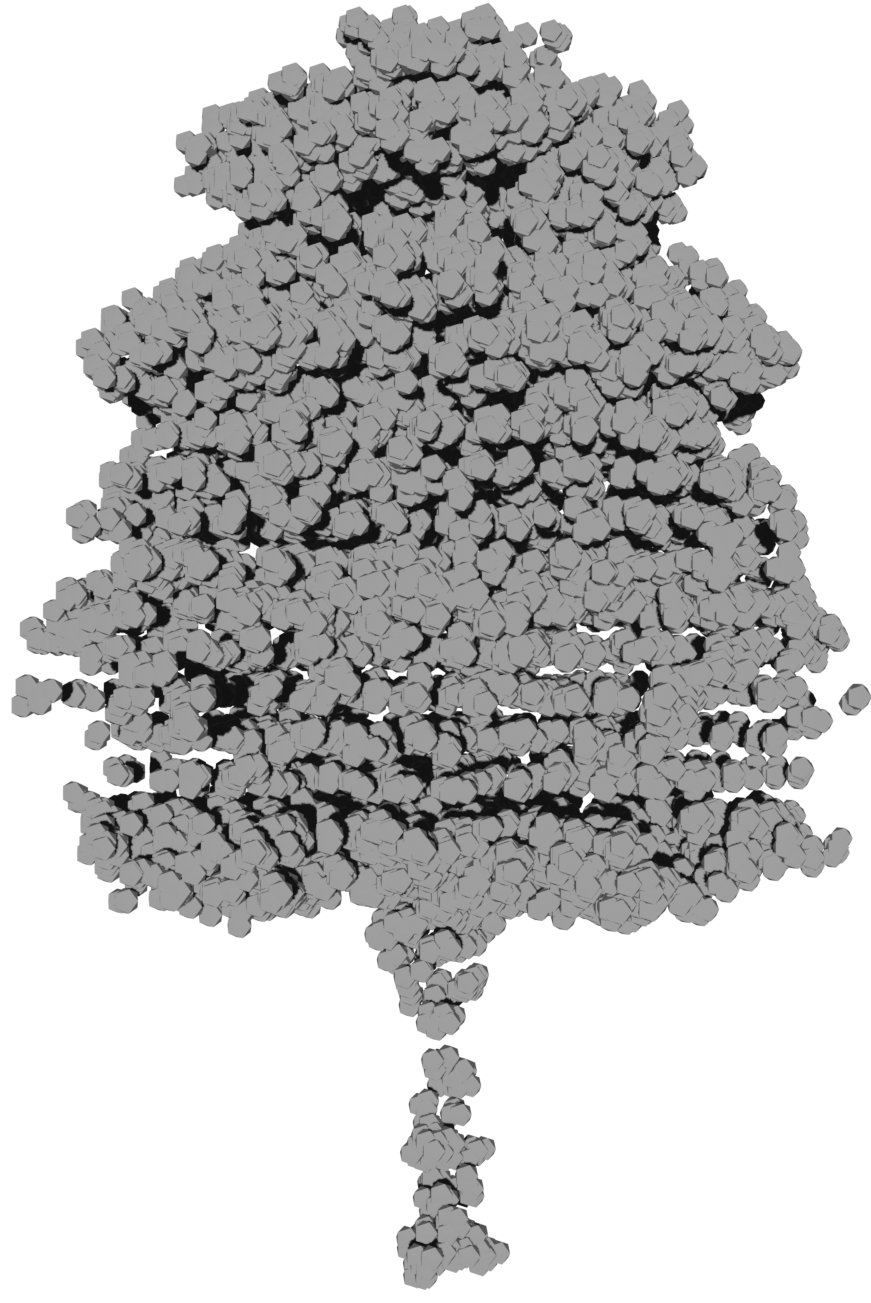}\rule{0pt}{0.18\columnwidth}} \\
\midrule
\textbf{2} &
\parbox{\linewidth}{\centering\includegraphics[width=\linewidth,height=0.18\columnwidth,keepaspectratio]{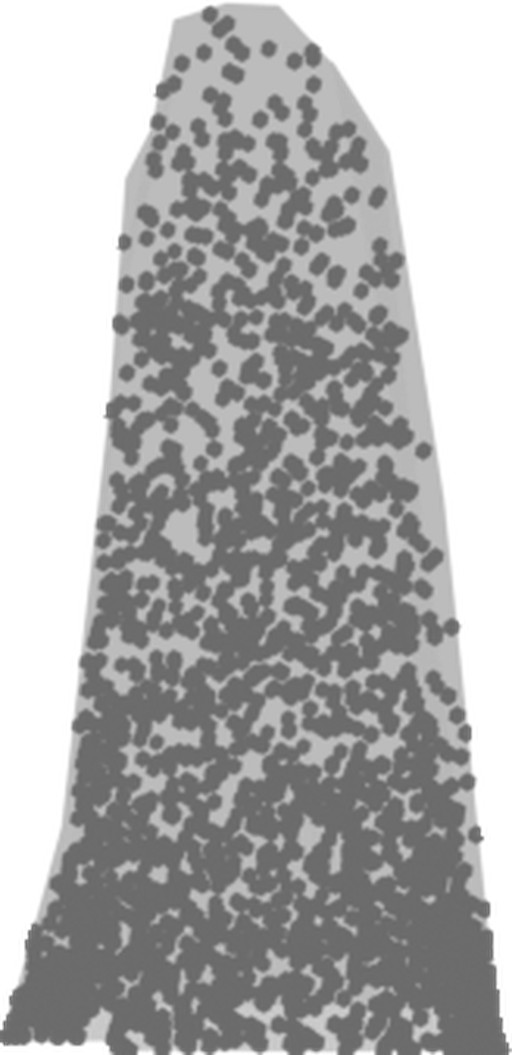}\rule{0pt}{0.18\columnwidth}} &
\parbox{\linewidth}{\centering\includegraphics[width=\linewidth,height=0.18\columnwidth,keepaspectratio]{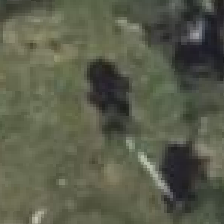}\rule{0pt}{0.18\columnwidth}} &
\parbox{\linewidth}{\centering\includegraphics[width=\linewidth,height=0.18\columnwidth,keepaspectratio]{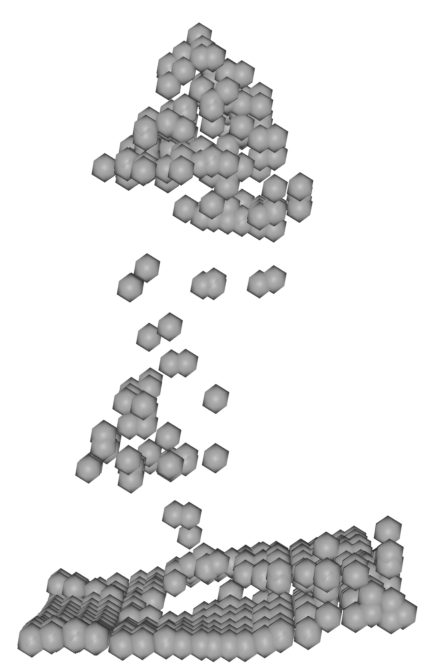}\rule{0pt}{0.18\columnwidth}} &
\parbox{\linewidth}{\centering\includegraphics[width=\linewidth,height=0.18\columnwidth,keepaspectratio]{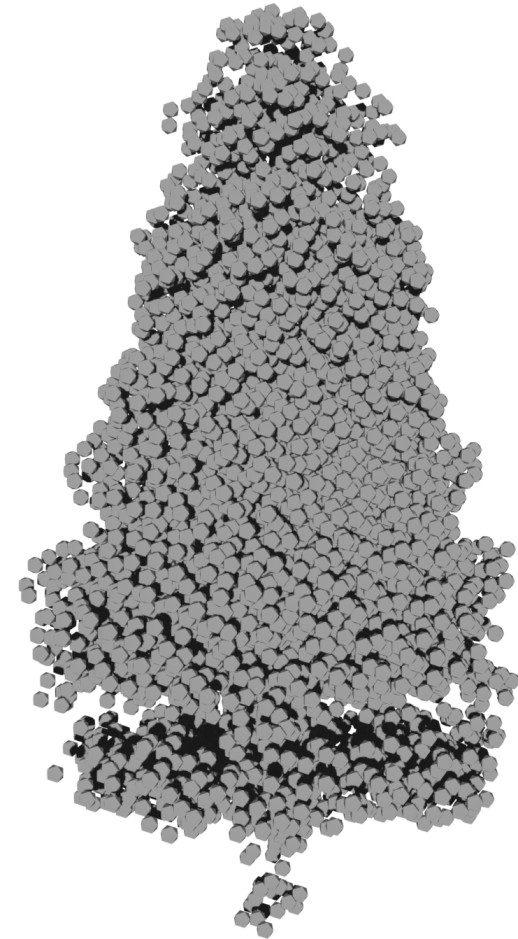}\rule{0pt}{0.18\columnwidth}} \\
\bottomrule
\end{tabular}
}
\caption{Qualitative comparison of our reconstructions with IGN LiDAR and orthophoto data in the French Pyrenees. Point cloud density from aerial LiDAR is 37 and 25 pts/m$^2$, respectively.}
\label{tab:lidar_comparison}
\end{table}

\subsection{Computational Efficiency}
In addition to accuracy and realism, our method achieves high computational efficiency. A full pipeline run, including postprocessing, takes on average 0.3\,s per tree on a single GPU. 
Each reconstructed tree requires only $\sim$0.2\,MB of memory on disk, compared to tens of MBs for high-detailed meshes. This is an order of magnitude faster and lighter than the high-detail trees from Grammatikaki et al.~\cite{Grammatikaki09052025}, the only prior non-network method using the same DSM and orthophoto inputs to reconstruct trees, reported render times of up to 15\,s per tree for 30M-triangle meshes and storage requirements in the tens of MB range. 
In contrast, TreeON performs end-to-end reconstruction and directly predicts a compact 6k-point colored point cloud at mapping-relevant resolution, rather than relying on a staged classification–procedural pipeline.
As both approaches rely on the same Grove-generated source meshes, subsampling effects are already reflected in Table~\ref{tab:ablation_metrics_split}, enabling a fair comparison at mapping-relevant resolution.

\section{Limitations}

A key limitation of our approach, as with any supervised ML method, is its dependence on the quality and representativeness of the training data.
In our case, synthetic meshes define which geometric structures the model can learn to reconstruct. 
If certain species, sizes, or crown shapes are underrepresented, the model may be biased toward more common patterns and struggle with rare or unseen structures. 
An instance of this limitation is the seasonal bias in our dataset: 
Most samples are drawn from summer orthophotos, which expose the model mainly to trees with full foliage under consistent lighting conditions. 
This reduces robustness to other seasonal appearances, such as leafless winter trees or dead trees (a reconstruction example is shown in Table~\ref{tab:landmark_trees}, case 4 on the right), snow cover, or different vegetation stages, that are rarely included in the training data. 
While our current dataset does not account for this, seasonal variability could in principle be simulated directly in our training pipeline and learned by the model. Exploring such synthetic augmentation would be an interesting direction for future work, enabling the model to handle leafless or snow-covered trees.

Another challenge is the difficulty of quantitatively assessing generalizability to real-world data. While we provide qualitative examples and landmark tree reconstructions, a systematic benchmark on real orthophotos and DSMs is challenging due to the lack of high-quality paired ground truth. This underscores the motivation for using synthetic training data in the first place, but it also indicates that evaluating domain transfer remains only partially addressed.

\section{Conclusion}
We introduced TreeON, a neural framework for reconstructing detailed 3D tree point clouds from sparse geospatial inputs. DSMs provide strong geometric cues, while orthophotos complement them with appearance and shadow information. Combining both modalities with occupancy, shadow, and silhouette supervision achieves higher reconstruction quality than single-image or upsampling methods.

Where previous work typically fulfills only some requirements for rural 3D mapping: visual appeal, structural plausibility, efficiency, and automation --- our framework addresses all four simultaneously. This is enabled by three key ingredients: (1) the strong geometric foundation from DSMs enriched by orthophoto-based appearance and shadow cues, (2) a multi-loss training strategy combining occupancy, shadow, and silhouette supervision, and (3) a compact implicit representation that allows fast inference and lightweight outputs.

The method provides structurally plausible reconstructions on a large scale, with light inference times (0.3s per tree) and lightweight outputs (0.2MB). This combination of quality and efficiency makes it particularly well-suited for integration into digital 3D maps. It effectively bridges the gap between oversimplified 2.5D surfaces and more data-rich, computationally intensive reconstructions, prioritizing visual plausibility over biological fidelity.

\section*{Acknowledgments}
The authors acknowledge TU Wien Bibliothek for financial support through its Open Access Funding Programme. The VRVis GmbH is funded by BMIMI, BMWET, Tyrol, Vorarlberg and Vienna Business Agency in the scope of COMET - Competence Centers for Excellent Technologies (904918, 911654) which is managed by FFG.
This research has been funded by WWTF project ICT22-055 - Instant Visualization and Interaction for Large Point Clouds, 
and Grant PID2021-122136OB-C21 funded by MICIU/AEI/10.13039/501100011033 and ERDF/EU.

\printbibliography

\printbibliography

@String{cgforum = "Computer Graphics Forum"}

@String{tog = "ACM TOG"}

@inproceedings{fan2017point,
  title={A point set generation network for 3D object reconstruction from a single image},
  author={Fan, Haoqiang and Su, Hao and Guibas, Leonidas J},
  booktitle={CVPR},
  year={2017}
}

@inproceedings{mescheder2019occupancy,
  title={Occupancy networks: Learning 3D reconstruction in function space},
  author={Mescheder, Lars and Oechsle, Michael and Niemeyer, Michael and Nowozin, Sebastian and Geiger, Andreas},
  booktitle={CVPR},
  year={2019}
}

@article{shlyakhter_reconstructing_2001,
	title = {Reconstructing {3{D}} tree models from instrumented photographs},
	volume = {21},
	issn = {02721716},
	doi = {10.1109/38.920627},
	number = {1},
	journal = ieee_cg_a,
	author = {Shlyakhter, I. and Rozenoer, M. and Dorsey, J. and Teller, S.},
	year = {2001},
	pages = {53--61}
}

@article{bradley_image-based_2013,
	title = {Image-based reconstruction and synthesis of dense foliage},
	volume = {32},
	issn = {0730-0301, 1557-7368},
	doi = {10.1145/2461912.2461952},
	number = {4},
    journal = {ACM Trans. Graph.},
	author = {Bradley, Derek and Nowrouzezahrai, Derek and Beardsley, Paul},
	year = {2013},
	pages = {1--10}
}

@article{tan_image-based_2007,
    author = {Tan, Ping and Zeng, Gang and Wang, Jingdong and Kang, Sing Bing and Quan, Long},
    year = {2007},
    pages = {87},
    title = {Image-based tree modeling},
    volume = {26},
    journal = {ACM Trans. Graph.}}

@article{argudo_single-picture_2016,
    title = {Single-picture reconstruction and rendering of trees for plausible vegetation synthesis},
    journal = {Computers \& Graphics},
    volume = {57},
    pages = {55-67},
    year = {2016},
    issn = {0097-8493},
    doi = {https://doi.org/10.1016/j.cag.2016.03.005},
    author = {Oscar Argudo and Antonio Chica and Carlos And\'ujar},
}

@article{inverse_procedural_modelling,
    author = {Stava, Ondrej and Pirk, Sören and Kratt, Julian and Chen, Baoquan and Mech, Radomir and Deussen, Oliver and Benes, Bedrich},
    year = {2014},
    pages = {},
    title = {Inverse Procedural Modelling of Trees},
    volume = {33},
    journal = cgforum,
    doi = {10.1111/cgf.12282}
}

@article{ARGUDO201823,
    title = {Segmentation of aerial images for plausible detail synthesis},
    journal = {Computers \& Graphics},
    volume = {71},
    pages = {23-34},
    year = {2018},
    issn = {0097-8493},
    doi = {https://doi.org/10.1016/j.cag.2017.11.004},
    author = {Oscar Argudo and Marc Comino and Antonio Chica and Carlos And\'ujar and Felipe Lumbreras}
}

@Article{rs16030524,
    AUTHOR = {Gong, Haoyu and Sun, Qian and Fang, Chenrong and Sun, Le and Su, Ran},
    TITLE = {Tree{D}etector: {U}sing {D}eep {L}earning for the {L}ocalization and {R}econstruction of {U}rban {T}rees from {H}igh-{R}esolution {R}emote {S}ensing {I}mages},
    JOURNAL = {Remote Sensing},
    VOLUME = {16},
    YEAR = {2024},
    NUMBER = {3},
    ARTICLE-NUMBER = {524},
    ISSN = {2072-4292},
    DOI = {10.3390/rs16030524}
}

@String{tog = "ACM Transactions on Graphics"}

@String{ieee_cg_a = "IEEE Computer Graphics and Applications"}

@misc{datagv,
    author = {{Offene Daten Österreichs}},
    title = {Offene {Daten} {Ö}sterreichs},
    howpublished = {\url{https://www.data.gv.at}},
    year = {2025},
    note = {Accessed: 2025-11-03}
}

@misc{basemapat,
    author = {{basemap.at}},
    title = {basemap.at},
    howpublished = {\url{https://basemap.at/}},
    year = {2025},
    note = {Accessed: 2025-11-03}
}

@INPROCEEDINGS{he2015deepresiduallearningimage,

  author={He, Kaiming and Zhang, Xiangyu and Ren, Shaoqing and Sun, Jian},

  booktitle={2016 IEEE Conference on Computer Vision and Pattern Recognition (CVPR)}, 

  title={Deep Residual Learning for Image Recognition}, 

  year={2016},

  volume={},

  number={},

  pages={770-778},

  doi={10.1109/CVPR.2016.90}}

@inproceedings{liu2019softras,
  title={Soft Rasterizer: A Differentiable Renderer for Image-based 3D Reasoning},
  author={Liu, Shichen and Li, Tianye and Chen, Weikai and Li, Hao},
  booktitle={Proceedings of the IEEE/CVF International Conference on Computer Vision},
  pages={7708--7717},
  year={2019}
}

@article{sheppard2005landscape,
    title = {Landscape visualisation and climate change: the potential for influencing perceptions and behaviour},
    journal = {Environmental Science \& Policy},
    volume = {8},
    number = {6},
    pages = {637-654},
    year = {2005},
    issn = {1462-9011},
    doi = {https://doi.org/10.1016/j.envsci.2005.08.002},
    author = {Stephen R.J. Sheppard}
}

@article{xu_knowledge_2007,
	title = {Knowledge and heuristic-based modeling of laser-scanned trees},
	volume = {26},
	issn = {0730-0301, 1557-7368},
	number = {4},
	journal = tog,
	author = {Xu, Hui and Gossett, Nathan and Chen, Baoquan},
	year = {2007},
	pages = {19},
    doi = {10.1145/1289603.1289610}
}

@article{Hu,
author = {Hu, Shaojun and Li, Zhengrong and Zhiyi, Zhang and He, Dongjian and Wimmer, Michael},
year = {2017},
month = {05},
pages = {},
title = {Efficient Tree Modeling from Airborne LiDAR Point Clouds},
volume = {67},
journal = {Computers \& Graphics},
doi = {10.1016/j.cag.2017.04.004}
}

@article{du_adtree_2019,
	title = {{AdTree}: {Accurate}, {Detailed}, and {Automatic} {Modelling} of {Laser}-{Scanned} {Trees}},
	volume = {11},
	copyright = {http://creativecommons.org/licenses/by/3.0/},
	issn = {2072-4292},
	shorttitle = {{AdTree}},
	url = {https://www.mdpi.com/2072-4292/11/18/2074},
	doi = {10.3390/rs11182074},
	language = {en},
	number = {18},
	urldate = {2025-08-13},
	journal = {Remote Sensing},
	author = {Du, Shenglan and Lindenbergh, Roderik and Ledoux, Hugo and Stoter, Jantien and Nan, Liangliang},
	month = jan,
	year = {2019},
	note = {Publisher: Multidisciplinary Digital Publishing Institute},
	pages = {2074},
}

@article{Grammatikaki09052025,
author = {Angeliki Grammatikaki and Johannes Eschner and Florian Ledermann and Oscar Argudo and Manuela Waldner},
title = {How to represent landmark trees in digital 3D maps? An automated workflow and user study},
journal = {Cartography and Geographic Information Science},
volume = {0},
number = {0},
pages = {1--18},
year = {2025},
publisher = {Taylor \& Francis},
doi = {10.1080/15230406.2025.2489543},


url = { 
    
        https://doi.org/10.1080/15230406.2025.2489543
    
    

}

}

@techreport{AustriaForestReport2023,
  title        = {Austrian Forest Report 2023: We Take Care of the Forest},
  author       = {{Federal Ministry of Agriculture, Forestry, Regions and Water Management}},
  month        = aug,
  year         = {2023},
  institution  = {Federal Ministry of Agriculture, Forestry, Regions and Water Management},
  address      = {Vienna, Austria}
}

@misc{grove,
    author = {{The Grove 3D}},
    title = {{The Grove 3D}},
    howpublished = {\url{https://www.thegrove3d.com/}},
    year = {2025},
    note = {Accessed: 2025-03-13}
}

@inproceedings{zhang2018perceptual,
  title={The Unreasonable Effectiveness of Deep Features as a Perceptual Metric},
  author={Zhang, Richard and Isola, Phillip and Efros, Alexei A and Shechtman, Eli and Wang, Oliver},
  booktitle={CVPR},
  year={2018}
}

@article{Xu,
author = {Xu, Ling and Mould, David},
title = {Procedural Tree Modeling with Guiding Vectors},
journal = {Computer Graphics Forum},
volume = {34},
number = {7},
pages = {47-56},
doi = {https://doi.org/10.1111/cgf.12744},
url = {https://onlinelibrary.wiley.com/doi/abs/10.1111/cgf.12744},
eprint = {https://onlinelibrary.wiley.com/doi/pdf/10.1111/cgf.12744},
year = {2015}
}

@article{Jaeger,
author = {J{\ae}ger, Peter and Nilsson, Niels and Palamas, George},
year = {2021},
month = {03},
pages = {},
title = {Can't see the Forest for the Trees: Perceiving Realism of Procedural Generated Trees in First-Person Games},
journal = {EAI Endorsed Transactions on Creative Technologies},
doi = {10.4108/eai.16-3-2021.169029}
}

@misc{yang2022rulebasedproceduraltreemodeling,
      title={Rule-based Procedural Tree Modeling Approach}, 
      author={Yinhui Yang and Rui Wang and Yuchi Huo},
      year={2022},
      eprint={2204.03237},
      archivePrefix={arXiv},
      primaryClass={cs.GR},
      url={https://arxiv.org/abs/2204.03237}, 
}

@misc{alonso2025moreefficientpointcloud,
      title={Less is More: Efficient Point Cloud Reconstruction via Multi-Head Decoders}, 
      author={Pedro Alonso and Tianrui Li and Chongshou Li},
      year={2025},
      eprint={2505.19057},
      archivePrefix={arXiv},
      primaryClass={cs.CV},
      url={https://arxiv.org/abs/2505.19057}, 
}

@misc{achlioptas2018learningrepresentationsgenerativemodels,
      title={Learning Representations and Generative Models for 3D Point Clouds}, 
      author={Panos Achlioptas and Olga Diamanti and Ioannis Mitliagkas and Leonidas Guibas},
      year={2018},
      eprint={1707.02392},
      archivePrefix={arXiv},
      primaryClass={cs.CV},
      url={https://arxiv.org/abs/1707.02392}, 
}

@article{ZHANG2022103551,
title = {Dual adversarial model: Exploring low-dimensional space features for point clouds generating and completing},
journal = {Computer Vision and Image Understanding},
volume = {223},
pages = {103551},
year = {2022},
issn = {1077-3142},
doi = {https://doi.org/10.1016/j.cviu.2022.103551},
url = {https://www.sciencedirect.com/science/article/pii/S1077314222001291},
author = {Yuhang Zhang and Zhenwei Miao and Tiebin Mi and Jie Li and Robert C. Qiu},
}

@inproceedings{Sokolova,
author = {Sokolova, Marina and Japkowicz, Nathalie and Szpakowicz, Stan},
year = {2006},
month = {01},
pages = {1015-1021},
title = {Beyond Accuracy, F-Score and ROC: A Family of Discriminant Measures for Performance Evaluation},
volume = {Vol. 4304},
isbn = {978-3-540-49787-5},
journal = {AI 2006: Advances in Artificial Intelligence, Lecture Notes in Computer Science},
doi = {10.1007/11941439_114}
}

@INPROCEEDINGS{10350393,

  author={Llull, Cristián and Baloian, Nelson and Bustos, Benjamin and Kupczik, Kornelius and Sipiran, Ivan and Baloian, Andrés},

  booktitle={2023 IEEE/CVF International Conference on Computer Vision Workshops (ICCVW)}, 

  title={Evaluation of 3D Reconstruction for Cultural Heritage Applications}, 

  year={2023},

  volume={},

  number={},

  pages={1634-1643},

  doi={10.1109/ICCVW60793.2023.00179}}

@dataset{grammatikaki_2024_nsj20-6ka24,
  author       = {Grammatikaki, Angeliki},
  title        = {Landmark trees in Austria},
  month        = mar,
  year         = 2024,
  publisher    = {TU Wien},
  version      = {1.0.0},
  doi          = {10.48436/nsj20-6ka24},
  url          = {https://doi.org/10.48436/nsj20-6ka24},
}

@article{snavely_photo_2006,
	title = {Photo tourism: exploring photo collections in {3D}},
	volume = {25},
	issn = {0730-0301},
	shorttitle = {Photo tourism},
	url = {https://dl.acm.org/doi/10.1145/1141911.1141964},
	doi = {10.1145/1141911.1141964},
	number = {3},
	urldate = {2025-08-13},
	journal = {ACM Trans. Graph.},
	author = {Snavely, Noah and Seitz, Steven M. and Szeliski, Richard},
	month = jul,
	year = {2006},
	pages = {835--846},
}

@article{indirabai_terrestrial_2019,
	title = {Terrestrial laser scanner based {3D} reconstruction of trees and retrieval of leaf area index in a forest environment},
	volume = {53},
	issn = {1574-9541},
	url = {https://www.sciencedirect.com/science/article/pii/S1574954118303182},
	doi = {https://doi.org/10.1016/j.ecoinf.2019.100986},
	journal = {Ecological Informatics},
	author = {Indirabai, Indu and Nair, M. V. Harindranathan and Jaishanker, R. Nair and Nidamanuri, Rama Rao},
	year = {2019},
	pages = {100986},
}

@article{you_tree_2021,
	title = {Tree {Extraction} from {Airborne} {Laser} {Scanning} {Data} in {Urban} {Areas}},
	volume = {13},
	doi = {10.3390/rs13173428},
	journal = {Remote Sensing},
	author = {You, Hangkai and Li, Shihua and Xu, Yifan and He, Ze and Wang, Di},
	month = aug,
	year = {2021},
	pages = {3428},
}

@article{choudhury_photogrammetry_2019,
	title = {Photogrammetry and {Remote} {Sensing} for the identification and characterization of trees in urban areas.},
	volume = {1249},
	doi = {10.1088/1742-6596/1249/1/012008},
	journal = {Journal of Physics: Conference Series},
	author = {Choudhury, Md and Costanzini, Sofia and Despini, Francesca and Rossi, Paolo and Galli, Andrea and Marcheggiani, Ernesto and Teggi, Sergio},
	month = may,
	year = {2019},
	pages = {012008},
}

@article{bostrom_urban_2006,
	title = {Urban {3D} modelling using terrestrial laser scanners},
	volume = {36},
	author = {Bostr{\"o}m, G. and Fiocco, M. and Gon{\c{c}}alves, Joao and Sequeira, V\'{i}tor},
	month = jan,
  journal={International Archives of Photogrammetry, Remote Sensing and Spatial Information Sciences},
  volume={36},
  number={Part5},
  pages={279--284},
	year = {2006},
}

@InProceedings{10.1007/978-3-642-60243-6_12,
author="Mayer, Helmut
and Mayr, Wilhelm",
editor="F{\"o}rstner, Wolfgang
and Buhmann, Joachim M.
and Faber, Annett
and Faber, Petko",
title="Automatic Extraction of Deciduous Trees from High Resolution Aerial Imagery",
booktitle="Mustererkennung 1999",
year="1999",
publisher="Springer Berlin Heidelberg",
address="Berlin, Heidelberg",
pages="102--110",
isbn="978-3-642-60243-6"
}

@article{HIRSCHMUGL2007533,
  title = {Single tree detection in very high resolution remote sensing data},
  author = {Hirschmugl, Manuela and Ofner, Martin and Raggam, Johann and Schardt, Mathias},
  journal = {Remote Sensing of Environment},
  volume = {110},
  number = {4},
  pages = {533--544},
  year = {2007},
  note = {ForestSAT Special Issue},
  issn = {0034-4257},
  doi = {10.1016/j.rse.2007.02.029}
}

@misc{zhou2023deeptreemodelingtreessituated,
      title={DeepTree: Modeling Trees with Situated Latents}, 
      author={Xiaochen Zhou and Bosheng Li and Bedrich Benes and Songlin Fei and Sören Pirk},
      year={2023},
      eprint={2305.05153},
      archivePrefix={arXiv},
      primaryClass={cs.LG},
      url={https://arxiv.org/abs/2305.05153}, 
}

@inproceedings{10.1145/3721250.3743024,
author = {Todd, Grace and Bailey, Mike},
title = {Foliager: Procedural Forest Generation from Natural Language Using Scientific Data and AI},
year = {2025},
isbn = {9798400715495},
publisher = {Association for Computing Machinery},
address = {New York, NY, USA},
url = {https://doi.org/10.1145/3721250.3743024},
doi = {10.1145/3721250.3743024},
booktitle = {Proceedings of the Special Interest Group on Computer Graphics and Interactive Techniques Conference Posters},
articleno = {16},
numpages = {2},
location = {
},
series = {SIGGRAPH Posters '25}
}

@misc{huang2023evaluatingpointcloudindividual,
      title={Evaluating the point cloud of individual trees generated from images based on Neural Radiance fields (NeRF) method}, 
      author={Hongyu Huang and Guoji Tian and Chongcheng Chen},
      year={2023},
      eprint={2312.03372},
      archivePrefix={arXiv},
      primaryClass={cs.CV},
      url={https://arxiv.org/abs/2312.03372}, 
}

@misc{lee2024treedfusionsimulationreadytree,
      title={Tree-D Fusion: Simulation-Ready Tree Dataset from Single Images with Diffusion Priors}, 
      author={Jae Joong Lee and Bosheng Li and Sara Beery and Jonathan Huang and Songlin Fei and Raymond A. Yeh and Bedrich Benes},
      year={2024},
      eprint={2407.10330},
      archivePrefix={arXiv},
      primaryClass={cs.CV},
      url={https://arxiv.org/abs/2407.10330}, 
}

@INPROCEEDINGS{10656708,

  author={Li, Yuan and Liu, Zhihao and Benes, Bedrich and Zhang, Xiaopeng and Guo, Jianwei},

  booktitle={2024 IEEE/CVF Conference on Computer Vision and Pattern Recognition (CVPR)}, 

  title={SVDTree: Semantic Voxel Diffusion for Single Image Tree Reconstruction}, 

  year={2024},

  volume={},

  number={},

  pages={4692-4702},

  doi={10.1109/CVPR52733.2024.00449}}

@ARTICLE{10950450,

  author={Zhou, Xiaochen and Li, Bosheng and Benes, Bedrich and Habib, Ayman and Fei, Songlin and Shao, Jinyuan and Pirk, Sören},

  journal={IEEE Transactions on Geoscience and Remote Sensing}, 

  title={TreeStructor: Forest Reconstruction With Neural Ranking}, 

  year={2025},

  volume={63},

  number={},

  pages={1-19},

  doi={10.1109/TGRS.2025.3558312}}

@misc{yu2023inpaint,
      title={Inpaint {A}nything: {S}egment {A}nything {M}eets {I}mage {I}npainting}, 
      author={Tao Yu and Runseng Feng and Ruoyu Feng and Jinming Liu and Xin Jin and Wenjun Zeng and Zhibo Chen},
      year={2023},
      journal={arXiv preprint arXiv:2304.06790},
      doi = {10.48550/arXiv.2304.06790},
    eprint={2304.06790},
}

@inproceedings{du_modeling_2024,
	address = {Seattle, WA, USA},
	title = {Modeling {Detailed} {Human} {Geometry} with {Adaptive} {Local} {Refinement}},
	copyright = {https://doi.org/10.15223/policy-029},
	isbn = {9798350365474},
	url = {https://ieeexplore.ieee.org/document/10678025/},
	doi = {10.1109/CVPRW63382.2024.00571},
	language = {en},
	urldate = {2025-08-20},
	booktitle = {2024 {IEEE}/{CVF} {Conference} on {Computer} {Vision} and {Pattern} {Recognition} {Workshops} ({CVPRW})},
	publisher = {IEEE},
	author = {Du, Bang and Chen, Kunyao and Zhang, Haochen and Yin, Fei and Wu, Baichuan and Nguyen, Truong},
	month = jun,
	year = {2024},
	pages = {5620--5630},
}

@misc{wang2024occgengenerativemultimodal3d,
      title={OccGen: Generative Multi-modal 3D Occupancy Prediction for Autonomous Driving}, 
      author={Guoqing Wang and Zhongdao Wang and Pin Tang and Jilai Zheng and Xiangxuan Ren and Bailan Feng and Chao Ma},
      year={2024},
      eprint={2404.15014},
      archivePrefix={arXiv},
      primaryClass={cs.CV},
      url={https://arxiv.org/abs/2404.15014}, 
}

@InProceedings{10.1007/978-3-030-11680-4_31,
author="Morales, Giorgio
and Huam{\'a}n, Samuel G.
and Telles, Joel",
editor="Lossio-Ventura, Juan Antonio
and Mu{\~{n}}ante, Denisse
and Alatrista-Salas, Hugo",
title="Shadow Removal in High-Resolution Satellite Images Using Conditional Generative Adversarial Networks",
booktitle="Information Management and Big Data",
year="2019",
publisher="Springer International Publishing",
address="Cham",
pages="328--340",
isbn="978-3-030-11680-4"
}

@misc{dpms,
      title={Diffusion Probabilistic Models for 3D Point Cloud Generation}, 
      author={Shitong Luo and Wei Hu},
      year={2021},
      eprint={2103.01458},
      archivePrefix={arXiv},
      primaryClass={cs.CV},
      url={https://arxiv.org/abs/2103.01458}, 
}

@inproceedings{tmnet,
  title={Deep Mesh Reconstruction from Single RGB Images via Topology Modification Networks},
  author={Pan, Junyi and Han, Xiaoguang and Chen, Weikai and Tang, Jiapeng and Jia, Kui},
  booktitle={Proceedings of the IEEE International Conference on Computer Vision},
  pages={9964--9973},
  year={2019}
}

@misc{pugeonet,
      title={PUGeo-Net: A Geometry-centric Network for 3D Point Cloud Upsampling}, 
      author={Yue Qian and Junhui Hou and Sam Kwong and Ying He},
      year={2020},
      eprint={2002.10277},
      archivePrefix={arXiv},
      primaryClass={cs.CV},
      url={https://arxiv.org/abs/2002.10277}, 
}

@INPROCEEDINGS{repkpu,
  author={Rong, Yi and Zhou, Haoran and Xia, Kang and Mei, Cheng and Wang, Jiahao and Lu, Tong},
  booktitle={2024 IEEE/CVF Conference on Computer Vision and Pattern Recognition (CVPR)}, 
  title={RepKPU: Point Cloud Upsampling with Kernel Point Representation and Deformation}, 
  year={2024},
  volume={},
  number={},
  pages={21050-21060},
  doi={10.1109/CVPR52733.2024.01989}}

@misc{openlrm,
      title={LRM: Large Reconstruction Model for Single Image to 3D}, 
      author={Yicong Hong and Kai Zhang and Jiuxiang Gu and Sai Bi and Yang Zhou and Difan Liu and Feng Liu and Kalyan Sunkavalli and Trung Bui and Hao Tan},
      year={2024},
      eprint={2311.04400},
      archivePrefix={arXiv},
      primaryClass={cs.CV},
      url={https://arxiv.org/abs/2311.04400}, 
}

@misc{atlasnet,
      title={{AtlasNet}: A Papier-M\^ach\'e Approach to Learning 3D Surface Generation}, 
      author={Thibault Groueix and Matthew Fisher and Vladimir G. Kim and Bryan C. Russell and Mathieu Aubry},
      year={2018},
      eprint={1802.05384},
      archivePrefix={arXiv},
      primaryClass={cs.CV},
      url={https://arxiv.org/abs/1802.05384}, 
}

@misc{IGN_LiDARHD,
  author       = {{IGN}},
  title        = {{LiDAR HD} {G}\'eoservices},
  year         = {2025},
  howpublished = {\url{https://geoservices.ign.fr/lidarhd}},
  note         = {Institut national de l'information g\'eographique et foresti\`ere (IGN), France}
}

@article{Quan31122023,
author = {Ying Quan and Mingze Li and Yuanshuo Hao and Jianyang Liu and Bin Wang},
title = {Tree species classification in a typical natural secondary forest using UAV-borne LiDAR and hyperspectral data},
journal = {GIScience \& Remote Sensing},
volume = {60},
number = {1},
pages = {2171706},
year = {2023},
publisher = {Taylor \& Francis},
doi = {10.1080/15481603.2023.2171706},
url = { 
    
        https://doi.org/10.1080/15481603.2023.2171706
    
    

},

}

@Article{rs16203836,
AUTHOR = {Huang, Yunmei and Ou, Botong and Meng, Kexin and Yang, Baijian and Carpenter, Joshua and Jung, Jinha and Fei, Songlin},
TITLE = {Tree Species Classification from {UAV} Canopy Images with Deep Learning Models},
JOURNAL = {Remote Sensing},
VOLUME = {16},
YEAR = {2024},
NUMBER = {20},
ARTICLE-NUMBER = {3836},
url = {https://www.mdpi.com/2072-4292/16/20/3836},
ISSN = {2072-4292},
DOI = {10.3390/rs16203836}
}

@article{FASSNACHT201664,
title = {Review of studies on tree species classification from remotely sensed data},
journal = {Remote Sensing of Environment},
volume = {186},
pages = {64-87},
year = {2016},
issn = {0034-4257},
doi = {https://doi.org/10.1016/j.rse.2016.08.013},
url = {https://www.sciencedirect.com/science/article/pii/S0034425716303169},
author = {Fabian Ewald Fassnacht and Hooman Latifi and Krzysztof Stereńczak and Aneta Modzelewska and Michael Lefsky and Lars T. Waser and Christoph Straub and Aniruddha Ghosh}
}

@misc{peng2020convolutionaloccupancynetworks,
      title={Convolutional Occupancy Networks}, 
      author={Songyou Peng and Michael Niemeyer and Lars Mescheder and Marc Pollefeys and Andreas Geiger},
      year={2020},
      eprint={2003.04618},
      archivePrefix={arXiv},
      primaryClass={cs.CV},
      url={https://arxiv.org/abs/2003.04618}, 
}

@inproceedings{10.1145/1278780.1278807,
author = {Bridson, Robert},
title = {Fast Poisson disk sampling in arbitrary dimensions},
year = {2007},
isbn = {9781450347266},
publisher = {Association for Computing Machinery},
address = {New York, NY, USA},
url = {https://doi.org/10.1145/1278780.1278807},
doi = {10.1145/1278780.1278807},
booktitle = {ACM SIGGRAPH 2007 Sketches},
pages = {22–es},
location = {San Diego, California},
series = {SIGGRAPH '07}
}

@article{Mirjalili_2019,
   title={Color-difference formula for evaluating color pairs with no separation: ΔENS},
   volume={36},
   ISSN={1520-8532},
   url={http://dx.doi.org/10.1364/JOSAA.36.000789},
   DOI={10.1364/josaa.36.000789},
   number={5},
   journal={Journal of the Optical Society of America A},
   publisher={Optica Publishing Group},
   author={Mirjalili, Fereshteh and Luo, Ming Ronnier and Cui, Guihua and Morovic, Jan},
   year={2019},
   month=apr, pages={789} }

@article{Luo,
author = {Luo, Ming and Cui, Guihua and Rigg, B.},
year = {2001},
month = {10},
pages = {340 - 350},
title = {The development of the CIE 2000 colour‐difference formula: CIEDE2000},
volume = {26},
journal = {Color Research \& Application},
doi = {10.1002/col.1049}
}

@misc{tancik2020fourierfeaturesletnetworks,
      title={Fourier Features Let Networks Learn High Frequency Functions in Low Dimensional Domains}, 
      author={Matthew Tancik and Pratul P. Srinivasan and Ben Mildenhall and Sara Fridovich-Keil and Nithin Raghavan and Utkarsh Singhal and Ravi Ramamoorthi and Jonathan T. Barron and Ren Ng},
      year={2020},
      eprint={2006.10739},
      archivePrefix={arXiv},
      primaryClass={cs.CV},
      url={https://arxiv.org/abs/2006.10739}, 
}

@misc{ba2016layernormalization,
      title={Layer Normalization}, 
      author={Jimmy Lei Ba and Jamie Ryan Kiros and Geoffrey E. Hinton},
      year={2016},
      eprint={1607.06450},
      archivePrefix={arXiv},
      primaryClass={stat.ML},
      url={https://arxiv.org/abs/1607.06450}, 
}

@article{dropout,
  author  = {Nitish Srivastava and Geoffrey Hinton and Alex Krizhevsky and Ilya Sutskever and Ruslan Salakhutdinov},
  title   = {Dropout: A Simple Way to Prevent Neural Networks from Overfitting},
  journal = {Journal of Machine Learning Research},
  year    = {2014},
  volume  = {15},
  number  = {56},
  pages   = {1929--1958},
  url     = {http://jmlr.org/papers/v15/srivastava14a.html}
}

@misc{ioffe2015batchnormalizationacceleratingdeep,
      title={Batch Normalization: Accelerating Deep Network Training by Reducing Internal Covariate Shift}, 
      author={Sergey Ioffe and Christian Szegedy},
      year={2015},
      eprint={1502.03167},
      archivePrefix={arXiv},
      primaryClass={cs.LG},
      url={https://arxiv.org/abs/1502.03167}, 
}

@inproceedings{Maas2013RectifierNI,
  title={Rectifier Nonlinearities Improve Neural Network Acoustic Models},
  author={Andrew L. Maas},
  year={2013},
  url={https://api.semanticscholar.org/CorpusID:16489696}
}

@misc{qi2017pointnetdeeplearningpoint,
      title={PointNet: Deep Learning on Point Sets for 3D Classification and Segmentation}, 
      author={Charles R. Qi and Hao Su and Kaichun Mo and Leonidas J. Guibas},
      year={2017},
      archivePrefix={arXiv},
      primaryClass={cs.CV},
      url={https://arxiv.org/abs/1612.00593}, 
}

@misc{eschernet,
      title={EscherNet: A Generative Model for Scalable View Synthesis}, 
      author={Xin Kong and Shikun Liu and Xiaoyang Lyu and Marwan Taher and Xiaojuan Qi and Andrew J. Davison},
      year={2024},
      archivePrefix={arXiv},
      primaryClass={cs.CV},
      url={https://arxiv.org/abs/2402.03908}, 
}

@misc{trellis,
      title={Structured 3D Latents for Scalable and Versatile 3D Generation}, 
      author={Jianfeng Xiang and Zelong Lv and Sicheng Xu and Yu Deng and Ruicheng Wang and Bowen Zhang and Dong Chen and Xin Tong and Jiaolong Yang},
      year={2025},
      archivePrefix={arXiv},
      primaryClass={cs.CV},
      url={https://arxiv.org/abs/2412.01506}, 
}

@misc{3dtopiaxl,
      title={3DTopia-XL: Scaling High-quality 3D Asset Generation via Primitive Diffusion}, 
      author={Zhaoxi Chen and Jiaxiang Tang and Yuhao Dong and Ziang Cao and Fangzhou Hong and Yushi Lan and Tengfei Wang and Haozhe Xie and Tong Wu and Shunsuke Saito and Liang Pan and Dahua Lin and Ziwei Liu},
      year={2025},
      eprint={2409.12957},
      archivePrefix={arXiv},
      primaryClass={cs.CV},
      url={https://arxiv.org/abs/2409.12957}, 
}

@article{cielab,
author = {Kim, Dong-Ho},
year = {1997},
month = {01},
pages = {},
title = {New Weighting Functions for the Modified CIELAB Colour-Difference Formulae},
volume = {9},
journal = {Textile Coloration and Finishing}
}

@misc{prolificdreamer,
      title={ProlificDreamer: High-Fidelity and Diverse Text-to-3D Generation with Variational Score Distillation}, 
      author={Zhengyi Wang and Cheng Lu and Yikai Wang and Fan Bao and Chongxuan Li and Hang Su and Jun Zhu},
      year={2023},
      archivePrefix={arXiv},
      primaryClass={cs.LG},
      url={https://arxiv.org/abs/2305.16213}, 
}

@misc{magic3d,
      title={Magic3D: High-Resolution Text-to-3D Content Creation}, 
      author={Chen-Hsuan Lin and Jun Gao and Luming Tang and Towaki Takikawa and Xiaohui Zeng and Xun Huang and Karsten Kreis and Sanja Fidler and Ming-Yu Liu and Tsung-Yi Lin},
      year={2023},
      archivePrefix={arXiv},
      primaryClass={cs.CV},
      url={https://arxiv.org/abs/2211.10440}, 
}

@misc{nerfdiff,
      title={NerfDiff: Single-image View Synthesis with NeRF-guided Distillation from 3D-aware Diffusion}, 
      author={Jiatao Gu and Alex Trevithick and Kai-En Lin and Josh Susskind and Christian Theobalt and Lingjie Liu and Ravi Ramamoorthi},
      year={2023},
      archivePrefix={arXiv},
      primaryClass={cs.CV},
      url={https://arxiv.org/abs/2302.10109}, 
}

@misc{3dscenesurvey,
      title={3D Scene Generation: A Survey}, 
      author={Beichen Wen and Haozhe Xie and Zhaoxi Chen and Fangzhou Hong and Ziwei Liu},
      year={2025},
      archivePrefix={arXiv},
      primaryClass={cs.CV},
      url={https://arxiv.org/abs/2505.05474}, 
}

@misc{zero123plus,
      title={Zero123++: a Single Image to Consistent Multi-view Diffusion Base Model}, 
      author={Ruoxi Shi and Hansheng Chen and Zhuoyang Zhang and Minghua Liu and Chao Xu and Xinyue Wei and Linghao Chen and Chong Zeng and Hao Su},
      year={2023},
      archivePrefix={arXiv},
      primaryClass={cs.CV}
}

@misc{dreamfusion,
      title={DreamFusion: Text-to-3D using 2D Diffusion}, 
      author={Ben Poole and Ajay Jain and Jonathan T. Barron and Ben Mildenhall},
      year={2022},
      archivePrefix={arXiv},
      primaryClass={cs.CV},
      url={https://arxiv.org/abs/2209.14988}, 
}

\end{document}